\documentclass[12pt]{article}
\usepackage{amsmath,amssymb,amsfonts,amsthm}
\usepackage[dvips]{graphicx} 
\title  {
        The scaling limit of the incipient infinite cluster \\ 
        in high-dimensional percolation. \\
	II. Integrated super-Brownian excursion
        }

\author {Takashi Hara\thanks{
        Department of Mathematics,
        Tokyo Institute of Technology,
        Oh-Okayama, Meguro-ku, Tokyo 152-8551,
        Japan,
        {\tt hara@ap.titech.ac.jp}.} 
        \; and 
        Gordon Slade\thanks{
        Department of Mathematics and Statistics, 
        McMaster University, 
        Hamilton, ON, Canada L8S 4K1,
        {\tt slade@math.mcmaster.ca}.
	Address after July 1, 1999:  Department of Mathematics, University
	of British Columbia, Vancouver, BC, Canada V6T 1Z2,
	{\tt slade@math.ubc.ca}.}
        }   
	     
\date{May 14, 1999}

\def\UseSection{
        \numberwithin{equation}{section}
	\theoremstyle{plain}
        \newtheorem{theorem}    {Theorem}[section]
        \DefineTheorems 
}

\def\DefineTheorems{
	\newtheorem{lemma}      [theorem] {Lemma}
	\newtheorem{prop}       [theorem] {Proposition}
	\newtheorem{cor}        [theorem] {Corollary}

	\theoremstyle{definition}
	\newtheorem{defn}       [theorem] {Definition}

	\theoremstyle{definition}

}

\UseSection  

\newcommand{\bt}   {\begin{theorem}}
\newcommand{\et}   {\end  {theorem}}
\newcommand{\bl}   {\begin{lemma}}
\newcommand{\el}   {\end  {lemma}}
\newcommand{\bp}   {\begin{prop}}
\newcommand{\ep}   {\end  {prop}}
\newcommand{\bc}   {\begin{cor}}
\newcommand{\ec}   {\end  {cor}}
\newcommand{\bd}   {\begin{defn}}
\newcommand{\ed}   {\end  {defn}}

\newcommand{\ba}   {\begin{array}}
\newcommand{\ea}   {\end  {array}}
\newcommand{\be}   {\begin{enumerate}}
\newcommand{\ee}   {\end  {enumerate}}
\newcommand{\bi}   {\begin{itemize}}
\newcommand{\ei}   {\end  {itemize}}

\def\eq#1\en{\begin{equation}#1\end{equation}}  
\def\eqsplit#1\ensplit{
	\begin{equation}\begin{split}#1\end{split}\end{equation}
	}
\def\eqalign#1\enalign{
	\begin{align}#1\end{align}
	}
\newcommand{\eqarrstar} {\begin{eqnarray*}} 
\newcommand{\enarrstar} {\end{eqnarray*}} 
\newcommand{\eqarray}   {\begin{eqnarray}} 
\newcommand{\enarray}   {\end{eqnarray}} 
\newcommand{\nnb}	{\nonumber \\} 

\newcommand{\lbeq}[1]  {\label{eq:#1}}
\newcommand{\refeq}[1] {\eqref{eq:#1}}    

\newcommand{\Cbold} {{\mathbb C}}  
  
\newcommand{\Ebold} {{\mathbb E}}

\newcommand{\Rbold} {{\mathbb R}}

\newcommand{\Zbold} {{\mathbb Z}}

\newcommand{\Dhat} {{\hat{D} }}  
\newcommand{\Fhat} {{\hat{F} }}  
 
\newcommand{\ghat}  {{ \hat{g}  }}

\newcommand{\tauhat}{{ \hat{\tau}  }}
\newcommand{\Pihat} {\mbox{${\hat{\Pi}}$}}  
\newcommand{\Rd}    {{ {\Rbold}^d}}
\newcommand{\Zd}    {{ {\Zbold}^d }}

\newcommand{\Ind}  { {\rm I} } 

\newcommand{\nin}  {{ \not\in }}

\newcommand{\expec}[1]	{\left \langle #1 \right \rangle}

\newcommand{\AND}       {\;\&\;}
\newcommand{\IN}        {\mbox{ {\rm in} }}
\newcommand{\INSIDE}    {\mbox{ occurs in }}
\newcommand{\ON}        {\mbox{ occurs on }}

\newcommand{\conn}      {\longleftrightarrow }
\newcommand{\nc}        { \conn  {\hspace{-3.0ex} /} \hspace{1.8ex}   }
\newcommand{\ct}[1]     { \stackrel{#1} {\conn} }
\newcommand{\dbc}{\Longleftrightarrow}

\renewcommand{\to}      {\rightarrow}

\newcommand{\prob}[1]   {  {  P ( #1 ) }  }

\newcommand{\event}[1] {\left \{ #1 \right \} }

\newcommand{\Ctilde}    {\tilde{C}}

\renewcommand{\proof} {\noindent {\bf Proof}. \hspace{2mm}}

\newcommand{\okone}   {o_k(1)}

\begin{document}

\maketitle

\begin{abstract}
For independent nearest-neighbour
bond percolation on $\Zd$ with $d \gg 6$,
we prove that the incipient infinite cluster's
two-point function and three-point function converge to those of 
integrated super-Brownian excursion (ISE) in the scaling limit.   
The proof is based on an extension of
the new expansion for percolation derived in a previous paper,
and involves treating the magnetic field as a complex variable.
A special case of
our result for the two-point function implies that the
probability that the cluster of the origin consists of $n$ sites,
at the critical point,
is given by a multiple of $n^{-3/2}$, plus an error term of order
$n^{-3/2-\epsilon}$ with $\epsilon >0$.  This is a strong statement
that the critical exponent $\delta$ is given by $\delta =2$.
\end{abstract}

\tableofcontents

\section{Introduction}

\subsection{The incipient infinite cluster}

We consider independent nearest-neighbour bond percolation on $\Zd$.
For $x = (x_1,\ldots , x_d) \in \Zd$, we write $\| x\|_1 = \sum_{j=1}^d |x_j|$.
A nearest-neighbour {\em bond}\/ is a pair
$\{x,y\}$ of sites in $\Zd$ with $\| x-y \|_1 =1$.  To each bond,
we associate an independent 
Bernoulli random variable $n_{\{x,y\}}$, which takes  
the value 1 with probability $p$, and the value
0 with probability $1-p$.  A bond $\{x,y\}$ 
is said to be {\em occupied}\/ if $n_{\{x,y\}}=1$, and {\em vacant}\/ if
$n_{\{x,y\}}=0$. We say that sites $u,v \in \Zd$ are {\em connected}\/
if there is a lattice path from $u$ to $v$ consisting of occupied bonds.  
In this system, a phase transition occurs, for $d \geq 2$, 
in the sense that there is a critical value $p_c=p_c(d) \in (0,1)$,
such that for $p<p_c$ there is almost surely no infinite connected
cluster of occupied bonds, whereas for $p>p_c$ there is almost surely
a unique infinite connected cluster of occupied bonds (percolation occurs).

It is widely believed that there is no infinite cluster when $p=p_c$,
but, more than 
four decades after the mathematical study of percolation was initiated
by Broadbent and Hammersley \cite{BH57}, 
and after considerable study 
\cite{Grim89,Grim97,Hugh96,Kest82},
there is still no general proof that this is the case.
The absence of an infinite cluster at $p_c$ has been proved
only for $d=2$ (see \cite{Grim89} and references therein)
and, in high dimensions, for $d \geq 19$, 
and for $d>6$ for sufficiently ``spread-out'' models having long finite range
\cite{AN84,HS90a,HS94,Newm86}.  
We focus in this paper on the 
high-dimensional case, where the absence of percolation at $p_c$ has been
established.  This presents
a picture where at $p_c$ there are extensive connections
present, on all length scales, but no infinite cluster.  However, the
slightest increase in $p$ will lead to the formation of an infinite
cluster.  This inchoate state of affairs at $p_c$ is often represented
by an appeal to the notion of the ``incipient infinite cluster.''

The incipient infinite cluster is a concept admitting various
interpretations.  A construction of the incipient infinite cluster as 
an infinite cluster
in the 2-dimensional lattice ${\mathbb Z}^2$ has been 
carried out by Kesten \cite{Kest86},
and, for an inhomogeneous 2-dimensional model, by Chayes, Chayes
and Durrett \cite{CCD87}.  
Such constructions are necessarily singular with
respect to the original percolation model, which has no infinite cluster
at $p_c$.  Our point of view is to regard the incipient infinite cluster
as a cluster in $\Rd$ arising in the scaling limit.  More precisely, at $p_c$
we condition the size of the cluster of the origin to be $n$,
scale space by $n^{-1/4}$ (for $d>6$), and examine the
cluster in the limit $n \to \infty$.  

In this paper, we obtain strong evidence for our conjecture
that this scaling limit is integrated super-Brownian excursion
(ISE) for $d>6$.  
Our conjecture has been discussed in \cite{HS98a}
(see also \cite{DS97,Slad99}).
ISE can be regarded as the law
of a random probability measure on $\Rd$, which is almost surely
supported on a 
compact subset of $\Rd$.  On the scale of the lattice,
this compact set
corresponds to an infinite cluster, and we therefore regard the limiting
object as the scaling limit of the incipient infinite cluster.

In addition to providing the law of a random probability
measure on $\Rd$, ISE contains
more detailed information including the structure of all paths joining
pairs of points in the cluster.  This is consistent with the interesting
recent approach of \cite{Aize96,Aize97,Aize97a,AB98} 
defining the scaling limit in terms of a collection of continuous paths.
In their work, the continuous paths correspond to the occupied paths
within all clusters within a large box.  Another approach to the
incipient infinite cluster is to study the largest clusters present within
a large lattice box without taking a scaling limit, as in the work
of \cite{Aize97,BCKS98a}.  In constrast, 
our focus here is on a single
percolation cluster, rather than on many clusters.  

In general dimensions, the
appropriate spatial scaling of the lattice is presumably $n^{-1/D_H}$,
where $D_H$ is the Hausdorff dimension of the incipient infinite cluster.
We will scale space by $n^{-1/4}$ in high dimensions, consistent
with the belief that the Hausdorff dimension of the incipient infinite
cluster is 4 for $d>6$.  
The {\em upper critical dimension}\/ 6 was
identified twenty-five years ago by Toulouse \cite{Toul74}
as the dimension above which the behaviour of
percolation models near $p_c$ should no longer exhibit the dimension-dependence
typical of lower dimensions, adopting instead
the behaviour associated with percolation on trees.   

ISE is defined by conditioning super-Brownian motion to have total mass 1.
Super-Brownian motion is a basic example in the theory of superprocesses,
modelling a branching Brownian motion in which
branching occurs on all (arbitrarily short) length scales.
Discussion of ISE can be found in  
\cite{Aldo93,BCHS99,DP97,LeGa99b,LeGa99a}.
ISE is almost surely supported on a set of
Hausdorff dimension $4$, for dimensions $d \geq 4$, consistent
with $D_H=4$ for $d>6$.

Our results concern scaling limits
of the two-point and three-point functions, at the critical point.  
Fix $p=p_c$ and $x,y \in \Rd$.  
We prove that in sufficiently high dimensions, the probability that
a site $\lfloor xn^{1/4} \rfloor \in \Zd$ 
is connected to the origin, conditional on the cluster 
size being $n$, corresponds, in the scaling limit, to the mean mass density
function of ISE at $x$.  
This will be stated more precisely in Theorem~\ref{thm-2pt} below.
An immediate consequence is that the probability at $p_c$ that the cluster
of the origin consists of $n$ sites is given by a constant multiple
of $n^{-3/2}$, 
plus an error $O(n^{-3/2-\epsilon})$ with $\epsilon >0$.  This probability
is believed in general to behave as $n^{-1-1/\delta}$, so we have a proof that
$\delta =2$ in high dimensions.

We also prove that in sufficiently high dimensions, 
the probability that the origin is connected to 
sites $\lfloor xn^{1/4} \rfloor$ and 
$\lfloor yn^{1/4} \rfloor$, conditional on the cluster having 
size $n$, corresponds, 
in the scaling limit, to the joint mean mass density function for 
ISE at $x,y$.  A precise statement will be given
in Theorem~\ref{thm-3pt} below.

The proof of these results is based on the fact that two of the
standard critical exponents for percolation, $\eta$ and $\delta$,
{\em jointly}\/ take their mean-field values $\eta =0$ and $\delta =2$
in high dimensions.  Such joint behaviour was proven in 
a previous paper \cite{HS98b},
which we will refer to as I.  We will prove a stronger statement
concerning this joint behaviour, for the
nearest-neighbour model, in this paper.
The connection of ISE as a scaling limit with these mean-field values for
the critical exponents indicates a universal aspect to the
occurrence of ISE as scaling limit.  This connection is discussed
in more detail in \cite{DS97,Slad99}.

The upper critical dimension arises in this work as the dimension above
which there is generically no intersection between a 4-dimensional ISE
cluster and a 2-dimensional Brownian ``backbone.''  This allows for
an understanding of the critical dimension 6 as arising as $4+2$.
The interplay 
between the backbone and cluster gives rise to triangle diagrams
in bounds, as first observed in \cite{AN84}.  
This is in contrast to
the situation for lattice trees, where the critical dimension 8
can be understood as $4+4$, corresponding to the dimension above which
there is generically no intersection of two 4-dimensional clusters.
For lattice trees, square diagrams arise instead of triangle diagrams.
It was shown in I that square diagrams also can arise for percolation,
but they occur in conjunction with factors of the magnetization in
a manner consistent with the upper critical dimension being 6 rather than 8.

Our results for the two- and three-point functions
are restricted to sufficiently high dimensions
(we have not computed {\em how}\/ high is sufficient), rather than to
$d>6$, in part because
we use an expansion method for which the inverse
dimension serves as a small parameter ensuring convergence.
There is an alternate small parameter that has been used in lace expansion
methods in the past, which removes the need for the spatial dimension to
serve also as a small parameter, and allows for results in all
dimensions $d>6$.  This involves the introduction of spread-out models,
in which the nearest-neighbour bonds used above are enriched to a set of
bonds of the form $\{x,y\}$ with $0<\|x-y\|_\infty \leq L$.
The parameter $L$ is large, and $L^{-1}$  
serves as a small parameter to make the lace expansion converge.
The conventional wisdom, and an assertion of the hypothesis of universality,
is that in any
dimension $d$ the spread-out models have identical critical behaviour
for all finite $L \geq 1$, and for any choice of norm which respects the lattice
symmetries.  

At present, our method is not adequate to prove that the scaling limit
of the probability of a connection of two points, or three points, is
the corresponding ISE density for sufficiently spread-out models in all
dimensions $d>6$.  This is due to a difficulty related to
the occurrence of square diagrams mentioned above, and discussed 
further in Section~\ref{sec-discussion}.
This difficulty prevents us from handling dimensions 
above but near 6 in such detail.  Nevertheless,
the results of I do suggest that ISE occurs as 
the scaling limit of the incipient infinite cluster,
for sufficiently spread-out models in all dimensions $d>6$, and we regard
this difficulty as being of technical, rather than physical, origin.

It is interesting to compare our results with
those of Aizenman \cite{Aize97} for $d>6$ 
(for related work in the physics literature, see
\cite{AGK84,AGNW84,Coni85}).
Aizenman's results are based on
the assumption that at $p_c$ the probability that $0$ and $x$ are 
connected is comparable to $|x|^{2-d}$.  This is a plausible statement that
the critical exponent $\eta$ is equal to zero, but it remains unproved
in this form, and requires more
than the results for the Fourier transform of the two-point function
obtained in \cite{HS90a} and improved in this paper.
(We intend to return to this issue in a future publication.)  
Given the assumption, Aizenman proves that for percolation
on a lattice with $d>6$ and with small spacing $a$,
in a window of fixed size in the continuum, the largest clusters
have size of order $a^{-4}$, and there are on the order
of $a^{6-d}$ of these maximal clusters.  
Our results suggest that, 
for $d>6$,
such a cluster of size $n=a^{-4}$ in a lattice with spacing
$a=n^{-1/4}$ will typically be an ISE cluster, in the limit $n \to \infty$.  

Our method of proof involves an extension of the expansion methods
derived in I.  As in I, a double expansion
will be used.  Our analysis is based in part on the corresponding analysis
for lattice trees, for which a double lace expansion was performed in
\cite{HS92c}, and 
for which a proof that the scaling limit is ISE in high dimensions
was given in \cite{DS97,DS98}.
We will also make use
of the infrared bound proved in 
\cite{HS90a}, and of its consequence that,
for example, the triangle condition of 
\cite{AN84} holds in high dimensions.

It would be of interest to extend the methods of Nguyen and Yang 
\cite{NY93,NY95} to draw connections between the scaling limit of
critical oriented percolation and super-Brownian motion, above the
upper critical dimension $d+1=5$.  There is work
in progress by Derbez, van der Hofstad and Slade on this problem.
Recent work of Durrett and Perkins \cite{DP98}, reviewed in
\cite{CDP99}, proves 
convergence of the critical contact process to super-Brownian motion
for all dimensions $d \geq 2$,
in the limit of an infinite range interaction.  This is a mean-field
limit, in contrast to finite-range oriented percolation, for which
mean-field behaviour is expected to hold only for $d+1 >5$.

The results obtained in this paper were announced in 
\cite{HS98a}.

Throughout this paper, we will use, for example, (I.1.1) to denote
Equation~(1.1) of I.

\subsection{Main results}
\label{sec-results}

Consider independent nearest-neighbour bond percolation  
on $\Zd$ with bond density $p$.  
Let $C(0)$ denote the random set of sites connected to $0$,
and let $|C(0)|$ denote its cardinality.  Let
\eq
	\tau_p(0,x; n) = P_{p}\left( C(0) \ni x , |C(0)|=n \right)
\en 
denote the probability  
that the origin is connected to $x$ by a cluster containing $n$ sites.
Then
\eq
	q_n(x) = \frac{\tau_p(0,x; n)}{\sum_x \tau_p(0,x; n)}
	= \frac{\tau_p(0,x; n)}{n P_{p}( |C(0)|=n ) }
	= \frac{1}{n} P_{p}\left( C(0) \ni x  | \, |C(0)|=n \right)
\en 
defines a $p$-dependent
probability measure on $\Zd$.
We will work with Fourier transforms, and for a summable function
$f$ on $\Zd$ define
\eq
	\hat{f}(k) = \sum_{x \in \Zd} f(x) e^{ik\cdot x}, \quad
	k \in [-\pi,\pi]^d.
\en

For $k \in \Rd$, define
\eq
\lbeq{Ak}
        \hat{A}^{(2)}(k) = \int_0^\infty t e^{-t^2/2} e^{-k^2 t/2} dt .
\en 
This is the Fourier integral transform 
$\hat{A}^{(2)}(k) = \int_\Rd A^{(2)}(x)e^{ik\cdot x}d^dx$
of the mean mass density function
\eq
	A^{(2)}(x) 
	= \int_0^\infty t e^{-t^2/2} (2\pi t)^{-d/2} e^{-x^2/2t} dt ,
	\quad x \in \Rd
\en 
for ISE.  Aspects of this formula are discussed
in \cite{Aldo93,DS97,Aldo93a,LeGa93,Slad99}.
The following theorem shows that in the scaling limit, the 
Fourier tranform of the two-point function
of the incipient infinite cluster converges to the Fourier
transform of the ISE two-point function, in
sufficiently high dimensions.
The scaling of $k$ in the theorem
corresponds to scaling the lattice spacing by a multiple of $n^{-1/4}$.

\begin{theorem}
\label{thm-2pt}
Let $p=p_c$ and $k \in \Rd$.  Fix any $\epsilon < \frac{1}{2}$.
For $d$ sufficiently large,
there are positive constants $C,D$ (depending on $d$) such that
\eq
	\hat{\tau}_{p_c}(kD^{-1}n^{-1/4};n) = 
	\frac{C}{\sqrt{8\pi n}} \hat{A}^{(2)}(k)[1+O(n^{-\epsilon})].
\en 
In particular, 
\eq 
\lbeq{delta2}
	P_{p_c}(|C(0)|=n) 
	= \frac{1}{n} \hat{\tau}_{p_c}(0;n)
	= \frac{C}{\sqrt{8\pi}} \frac{1}{n^{3/2}}
	[1+O(n^{-\epsilon})]
\en 
and 
\eq 
\lbeq{qn2}
	\lim_{n \to \infty} \hat{q}_n(kD^{-1}n^{-1/4}) = \hat{A}^{(2)}(k).
\en 
\end{theorem}

Equation~\refeq{delta2} asserts that $\delta =2$, where $\delta$ is
the critical exponent in the conjectured relation $P_{p_c}(|C(0)|=n) \approx
n^{-1-1/\delta}$.  A weaker statement that $\delta =2$ in high dimensions,
as an asymptotic statement for a generating function, was proved in I.
Prior to this, a weaker statement involving upper and lower bounds on
the generating function was obtained in
\cite{BA91,HS90a}.
The convergence of Fourier tranforms in 
\refeq{qn2} is equivalent to the assertion that
the probability measure on $\Rd$ placing a point mass $q_n(x)$ at 
$xD^{-1}n^{-1/4}$, for each $x \in \Zd$, converges weakly to the
measure $A^{(2)}(x)d^dx$.

We now consider the three-point function.  Let 
\eq
	\tau^{(3)}_p(0,x,y; n) = P_p( C(0) \ni x,y, \, |C(0)|=n).
\en 
For $k,l \in [-\pi,\pi]^d$, define
\eq 
\lbeq{t3def}
        \hat{\tau}^{(3)}_p(k,l;n)
        = \sum_{x,y} \tau^{(3)}_p(0,x,y; n) e^{ik\cdot x}e^{il\cdot y}.
\en 
Observe that
\eq
\lbeq{t3hat0}
	\hat{\tau}^{(3)}_p(0,0;n)
        = \sum_{x,y}\tau_p^{(3)}(0,x,y;n) = n^2 P_p(|C(0)|=n).
\en 
We define a probability measure on ${\mathbb Z}^{d} \times {\mathbb Z}^{d}$ by
\eq 
        q_n^{(3)}(x,y) = \frac{\tau^{(3)}_p(0,x,y; n)}
        {\sum_{x,y}\tau^{(3)}_p(0,x,y; n)}
        = \frac{\tau^{(3)}_p(0,x,y; n)}
        {n^2 P_p ( |C(0)|=n ) }
        = \frac{1}{n^2} P_p (C(0) \ni x,y | \, |C(0)=n).
\en 

For $k,l \in \Rd$, let $\hat{A}^{(3)}(k,l)$
denote the Fourier transform of the ISE three-point 
function (with branch point integrated out):
\eq 
\lbeq{A3k}
        \hat{A}^{(3)}(k,l) = \int_0^\infty \int_0^\infty \int_0^\infty 
        \left( \sum_{j=1}^3 t_j \right)
         e^{-(\sum_{j=1}^3 t_j)^2/2} 
        e^{- ((k+l)^2t_1 + k^2 t_2 +l^2 t_3)/2} dt_1 \, dt_2 \, dt_3.
\en 
Aspects of this formula are discussed in 
\cite{Aldo93,Aldo93a,DS97,LeGa93,Slad99}.
This is the Fourier integral transform of 
\eq
	A^{(3)}(x,y) = \int_0^\infty \int_0^\infty \int_0^\infty \int_\Rd
        \left( \sum_{j=1}^3 t_j \right)
         e^{-(\sum_{j=1}^3 t_j)^2/2} 
        p_{t_1}(u) p_{t_2}(x-u) p_{t_3}(y-u) du \, dt_1 \, dt_2 \, dt_3 ,
\en
where $p_t(v) = (2\pi t)^{-d/2} e^{-v^2/2t}$.
The next theorem shows that in the scaling limit, the three-point function
of the incipient infinite cluster corresponds to that of ISE. 
The constants $C,D$ in the theorem are the same as those appearing in
Theorem~\ref{thm-2pt}.

\begin{theorem}
\label{thm-3pt}
Let $p=p_c$ and $k,l \in \Rd$.  Fix any $\epsilon < \frac{1}{2}$.
For $d$ sufficiently large, 
\eq 
\lbeq{tau3sim}
        \hat{\tau}^{(3)}_{p_c}(kD^{-1}n^{-1/4},lD^{-1}n^{-1/4};n)
        = 
        \frac{C}{\sqrt{8\pi}} n^{1/2}
        \hat{A}^{(3)}(k,l) [1 + O(n^{-\epsilon})].
\en 
In particular, 
\eq 
\lbeq{q3lim}
        \lim_{n \to \infty} 
        \hat{q}_n^{(3)}(kD^{-1}n^{-1/4},lD^{-1}n^{-1/4})
        = \hat{A}^{(3)}(k,l).
\en 
\end{theorem}

For $(k,l)=(0,0)$, \refeq{tau3sim} follows already from 
\refeq{delta2} and \refeq{t3hat0}, since $\hat{A}^{(3)}(0,0)=1$.
The convergence of Fourier transforms in \refeq{q3lim} is equivalent to
the assertion that the probability measure on $\Rd \times \Rd$ placing 
a point mass $q_n(x,y)$ at $(xD^{-1}n^{-1/4}, yD^{-1}n^{-1/4})$ converges
weakly to the measure $A^{(3)}(x,y)d^dxd^dy$.

We expect that Theorems~\ref{thm-2pt}--\ref{thm-3pt} should 
extend to general $m$-point functions, including all $m \geq 4$, but this
has not been proven.  This would essentially imply weak convergence
of the incipient infinite cluster to ISE. 
We now give a precise statement of our conjecture that this weak
convergence occurs for all $d>6$.  A corresponding statement has
been proved for lattice trees in high dimensions; see \cite{Slad99}.

Let $M_1(\Rd)$ denote the set of probability measures on $\Rd$, equipped
with the topology of weak convergence.  ISE can be regarded
as the law of a random
measure on $M_1(\Rd)$, {\it i.e.}, 
it is a measure $\mu_{\rm ISE}$ on $M_1(\Rd)$.
Given a site lattice animal $A$ containing $n$ sites, one of 
which is the origin, define the
probability measure $\mu_n^A \in M_1(\Rd)$ to assign mass $n^{-1}$ to
$xD^{-1} n^{-1/4}$, for each $x \in A$.  We define $\mu_n$ to be
the probability measure on $M_1(\Rd)$ which assigns probability
$P_{p_c}(C(0)=A \, | \, |C(0)|=n)$ to $\mu_n^A$, for each $A$ as above.
We then regard the limit of $\mu_n$, as $n \to \infty$, as the scaling limit
of the incipient infinite cluster.  Our conjecture is that
$\mu_n$ converges weakly to $\mu_{\rm ISE}$ for $d>6$.

\subsection{Generating functions}

The proofs of Theorems~\ref{thm-2pt} and \ref{thm-3pt} rely heavily
on generating functions, and we now describe this briefly for the
two-point function.  Define
\eq
\lbeq{Mzdef}
	\tau_{z,p}(0,x) = \sum_{n=1}^\infty \tau_p(0,x;n) z^{n},
	\quad |z| \leq 1.
\en 
The parameter $z$ is a complex variable.  
The generating function \refeq{Mzdef} converges absolutely if $|z|\leq 1$.
For $|z| <1$ and any $p \in [0,1]$, the Fourier transform $\hat{\tau}_{z,p}(k)$
exists since
\eq
\lbeq{tauhatexists}
	\sum_x \sum_{n=1}^\infty \tau_p(0,x;n) |z|^{n} = 
	\sum_{n=1}^\infty n P_p(|C(0)|=n) |z|^{n} 
	\leq \sum_{n=1}^\infty n |z|^{n}= \frac{|z|}{(1-|z|)^2} < \infty.
\en 
A similar estimate shows that the Fourier transform $\hat{\tau}_{z,p}(k)$
exists also for $|z|=1$ when $p<p_c$, using the fact that $P_p(|C(0)|=n)$
decays exponentially in the subcritical regime.

When $z \in [0,1]$, it is traditional to write $z = e^{-h}$, with $h$ playing
the role of a magnetic field, and we used $h$ as our variable in I.  
However, since we will now be allowing $z$ to be complex,
we will not adopt this notation here.  For $z \in [0,1]$, 
we introduce a probability distribution on sites by declaring 
a site to be ``green'' with probability $1-z$ and ``not green'' with
probability $z$.  These site variables are independent, and are independent
of the bond occupation variables.  
We use $G$ to denote the random set of green sites.
In this framework,
$\tau_{z,p}(0,x)$ 
is the probability that the origin is connected
to $x$ by a cluster of any finite size, but containing no green sites, i.e.,
\eq
\lbeq{taugreen}
	\tau_{z,p}(0,x) = \sum_{n=1}^\infty P_p(x \in C(0), |C(0)|=n) z^n
	= P_p(x \in C(0), C(0) \cap G = \emptyset , |C(0)|<\infty ),
\en
This well-known probabilistic
interpretation will play an important role in our analysis.

By Cauchy's theorem,
\eq
\lbeq{tauoint}
	\hat{\tau}_p(k;n) = \frac{1}{2\pi i} \oint_\Gamma \hat{\tau}_{z,p}(k)
	\frac{dz}{z^{n+1}},
\en 
where $\Gamma$ is a circle centred at the origin of any radius less than 1.
This is our basic formula for the analysis of $\hat{\tau}_{p_c}(k;n)$.
We obtain sufficient control of $\hat{\tau}_{z,p_c}(k)$ to allow for the
evaluation of the contour integral.  The leading behaviour of this
quantity, in the important limits $k \to 0$ and $z \to 1$, can be
anticipated in terms of critical exponents, as we now describe.

Assuming there is no infinite cluster at $p_c$, 
$\tau_{1,p}(0,x)$ is the probability that $0$ is connected to $x$,
for any $p \leq p_c$.
Since $\hat{\tau}_{p_c}(0;n) = nP_{p_c}(|C(0)|=n)$,
the conventional definitions 
\nocite{Grim89} of the critical exponents $\eta$ and $\delta$
(see \cite[Section~7.1]{Grim89})
suggest that
\eq
	\hat{\tau}_{1,p_c}(k) \sim \frac{c_1}{k^{2-\eta}}, \;\; \mbox{as} \;
	k \to 0; \quad \quad
	\hat{\tau}_{z,p_c}(0) \sim 
	c_2 \sum_{n=1}^\infty \frac{1}{n^{1/\delta}}z^{n}
	\sim \frac{c_3}{(1-z)^{1-1/\delta}}, \;\; 
	\mbox{as} \; z \uparrow 1.
\en 
In I, it was shown that for the nearest-neighbour model in sufficiently
high dimensions, and for the spread-out model with $d>6$ and $L$ sufficiently
large, the above relations hold {\em jointly}\/ and {\em asymptotically}\/
with the mean-field values $\eta =0$ and $\delta =2$,
in the sense that for $z \in [0,1]$ and $k \in [-\pi,\pi]^d$,
\eq
\lbeq{tauasyII}
	\hat{\tau}_{z,p_c}(k)
 	= \frac{C}{D^2 k^2 + 2^{3/2} (1-z)^{1/2} } 
	\left[ 1 + \epsilon(z,k) \right],
\en 
with
\eq
\lbeq{EbdfromI}
	|\epsilon(z,k)| \leq  \okone + o_z(1)  
\en
as $z \to 1$ and/or $k \to 0$.  Here, $\okone$ denotes a function  
of $k$ that goes to zero as $k$ approaches $0$, and $o_z(1)$ denotes
a function of {\em real}\/ $z \in [0,1)$ that goes to zero as $z \to 1$. 
This rules out the possibility of cross terms, such as $|k|(1-z)^{1/4}$,
in the leading behaviour of $\hat{\tau}_{z,p_c}(k)$.
Some such cross terms could possibly occur for $d<6$.
We rewrite \refeq{tauasyII} as
\eq
\lbeq{tauE}
	\hat{\tau}_{z,p_c}(k) 
	= \frac{C}{D^2k^{2} + 2^{3/2}(1-z)^{1/2}}
	+ E_z(k),
\en 
and improve \refeq{EbdfromI} for the nearest-neighbour model, in the
following theorem.  In both \refeq{tauE} and Theorem~\ref{thm-E}, 
the terms $k^2$ and $(1-z)^{1/2}$ should be regarded as being of roughly
the same size, as far as the critical behaviour is concerned.

\begin{theorem}
\label{thm-E}
Let $p=p_c$, and fix any $\epsilon \in (0, \frac{1}{2})$.  
For the nearest-neighbour model in sufficiently high dimensions,
\refeq{tauE} holds with
\eq
\lbeq{Ethmeq}
	\left| \frac{d}{dz}E_z(k) \right| \leq O\left( 
	\frac{1}{|1-z|^{3/2-\epsilon}} + \frac{k^2}{|1-z|^{3/2}} 
	+ \frac{k^{2+4\epsilon}}{|1-z|^2} \right)
\en
uniformly in small $k  \in [-\pi,\pi]^d$ and in complex $z$ with $|z|<1$.
The constants $C,D$ in the definition of $E_z(k)$ in \refeq{tauE} 
are those appearing in Theorem~\ref{thm-2pt}.
\end{theorem}

Integration of the bound of Theorem~\ref{thm-E} yields a bound
on $|E_z(k)|$ by $|1-z|$ times the right side of \refeq{Ethmeq}.

Define the magnetization
\eq
	M_{z,p} = P_p(C(0)\cap G \neq \emptyset) 
	= 1 - \sum_{n=1}^\infty P_p(|C(0)|=n) z^n ,
\en
and the susceptibility (the expected size of the
$G$-free cluster of the origin)
\eq
\lbeq{chidef}
	\chi_{z,p} = -z \frac{d}{dz} M_{z,p}
	= \sum_{n=1}^\infty n P_p(|C(0)|=n) z^n 
	= \sum_{n=1}^\infty n P_p(|C(0)|=n , C(0)\cap G \neq \emptyset)
	= \hat{\tau}_{z,p}(0) .
\en
Setting $k=0$ in \refeq{tauE}, it follows from Theorem~\ref{thm-E} that
\eq
\lbeq{chizpc}
	\chi_{z,p_c} = \frac{C}{2^{3/2}(1-z)^{1/2}} 
	+ O \left( \frac{1}{|1-z|^{1/2-\epsilon}}  \right)
\en
and, by integration (using $M_{1,p_c}=0$), that
\eq
\lbeq{Mzpc}
	M_{z,p_c} = 2^{-1/2} C (1-z)^{1/2} + O(|1-z|^{1/2+\epsilon}),
\en
uniformly in complex $z$ with $|z|<1$.  
The above two equations improve the asymptotic relations obtained in
I for positive $z$, not only by obtaining error estimates, but by obtaining
estimates valid for complex $z$.
Equations~\refeq{chizpc} and \refeq{Mzpc} are 
statements that the critical exponent $\delta$ in the
relation $M_{z,p_c} \approx (1-z)^{1/\delta}$ is equal to 2.  
These statements improve
the upper and lower bounds on $M_{z,p_c}$ with different constants, obtained
for nonnegative $z$ in the combined results of
\cite{BA91,HS90a}.

Write
\eq
	E_z(k) = \sum_{n=0}^\infty e_n(k)z^{n}.
\en 
To abbreviate the notation, we write
\eq
\lbeq{kappandef}
	\kappa_n = \frac{k}{Dn^{1/4}}.
\en 
We will show in Section~\ref{sec-reduction} that it follows
directly from Theorem~\ref{thm-E} that
\eq
\lbeq{enkapn}
	|e_n(\kappa_n)| \leq O(n^{-\epsilon -1/2}).
\en 
With \refeq{tauoint} and \refeq{tauE}, \refeq{enkapn} implies that
\eq
\lbeq{tau0oint}
	\hat{\tau}_p(\kappa_n;n) = \frac{C}{2\pi i} \oint_\Gamma
	\frac{1}{k^{2}n^{-1/2} + 2^{3/2}(1-z)^{1/2}}
	\frac{dz}{z^{n+1}} + O(n^{-\epsilon -1/2}).
\en 
The square root here is evaluated using the branch for which
$(1-z)^{1/2}>0$ for $z\in(-\infty,1)$.
The elementary contour integral in \refeq{tau0oint} can be 
analyzed by deforming the contour $\Gamma$ to the branch cut $[1,\infty)$.  
The asymptotic behaviour has been obtained in \cite[Lemma~1]{DS97},
and this can be extended to show that the first term on the right side
of \refeq{tau0oint} is equal to $C(8\pi n)^{-1/2}\hat{A}^{(2)}(k)
+O(n^{-3/2})$.
This proves Theorem~\ref{thm-2pt}, assuming the bound \refeq{enkapn}
on $e_n(\kappa_n)$.   

Although we have not carried out the detailed analysis, we
expect that our methods can be used to extend Theorems~\ref{thm-2pt}
and \ref{thm-3pt} to any $m$-point function, if we take $d$ sufficiently
large depending on $m$.  However, our method appears inadequate
at present to handle all $m$-point functions simultaneously in
any finite dimension.  This is connected with difficulties in
inverting generating functions.
Our reliance on complex analysis to invert generating functions
in Theorems~\ref{thm-2pt} and \ref{thm-3pt} is thus a serious hindrance.  
It may be that such difficulties can be avoided in some respects by
using the new approach to the lace expansion recently
formulated in \cite{HHS98}, which uses neither
generating functions nor complex analysis.  
However, an adaptation of this approach to percolation
would be nontrivial and would require new ideas.

\subsection{The infrared bound}

The proofs of Theorems~\ref{thm-2pt}--\ref{thm-E} rely on an expansion
whose convergence requires a small parameter.  As in \cite{HS90a},
the small parameter is defined in terms of
the critical triangle diagram, defined below.  
The triangle diagram is bounded using the infrared bound given
in the following theorem.  For the nearest-neighbour model,
this bound was obtained for dimensions $d \geq 48$
in \cite{HS90a}, but this was subsequently improved to all
dimensions $d \geq 19$ \cite{HS94}.

\begin{theorem}
\label{thm-irbd}
For $d \geq 19$, 
there is a constant $c>0$
(independent of $p,k$) such that for all $p<p_c$ and $k \in [-\pi,\pi]^d$,
\[
	0 \leq \hat{\tau}_{1,p}(k) \leq \frac{c}{k^2}.
\]
\end{theorem}

It follows from Theorem~\ref{thm-irbd} and the monotone convergence
theorem that the triangle
diagram
\eq
	\nabla(p) = \sum_{x,y} \tau_{1,p}(0,x)\tau_{1,p}(x,y)\tau_{1,p}(y,0)
	= \int_{[-\pi,\pi]^d} \hat{\tau}_{1,p}(k)^3 \frac{d^dk}{(2\pi)^d}
\en 
is finite at $p=p_c$, for $d \geq 19$.  It was shown
in \cite{HS90a}
that, moreover, $\nabla(p_c)-1$ is a small parameter for large $d$.  

A somewhat different statement of the infrared bound follows directly
from \refeq{tauasyII}, namely that for the nearest-neighbour model in
sufficiently high dimensions, or for sufficiently spread-out models
with $d>6$ and $L$ sufficiently large, there is a constant $c'>0$ such that
for $k \in [-\pi,\pi]^d$ and $z \in [0,1]$,
\eq
\lbeq{irbdz}
	\hat{\tau}_{z,p_c}(k) \leq \frac{c'}{k^2 + (1-z)^{1/2}} 
	\leq \frac{c'}{k^2}.
\en
For $z=1$, we are interpreting $\hat{\tau}_{1,p_c}(k)$ as 
the limit $\lim_{z \uparrow 1} \hat{\tau}_{z,p_c}(k)$.  
In Theorem~I.1.1, 
this limit was proven to exist,
to be finite for $k \neq 0$, and to obey
\eq
	\hat{\tau}_{1,p_c}(k) = \frac{C}{D^2 k^2} 
	[1+o_k(1)].
\en
This promotes Theorem~\ref{thm-irbd} to a statement at $p=p_c$, as opposed
to a bound uniform in $p<p_c$.
Some care is required in discussing the Fourier transform of the critical
two-point function, since it is expected that $\tau_{1,p_c}(0,x)$ 
decays like $|x|^{2-d}$ for $d>6$, hence is not summable, and hence the 
summation
defining its Fourier transform is problematic.  However, the identity
\eq
	\tau_{1,p_c}(0,x) = \lim_{z \uparrow 1}\tau_{z,p_c}(0,x)
	= \lim_{z \uparrow 1}
	\int_{[-\pi,\pi]^d}\hat{\tau}_{z,p_c}(k) 
	e^{-ik\cdot x} \frac{d^dk}{(2\pi)^d}
	=
	\int_{[-\pi,\pi]^d}\hat{\tau}_{1,p_c}(k)
	e^{-ik\cdot x}  \frac{d^dk}{(2\pi)^d},
\en
which follows from \refeq{irbdz} and the dominated convergence theorem,
justifies our interpretation of $\hat{\tau}_{1,p_c}(k)$ as the Fourier
transform of $\tau_{1,p_c}(0,x)$.

Setting $k=0$ in \refeq{irbdz} gives, for $z \in [0,1)$,
\eq
\lbeq{chizub}
	\hat{\tau}_{z,p_c}(0) = \chi_z \leq \frac{K}{(1-z)^{1/2}},
\en
where $K$ is a constant.  We will make use of the fact, proved in
Proposition~I.3.1, 
that $K$ may be taken to be independent of $d$. 
Throughout this paper, we will use $K$ to denote a generic positive
constant that is independent of $d$.  The value of $K$ may change from
line to line.  A bound of the form \refeq{chizub} was shown in 
\cite{BA91} to be a consequence of the triangle condition, and hence is valid
(at least) for $d \geq 19$, but the constant obtained in \cite{BA91} 
was not shown to be uniform in $d$.

We will also make use of the bound
\eq
\lbeq{pc}
	1 \leq 2dp_c  \leq 1 + O(d^{-1})
\en 
of \cite{HS90a}.
This was improved to $p_c = \frac{1}{2d} + (\frac{1}{2d})^2 + \frac{7}{2}
(\frac{1}{2d})^3 + O(\frac{1}{2d})^4$
in \cite{HS95}, but \refeq{pc} suffices for our
present needs.

\subsection{The backbone}
\label{sec-backbone}

The variables $t$ in \refeq{Ak} and $t_i$ in \refeq{A3k}
can be understood as time variables for Brownian paths, and it is
of interest to interpret our results in terms of a time variable.
For this, we introduce the notion of the {\em backbone}\/ of a cluster
containing two sites $x$ and $y$.  
This backbone depends on $x,y$.
We define the backbone to consist of those sites $u \in C(x)$ for
which there are disjoint self-avoiding walks 
consisting of occupied bonds from $x$ to $u$
and from $u$ to $y$.  
By {\it disjoint}, we mean paths which have no bond in common,
but they may have common sites.
The backbone can be depicted as a string of sausages
from $x$ to $y$, with all ``dangling ends'' removed.

We believe it likely that our methods can be extended and combined
with the methods of \cite{DS98}
to prove that (for high dimensions)
in a cluster of size
$n$, backbones joining sites 
$\lfloor xn^{1/4} \rfloor$ and $\lfloor yn^{1/4} \rfloor$ 
($x,y \in \Rd$)
typically consist of $O(n^{1/2})$ sites and converge in the scaling limit
to Brownian paths, with the Brownian time variable corresponding to
distance along the backbone.  But this study has not been carried out.
Such a result has been proved for high-dimensional lattice trees in
\cite[Theorem~1.2]{DS98}.  
In this interpretation, the integration
variables $t$ and $t_i$ appearing in $\hat{A}^{(2)}(k)$ and 
$\hat{A}^{(3)}(k,l)$
correspond to time intervals of backbone paths.

This concept of the backbone is relevant for an understanding of
the nature of the upper critical dimension 6.  In our expansion,
the leading behaviour corresponds to neglecting intersections between
a backbone and a percolation cluster.  Considering the backbone to
correspond to a 2-dimensional Brownian path, and the cluster to correspond
to a 4-dimensional ISE cluster, intersections will generically not occur
above $2+4=6$ dimensions.  This points to $d=6$ as the upper critical
dimension.

\subsection{Discussion of $6<d \leq 8$}
\label{sec-discussion}

As was 
mentioned above, we have not proved a version of Theorem~\ref{thm-2pt}
or \ref{thm-3pt} for sufficiently spread-out models in all dimensions
$d>6$.  Nevertheless, we believe the results remain true in this context,
and that the fact that \refeq{tauasyII} 
holds for sufficiently spread-out models
when $d>6$ provides strong evidence for this belief.
In this section, we give a heuristic discussion of 
where the method of this paper fails for $d$ near 6.  

Our method involves an expansion for the two-point function, with
terms in the expansion estimated using Feynman diagrams.
When $z=1$, all diagrams that 
occur can be bounded in terms of the triangle diagram,
as was done in \cite{HS90a}.
However, for real $z < 1$, or for complex $z$, new diagrams emerge.
This was already observed in I, where, among others, the diagram
\begin{center}
\setlength{\unitlength}{0.0005in}%
\begingroup\makeatletter\ifx\SetFigFont\undefined%
\gdef\SetFigFont#1#2#3#4#5{%
  \reset@font\fontsize{#1}{#2pt}%
  \fontfamily{#3}\fontseries{#4}\fontshape{#5}%
  \selectfont}%
\fi\endgroup%
\begin{picture}(1824,1095)(1489,-973)
\thinlines
\put(2401,-61){\line(-1, 0){900}}
\put(1501,-61){\line( 0,-1){900}}
\put(1501,-961){\line( 1, 0){900}}
\put(2401,-961){\line( 0, 1){900}}
\put(2401,-61){\line( 1, 0){900}}
\put(1501, 14){\makebox(0,0)[lb]{\smash{\SetFigFont{12}{14.4}{\rmdefault}{\mddefault}{\updefault}$0$}}}
\put(3301, 14){\makebox(0,0)[lb]{\smash{\SetFigFont{12}{14.4}{\rmdefault}{\mddefault}{\updefault}$G$}}}
\end{picture}
\end{center}
occurred.  The four lines in the square correspond to two-point functions,
while the line connecting to $G$ corresponds to a factor of the magnetization.
More precisely, the diagram represents
\eq
\lbeq{Msquare}
	M_{z,p_c} \sum_{w,x,y \in \Zd} \tau_{1,p_c}(0,w)\tau_{1,p_c}(w,x)
	\tau_{1,p_c}(x,y)\tau_{1,p_c}(y,0)
	= M_{z,p_c} \int_{[-\pi,\pi]^d}
	\frac{d^dk}{(2\pi)^d}  \left( \hat{\tau}_{1,p_c}(k) \right)^4,
\en
using the Parseval relation for the equality.  This is finite for $d>8$
by Theorem~\ref{thm-irbd} and the monotone convergence theorem.
We regard
the occurrence of the square diagram as connected with the 4-dimensional
character of the incipient infinite cluster, or, alternately, with
the fact that ISE has self-intersections for $d<8$.

If at least one line in the square diagram represented
$\tau_z(0,x)$, rather than $\tau_1(0,x)$, then \refeq{Msquare} would instead 
essentially be given by
\eq
\lbeq{P4h}
	M_{z,p_c} \int_{[-\pi,\pi]^d} \frac{d^dk}{(2\pi)^d}
	\left( \frac{1}{k^2} \right)^3 \frac{1}{k^2 + \sqrt{1-z}}.
\en
As $z \uparrow 1$, 
the integral is logarithmically divergent for $d=8$
and diverges as a multiple of $(1-z)^{(d-8)/4}$ for $d<8$.  Since 
$M_z$ vanishes like $(1-z)^{1/2}$ for $d>6$, \refeq{P4h} behaves
like $(1-z)^{(d-6)/4}$ for $6<d\leq 8$ (with a logarithmic factor
for $d=8$) and thus is harmless. 
In I, we were able to exploit this mechanism (or a related mechanism
involving the dominated convergence theorem), for real $z$, to
avoid divergent diagrams for all $d>6$.

However, for complex $z$, the probabilistic interpretation is lost
and our estimates rely more heavily on setting $z=1$ at intermediate
stages of a bound.  This destroys the above mechanism, and we are
unable to handle dimensions below 8.
In view of this inability to achieve an optimal result, and
the complicated nature of the diagrammatic estimates, we have
simplified the estimates in some places at the expense of losing
also dimensions above and near $d=8$.  For this gain in simplicity,
we lose sight of a good estimate for the dimension above which our
results could be proven for sufficiently spread-out models.   Henceforth,
we restrict attention to the nearest-neighbour model.

The above picture has an interesting parallel for the scaling behaviour
of large subcritical clusters, again heuristically.
For $p < p_c$, $P_p(|C(0)|=n)$ decays exponentially and the critical
$z_c(p)$ giving the radius of convergence of $\chi_{z,p}$ is now
strictly greater than 1.  In high dimensions,
we expect the Fourier transform of the subcritical two-point function
to behave like $(k^2 + (1-z/z_c(p))^{1/2})^{-1}$, and
the same Feynman diagrams to appear.  However, although $M_{1,p}=0$,
$M_{z_c(p),p} \neq 0$ at the critical $z$, and the mechanism outlined above
will not apply, leading to square diagrams.  In view of the fact that lace
expansion methods have been accurate predictors of the upper critical
dimension for self-avoiding walks, lattice trees and lattice animals,
and percolation, we interpret this as evidence for a Gaussian scaling
limit for subcritical percolation clusters conditioned to have size $n$,
with space scaled by $n^{-1/4}$, only for $d>8$ and not for
all $d>6$.  This is consistent
with the suggestion of \cite[page~62]{SA92}
that subcritical percolation
clusters scale like lattice animals rather than like
critical percolation clusters.

\subsection{Organization}

This paper is organized as follows.  In
Section~\ref{sec-reduction}, we show how the proof of 
Theorems~\ref{thm-2pt} and \ref{thm-E} 
can be reduced to \refeq{chizub}
together with several bounds on quantities arising in 
a double expansion.  These bounds are summarized in 
Lemmas~\ref{lem-Pidisk}--\ref{lem-Psidisk}.
Similarly, in Section~\ref{sec-3pt},
the proof of Theorem~\ref{thm-3pt} is reduced to bounds on
quantities arising in the expansions.
These bounds are summarized in Lemma~\ref{lem-Psi3disk}.
The first expansion, which is based on the expansion of I,
is described in Section~\ref{sec-expansion}. 
The necessary bounds on the first expansion are given in 
Section~\ref{sec-full.exp.bd}, where Lemmas~\ref{lem-Pidisk} and
\ref{lem-Fklb} are proved.
A second expansion is derived in Section~\ref{sec-Psidef}.
The necessary bounds on the second expansion are obtained in 
Sections~\ref{sec-Psidef2} and \ref{sec-Psider}, where Lemmas~\ref{lem-Psidisk}
and \ref{lem-Psi3disk} are proved.  

This paper can be read independently of I, apart from the fact that we
apply some methods of diagrammatic estimation  
from I in Sections~\ref{sec-full.exp.bd}, \ref{sec-Psidef2}
and \ref{sec-Psider}.

\section{Reduction of proof of Theorems~\protect\ref{thm-2pt} and 
\ref{thm-E}}
\label{sec-reduction}
\setcounter{equation}{0}

In this section, we 
fix $p=p_c$ and omit subscripts $p_c$
from the notation.  The purpose of this section is to reduce the proof
of Theorems~\ref{thm-2pt} and \ref{thm-E}
to several bounds on quantities arising
in our expansions.
These bounds are summarized in Lemmas~\ref{lem-Pidisk}--\ref{lem-Psidisk} 
below, and will
be proved in Sections~\ref{sec-full.exp.bd}, \ref{sec-Psidef2}
and \ref{sec-Psider}.  We will
also make use of the upper bound \refeq{chizub} on the susceptibility.

\subsection{Required bounds on the expansion}
\label{sec-reqbdexp}

In I, we derived two versions of an expansion, which we called the
one-$M$ and two-$M$ schemes.  The results of these expansions are 
given in (I.3.88) 
and 
(I.4.24).  In the two-$M$ scheme,
terms left unexpanded in the one-$M$ scheme were expanded.  In this paper,
we perform a more complete expansion, in which no terms are left
unexpanded.  To compare with the one- and two-$M$ schemes, the expansion 
in this paper could be called the infinite-$M$ scheme, though we will
not use this terminology.  The result of the expansion, which will
be obtained in Sections~\ref{sec-expansion} and \ref{sec-full.exp.bd},
is an identity
\eq
\lbeq{taugPix}
	\tau_z(0,x) = g_z(0,x) + \sum_{v \in \Zd} \Pi_z(0,v) \tau_z(v,x),
\en
valid for $p=p_c$ and $|z| <1$, with $g_z(0,x)$ and $\Pi_z(0,x)$
summable in $x$.  Taking the Fourier transform, and defining
$\hat{F}_z(k) = 1-\hat{\Pi}_z(k)$, this gives
\eq
\lbeq{taugPi}
	\hat{\tau}_{z}(k) = \frac{\hat{g}_{z}(k)}{1-\hat{\Pi}_{z}(k)}
	= \frac{\hat{g}_{z}(k)}{\hat{F}_{z}(k)}.
\en 
The quantities $\hat{g}_{z}(k)$ and $\hat{\Pi}_{z}(k)$ are almost identical
to each other
and will be shown in Section~\ref{sec-full.exp.bd} to be analytic in $|z|<1$.
We assume in this section that these quantities obey the bounds of
the following lemma.
In the statement of the lemma, $K$ denotes a constant 
which is independent of 
$d,z,k$.  The norm appearing in the lemma is defined, for
a power series $f(z) = \sum_{n=0}^\infty a_n z^n$, by
\eq
\lbeq{fnorm}
	\| f(z) \| = \sum_{n=0}^\infty |a_n| \, |z|^n.
\en 

\begin{lemma}
\label{lem-Pidisk}
There is a dimension $d_0$ such that for $d \geq d_0$ the following hold.
The quantities $\sum_x  \| g_{z}(0,x)\|$, 
$\sum_x x^2 \| g_{z}(0,x)\|$, $\sum_x \| \Pi_{z}(0,x)\|$, 
$\sum_x x^2 \| \Pi_{z}(0,x)\|$, $\sum_x x^4 \|\Pi_{z}(0,x)\|$ 
are bounded by $K$ uniformly in $|z|<1$.
In particular, this provides an extension, by continuity, of 
each of $\hat{\Pi}_{z}(k)$, 
$\hat{g}_{z}(k)$ and $\nabla_k^2 \hat{\Pi}_{z}(k)$
to $|z| = 1$.
In addition, $\hat{g}_{1}(0) = 1+O(d^{-1})$
and $-\nabla^2_k \hat{\Pi}_{1}(0) = 1+O(d^{-1})$.
\end{lemma}

We will also make use of the following auxiliary lemma, which guarantees
that for $|z|<1$ with $|z-1| \geq a >0$,
$\hat{F}_z(k)$ is bounded away from zero 
provided $d$ is sufficiently
large (depending on $a$).  This will be improved in \refeq{Flbgood}.
For the statement of the lemma, we define
\eq
	\hat{D}(k) = \frac{1}{d}\sum_{j=1}^d \cos k_j .
\en

\begin{lemma}
\label{lem-Fklb}
There is a dimension $d_0$ such that for $d \geq d_0$, 
$|z|<1$ and $k \in [-\pi,\pi]^d$,
\eq
\lbeq{FzpDlbd}
	|\hat{F}_{z}(k)| \geq -K d^{-1} 
	+ \frac{1}{2e}{\rm Re}[1-z\hat{D}(k)].
\en 
\end{lemma}

For $0 \leq z < 1$ we define $\hat{\Psi}_{z}(k)$ and $\hat{\Gamma}_{z}(k)$ by
\eqarray
\lbeq{FPsi}
	- z\frac{d}{dz} \hat{F}_{z}(k) & = & z\frac{d}{dz} \hat{\Pi}_{z}(k) 
	= \chi_z \hat{\Psi}_{z}(k) , \\
	\lbeq{Gammadef}
	z\frac{d}{dz} \hat{g}_{z}(k) & = & \chi_z \hat{\Gamma}_{z}(k) .
\enarray
Our second expansion will allow us to prove 
in Sections~\ref{sec-Psidef2}--\ref{sec-Psider} that these new quantities
are also analytic in $|z|<1$, for $d$ sufficiently large, and that
they obey the bounds given in the following lemma.

\begin{lemma}
\label{lem-Psidisk}
There is a dimension $d_0$ such that for $d \geq d_0$ the following hold.
\newline
(i)
The quantities
$\sum_x \| \Psi_{z}(x)\|$ and $\sum_x x^2 \| \Psi_{z}(x)\|$ 
are bounded by $K$ uniformly in $|z|<1$.
This provides an extension, by continuity, of $\hat{\Psi}_{z}(k)$ to
$|z|=1$.
Each of the above statements involving $\Psi$ is also true for $\Gamma$.
Also, $\hat{\Psi}_{1}(0) \geq K^{-1}$.

\noindent   (ii)
For all $|z|<1$ and $k \in [-\pi,\pi]^d$,
\eq
\lbeq{Psi1bd}
	\left\| \frac{d}{dz} \hat{\Psi}_{z}(k) \right\|
	\leq K \chi_{|z|}.
\en 
A similar bound holds also for $\Gamma$.
\end{lemma}

We will show in Section~\ref{sec-pfassl} that Lemmas~\ref{lem-Pidisk},
\ref{lem-Fklb}
and \ref{lem-Psidisk} are sufficient for the proof of Theorems~\ref{thm-2pt}
and \ref{thm-E}. 
In doing so, we will make use of the general power series method of
Section~\ref{sec-psm}.

We note the following consequence of Lemma~\ref{lem-Pidisk}
for future use.
Since $\hat{\tau}_1(0) = \chi_1$ is the expected size of the
connected cluster of the origin at the critical point, it is
infinite.  Since $\hat{g}_1(0) \in (0,\infty)$, it follows from 
\refeq{taugPi} that
\eq
\lbeq{F100}
	\hat{F}_1(0) = 1 - \hat{\Pi}_1(0) =  0.
\en

By Lemmas~\ref{lem-Pidisk} and \ref{lem-Psidisk}, we can define the constants:
\eqarray
	B_1  & = & - \nabla_k^2 \hat{\Pi}_{1}(0), \\
\lbeq{B2def}
	B_2^2 & = & 2\hat{g}_{1}(0) \hat{\Psi}_{1}(0), \\ 
\lbeq{Cdef}
	C & = & \hat{g}_{1}(0) B_2^{-1} 2^{3/2}, \\
\lbeq{Ddef}
	D^2 & = & (2d)^{-1} B_1 B_2^{-1} 2^{3/2}.
\enarray
By Lemmas~\ref{lem-Pidisk} and \ref{lem-Psidisk}, 
$B_2$ is bounded away from zero uniformly in $d$.
We define error terms for the numerator and denominator of \refeq{taugPi} by
\eqarray
\lbeq{Ef}
	\hat{g}_{z}(k) & = & \hat{g}_{1} (0) + E_g(z,k), \\
	\hat{F}_{z}(k) & = & B_1 \frac{k^2}{2d} + B_2 \sqrt{1-z} 
				+ E_F(z,k) 	\nonumber \\
\lbeq{EF}
	& \equiv & \hat{F}^{(0)}_{z}(k) + E_F(z,k).
\enarray
This gives \refeq{tauE}, namely
\eq
	\hat{\tau}_{z}(k) 
	= \frac{C}{D^2k^{2} + 2^{3/2}(1-z)^{1/2}}
	+ E_z(k),
\en
with $C$ and $D$ as in \refeq{Cdef}--\refeq{Ddef} and with
\eq
\lbeq{Eh}
	E_{z}(k) = \frac{E_g(z,k)}{\hat{F}_{z}(k)}
	- \frac{\hat{g}_{1}(0)E_F(z,k)}
	       {\hat{F}_{z}(k)\hat{F}^{(0)}_{z}(k)}.
\en 
We absorb the constant $\hat{g}_{1}(0)$ into $E_F$; this has no
effect on bounds.

We will show in Section~\ref{sec-Pidiski} that 
$\hat{g}_1(k) = \hat{\phi}_{h=0}(k)$ and 
$\hat{\Pi}_1(k) = \hat{\Phi}_{h=0}(k)$, 
where $\hat{\phi}_{h}(k)$ and $\hat{\Phi}_{h}(k)$ are the functions
occuring in the one-$M$ scheme in (I.3.88).  
This will follow
easily from the fact that the additional terms expanded beyond the
one-$M$ scheme all vanish when $z=1$.  
Therefore $\nabla_k^2 \hat{\Pi}_1(0) = \nabla_k^2 \hat{\Phi}_{h=0}(0)$.
Also, we will show in Section~\ref{sec-Psidef2} that
$\hat{\Psi}_1(0) = K_1 + K_2$, where $K_1$ and $K_2$ are the constants of
Propositions~I.5.1 
and I.5.2. 
In view of (I.5.3) 
and 
(I.5.10), 
the constants $C$ and $D^2$ of \refeq{Cdef}--\refeq{Ddef} 
are therefore the same as the constants
$C$ and $D^2$ appearing in the nearest-neighbour version
of Theorem~I.1.1.  

\subsection{A power series method}
\label{sec-psm}

As was argued at \refeq{enkapn}, to
prove Theorem~\ref{thm-2pt} it is sufficient to show that 
for any $\epsilon \in (0,\frac{1}{2})$ the coefficients
of the power series $E_{z}(k) = \sum_{n=0}^\infty e_n(k) z^n$ obey
\eq
\lbeq{ebd}
	|e_n(\kappa_n)| \leq O(n^{-1/2 - \epsilon}).
\en 
Lemma~\ref{lem-RLbound} below gives a general method which allows for the
transfer of bounds on a generating function to bounds on its coefficients.
This lemma, which is a special case of 
\cite[Lemma~3.2]{DS98}, incorporates improvements
found in \cite[Theorem~4]{FO90} 
to \cite[Lemma~6.3.3]{MS93}.
It will be our main tool in converting bounds on 
error terms, such as $E_z(k)$, into bounds on their coefficients of $z^n$.
The intuition behind the lemma is that if a power
series $f(z) = \sum_{n=0}^\infty a_n z^n$ has radius of convergence
$R>0$ and if $|f(z)|$ is bounded above by a multiple of $|R-z|^{-b}$
on the disk of radius $R$, with $b \geq 1$,
then $a_n$ should be on the order of $R^{-n}n^{b-1}$.

\begin{lemma}
\label{lem-RLbound}
(i) Let $f(z)=\sum_{n=0}^\infty a_n z^n$ have radius of convergence at least 
$R$, where $R>0$.  Suppose
that $|f(z)| \leq \mbox{const.} |1-z/R|^{-b}$ for $|z| < R$,
for some $b \geq 1$.
Then $|a_n | \leq \mbox{const.} R^{-n}n^{b  -1}$ if $b>1$
and $|a_n | \leq \mbox{const.} R^{-n}\log n$ if $b=1$, 
with the constants independent of $n$.

\smallskip \noindent
(ii) Let $j$ be a positive integer.  Suppose that 
$|\frac{d^j}{dz^j}f(z)| \leq \mbox{const.}|1-z/R|^{-b}$ for $|z| < R$, 
for some $b \geq 1$.
Then $|a_n| \leq \mbox{const.} R^{-n}n^{b-1 -j}$ if $b>1$
and $|a_n | \leq \mbox{const.} R^{-n}n^{-j}\log n$ if $b=1$.
\end{lemma}

In our applications of Lemma~\ref{lem-RLbound}, the hypotheses of the 
lemma will be supplied with the help of 
``fractional derivatives,'' which we now discuss.
Given $\epsilon > 0$, we define the $\epsilon^{{\rm th}}$ 
(fractional) derivative of $f$ by
\eq
	\delta_z^\epsilon f(z) = \sum_{n=1}^\infty n^\epsilon a_n z^n.
\en 
Note that for $\epsilon$ a positive integer, this gives 
$\left( z \frac{d}{dz} \right)^\epsilon$ rather than the
usual derivative.
The following is a restatement of \cite[Lemma~6.3.2]{MS93}.
The norm in the lemma is given by \refeq{fnorm}.

\begin{lemma}
\label{lem-Taylepsilon}
Let $\epsilon \in (0,1)$, $f (z) = \sum_{n=0}^\infty a_n z^n$, and 
$R >0$.  If $\| \delta_z^\epsilon f(R) \| < \infty$ (so in
particular $f(z)$ converges absolutely for $|z|\leq R$),
then for any $z$ with $|z| \leq R$,
\eq
\lbeq{Taylep0}
	|f(z) - f(R)  | \leq
	2^{1-\epsilon} \| \delta_z^\epsilon f(R) \| \, |1-z/R|^{\epsilon}.
\en 
\end{lemma}

The following lemma shows how bounds on the derivative of $f$, on the positive
axis, can be combined with Lemma~\ref{lem-Taylepsilon} to produce
bounds on $f$ in a complex disk.

\begin{lemma}
\label{lem-fd2step}
Let $\epsilon \in (0,1)$.
Suppose that $f(z) = \sum_{n=0}^\infty a_n z^n$ has radius of convergence
$R>0$, and that 
\eq
\lbeq{fdhyp}
	\| f'(z)\| \leq M_1 (1-z/R)^{\epsilon-1} 
	\quad \mbox{for $0 \leq z < R$.}
\en 
Then for any $\alpha \in (0, \epsilon)$, there is a constant $M_2$ (depending
only on $M_1, \alpha, \epsilon$), such that
\eq
\lbeq{fdcon}
	|f(z)-f(R)| \leq M_2 R |1-z/R|^{\alpha} \quad \mbox{for $|z| \leq R$.}
\en 
\end{lemma}

\proof
Fix $\alpha \in (0,\epsilon)$ and $z \in (0,R)$.
By \cite[Lemma~6.3.1]{MS93},
\eq
	\delta_z^\alpha f(z) = \frac{1}{\Gamma(2-\alpha)}  \int_0^\infty
	f'(z e^{-\lambda^{1/(1-\alpha)}})
	z e^{-\lambda^{1/(1-\alpha)}} d\lambda,
\en 
where $\Gamma$ denotes the gamma function.
Applying \refeq{fdhyp}, this gives
\eqarray
	\| \delta_z^\alpha f(z) \| 
	& \leq & \frac{|z|}{\Gamma(2-\alpha)} \int_0^\infty 
	\| f'(z e^{-\lambda^{1/(1-\alpha)}}) \|
	e^{-\lambda^{1/(1-\alpha)}} d\lambda 
	\nonumber \\
	& \leq & \frac{M_1 |z|}{\Gamma(2-\alpha)} \int_0^\infty 
	\frac{1}{|1-zR^{-1} e^{-\lambda^{1/(1-\alpha)}}|^{1-\epsilon}} 
	e^{-\lambda^{1/(1-\alpha)}} d\lambda  .
\enarray
Letting $z \to R$ gives
\eq
	\| \delta_z^\alpha f(R) \| 
	\leq \frac{M_1 R}{\Gamma(2-\alpha)} \int_0^\infty 
	\frac{1}{|1- e^{-\lambda^{1/(1-\alpha)}}|^{1-\epsilon}} 
	e^{-\lambda^{1/(1-\alpha)}} d\lambda  .
\en 
The integral is finite since $\alpha \in (0, \epsilon)$.  
Now \refeq{fdcon} follows from 
Lemma~\ref{lem-Taylepsilon}.
\qed

\subsection{Proof of Theorems~\protect\ref{thm-2pt} and 
\protect\ref{thm-E} assuming 
Lemmas~\protect\ref{lem-Pidisk}--\protect\ref{lem-Psidisk}}
\label{sec-pfassl}

In this section, we prove Theorems~\ref{thm-2pt} and \ref{thm-E},
assuming Lemmas~\ref{lem-Pidisk}--\ref{lem-Psidisk}.

To prove Theorem~\ref{thm-2pt}, it suffices to prove \refeq{ebd}.
In view of Lemma~\ref{lem-RLbound}(ii), to prove \refeq{ebd} it
suffices to show that we can bound 
$\left| \frac{d}{dz} E_{z}(\kappa) \right|$ by terms such as
$|1-z|^{\epsilon - 3/2}$ or $n^{-\epsilon}|1-z|^{-3/2}$,
uniformly in $|z| < 1$.
Denoting derivatives with respect to $z$ by primes, and omitting arguments
to simplify the notation, the derivative of \refeq{Eh} is
\eq
\lbeq{Ehder}
	E' = \frac{E_g'}{F} - \frac{E_g F'}{F^2} - \frac{E_F'}{F F^{(0)}}
	+ \frac{E_F F'}{F^2 F^{(0)}} + \frac{E_F (F^{(0)})'}{F (F^{(0)})^2}.
\en 
The right side can be bounded using the following two lemmas.
We assume in this section,
without further mention, that $d$ is sufficiently large.

\begin{lemma}
\label{lem-Flbmain}
For $k \in [-\pi, \pi]^{d}$ and ${\rm Re}(1-z) \geq 0$ (in particular,
for $|z| < 1$), we have 
\eq
	\left | \hat{F}^{(0)}_{z}(k) \right | 
	= \left| B_{1} \frac{k^{2}}{2d} +  B_2(1-z)^{1/2} \right|
	\geq 
	B_{1} \frac{k^{2}}{2d} + \frac{1}{\sqrt{2}} B_2|1-z|^{1/2} .
\en  
\end{lemma}

\proof
Write $a = B_{1} \frac{k^{2}}{2d}$ and $b = b_1+ib_2 = B_2(1-z)^{1/2}$.
Then $a \geq 0$, and since ${\rm Re}(1-z) \geq 0$, the principal value of
the argument of $(1-z)^{1/2}$ lies in $[-\frac{\pi}{4},\frac{\pi}{4}]$.
Hence $|b_2| \leq b_1$, and so $|b| \leq \sqrt{2}b_1$.
The lemma then follows from 
\eq
	|a+b| \geq {\rm Re}(a+b) = a + b_1 \geq a + \frac{1}{\sqrt{2}} |b|.
\en 
\qed

\begin{lemma}
\label{lem-Ebds}
Fix any $\epsilon \in (0, \frac{1}{2})$ and $k \in \Rd$.
There are positive constants $c$ (which may depend on $d, \epsilon$) 
and $K$ (independent of $d$)
such that for $|z| < 1$,  
\eqarray
\lbeq{EbdsEg}
	|E_g(z,\kappa_n)| & \leq & Kn^{-1/2} +K|1-z|^\epsilon, \\
\lbeq{EbdsEF}
	|E_F(z,\kappa_n)| & \leq & Kn^{-1/2-\epsilon}  
	+ K|1-z|^{\epsilon +1/2}, \\
\lbeq{EbdsF}
	|\hat{F}_z(\kappa_n)| & \geq & K^{-1} |1-z|^{1/2},  \\
\lbeq{Fprime}
	|\hat{F}'_z(\kappa_n)| & \leq & c|1-z|^{-1/2}, \\
\lbeq{EbdsEgprime}
	|E_g'(z,\kappa_n)| & \leq & 
	c|1-z|^{-1/2}, \\
\lbeq{EbdsEFprime}
	|E_F'(z,\kappa_n)| & \leq & c|1-z|^{\epsilon-1/2} 
	+cn^{-1/2}|1-z|^{-1/2}.
\enarray
The above bounds are valid for all $n$, except \refeq{EbdsF} which
is valid for $n$ sufficiently large.
\end{lemma}

Our proof of Lemma~\ref{lem-Ebds} also gives 
\refeq{EbdsEg}--\refeq{EbdsEFprime} with $\kappa_n$ replaced by 
$k \in [-\pi,\pi]^d$ on the left sides and $n^{-q}$ replaced by $k^{4q}$
on the right sides (with $k$ small for \refeq{EbdsF}).  
With Lemma~\ref{lem-Flbmain} and \refeq{Ehder}, 
this proves Theorem~\ref{thm-E}.

Using 
Lemma~\ref{lem-RLbound}(ii), it is straightforward to check that the
bounds of Lemmas~\ref{lem-Flbmain}--\ref{lem-Ebds}
are sufficient to prove \refeq{ebd}.  In the remainder of this section
we will prove Lemma~\ref{lem-Ebds}, assuming 
Lemmas~\ref{lem-Pidisk}--\ref{lem-Psidisk}. 
A basic mechanism in the proof is to use the bound \refeq{chizub}, which
asserts that
$\chi_z \leq K(1-z)^{-1/2}$ for $z \in [0,1)$, with
$K$ independent of $d$, together with Lemma~\ref{lem-fd2step}
to obtain bounds {\em valid in the disk}\/ $|z| < 1$.

The following lemma, which is an immediate consequence of \refeq{EbdsF}
and the uniform bound on $\hat{g}_z(0)$
of Lemma~\ref{lem-Pidisk}, promotes \refeq{chizub}
from a bound on $\chi_z = \hat{\tau}_z(0)$ for $z \in [0,1)$ to a bound
for all $|z|<1$.

\begin{lemma}
\label{lem-chizub}
There is a constant $K$ (independent of $d$) such that, for all
$|z|<1$,
\eq
	|\chi_z| \leq K|1-z|^{-1/2}.
\en
\end{lemma}
 
\medskip \noindent
{\bf Proof of \protect\refeq{EbdsEg}--\protect\refeq{EbdsEF}.}
We now prove the bounds \refeq{EbdsEg} and \refeq{EbdsEF} 
on the error terms defined in \refeq{Ef} and \refeq{EF}.  These error
terms can be written as
\eq
\lbeq{Efzk}
	E_g(z,k) 
	= \left[ \hat{g}_z(k) - \hat{g}_1 (k) \right] +
	\left[ \hat{g}_1(k) - \hat{g}_1 (0) \right] 
\en 
and
\eqarray
\lbeq{EFlong}
	E_F(z,k)  & = &  \left( \hat{\Pi}_z(0) - \hat{\Pi}_z(k) + 
	\frac{k^2}{2d}\nabla_k^2 \hat{\Pi}_z(0) \right)
		\nonumber \\ &&
	+ \frac{k^2}{2d} \left( 
	\nabla_k^2 \hat{\Pi}_1(0) -\nabla_k^2 \hat{\Pi}_z(0)\right)
	+ \left( \hat{F}_z(0) - B_2 \sqrt{1-z} \right)  .
\enarray
We will prove that there
is a positive constant $K$, independent of $d$, such that
for $|z| < 1$ and $k \in [-\pi,\pi]^d$, 
\eqarray
\lbeq{Efbd}
	|E_g(z, k)| & \leq & 
	K k^{2} + K |1-z|^\epsilon  , \\
\lbeq{EFzk}
	|E_F(z, k)| & \leq & K k^{2+4\epsilon} 
	+ K |1-z|^{\epsilon+1/2} . 
\enarray

We begin with \refeq{Efbd}, using \refeq{Efzk}.
By Lemma~\ref{lem-fd2step}, the first term on the right side of \refeq{Efzk}
is bounded above by $K|1-z|^\epsilon$ for $|z|<1$,  
provided that
$\| \frac{d}{dz} \hat{g}_z(k) \| \leq M_1 |1-z|^{\epsilon' -1}$ when
$z \in (0,1)$, for some $\epsilon' \in (\epsilon,1)$ and 
$M_1$ independent of $d$.  But this last bound follows from 
\refeq{chizub}, \refeq{Gammadef} and Lemma~\ref{lem-Psidisk}(i), 
with $\epsilon' = \frac{1}{2}$.
(Note that \refeq{chizub} implies $z^{-1}\chi_z = \|z^{-1}\chi_z \|
\leq K|1-z|^{-1/2}$
for $z \in [0,1)$, since $\chi_z$ is a power series with non-negative 
coefficients and has a simple zero at the origin.)

For the second term on the right side of \refeq{Efzk}, we use the
identity $\hat{g}_1(k) - \hat{g}_1 (0) = - \sum_x g_1(x)[1-\cos(k\cdot x)]$
and the fact that $|1-\cos (k\cdot x)| \leq \frac{1}{2}(k \cdot x)^2 
\leq \frac{1}{2} k^2x^2$.  This gives
\eq
	|\hat{g}_1(k) - \hat{g}_1 (0)| 
	\leq \frac{1}{2} k^2 \sum_x x^2 |g_1(x)|.
\en  
The sum on the right side is bounded by a $d$-independent constant,
by Lemma~\ref{lem-Pidisk}.  This completes the proof of \refeq{Efbd}.

We now turn to the proof of \refeq{EFzk}.  
To deal with the first term of \refeq{EFlong}, 
we use  
\eq
	\left| 1- \cos (k \cdot x) - \frac{(k \cdot x)^2}{2} \right|
	\leq \frac{1}{24} (k \cdot x)^{4} 
	\leq \frac{1}{24} k^{4} x^{4} . 
\en 
Since $\sum_x x^4 |\Pi_z(x)|$ is bounded by a $d$-independent constant
by Lemma~\ref{lem-Pidisk}, the first term of \refeq{EFlong} is bounded
above by $Kk^4$.  

We obtain a bound $K k^2 |1-z|^\epsilon$ 
for the second term of \refeq{EFlong}, 
by \refeq{chizub} and Lemmas~\ref{lem-fd2step} and \ref{lem-Pidisk}, 
since $\| \frac{d}{dz} \nabla^2
\hat{\Pi}_z(0) \|$ is bounded by $\| \nabla^2 \hat{\Psi}_z(0) \| \chi_z$
for $0<z<1$.
Now
\eq
	k^2 |1-z|^\epsilon \leq k^{2+4\epsilon} + |1-z|^{\epsilon +1/2},
\en 
which follows by considering separately the cases $k^2 \leq |1-z|^{1/2}$
and $|1-z|^{1/2} \leq k^2$.  This then gives terms of the appropriate
form for \refeq{EFzk}.

Finally, we consider the 
term $\hat{F}_z(0) - B_2 \sqrt{1-z}$ in \refeq{EFlong}.  By 
Lemma~\ref{lem-Pidisk} and \refeq{B2def}, 
$|\hat{F}_z(0) - B_2 \sqrt{1-z}|\leq K$.  Hence, for $|z| \leq 
\frac{1}{2}$, this is bounded above by $K|1-z|^{1/2+\epsilon}$.
We therefore restrict attention, in what follows, to $\frac{1}{2}<|z| <1$.
By \refeq{F100} and \refeq{FPsi}, 
\eq
\lbeq{Fh02}
	\hat{F}_z(0)^2 = \int_1^z \frac{d}{dz'} \hat{F}_{z'}(0)^2 dz'
	= \int_1^z 2 \hat{F}_{z'}(0) \frac{d}{dz'}\hat{F}_{z'}(0)  dz'
	= \int_z^1 2 \hat{g}_{z'}(0) (z')^{-1}
	\hat{\Psi}_{z'}(0)  dz' .
\en 
Together with \refeq{B2def}, this implies
\eq
\lbeq{FB22}
	\hat{F}_z(0)^2 - B_2^2 (1-z)
	= \int_z^1 2 \left( (z')^{-1} \hat{g}_{z'}(0) 
	\hat{\Psi}_{z'}(0) - \hat{g}_1(0) \hat{\Psi}_1(0) \right) dz' .
\en 
To bound this for $\frac{1}{2} \leq |z|<1$, we note that by \refeq{Efbd}, 
and by \refeq{Psi1bd} and Lemma~\ref{lem-fd2step},
\eqarray
	| \hat{g}_{z'}(0) - \hat{g}_1(0) | & \leq & K |1-z'|^\epsilon , \\
\lbeq{FB22bis}
	| \hat{\Psi}_{z'}(0) - \hat{\Psi}_1(0) | & \leq & K |1-z'|^\epsilon .
\enarray
With the uniform bounds on $\hat{g}_z(0)$ and $\hat{\Psi}_z(0)$ of
Lemmas~\ref{lem-Pidisk} and \ref{lem-Psidisk}, \refeq{FB22}--\refeq{FB22bis} 
imply that
\eq
	\hat{F}_z(0)^2 = B_2^2 (1-z) \left[ 1 + O(|1-z|^{\epsilon})\right],
\en 
with the error term bounded by a $d$-independent multiple of
$|1-z|^{\epsilon}$.  Hence,
as required,
\eq
\lbeq{FB}
	\left| \hat{F}_z(0) - B_2 \sqrt{1-z} \right|
	\leq K|1-z|^{1/2 +\epsilon}.
\en 
\qed

\medskip \noindent
{\bf Proof of \protect\refeq{EbdsF}.}
We will prove the following stronger statement than \refeq{EbdsF}:
There is a constant $K$ (independent of $d$)
such that for $|z| < 1$ and for $k^2$ sufficiently
small (depending on $d$),  
\eq
\lbeq{Flbgood}
	|\hat{F}_{z}(k)| \geq
	K^{-1}   \left ( \frac{k^{2}}{2 d} +  |1-z|^{1/2} \right ) .
\en 
To prove this, 
we consider separately the cases where $|1-z|$ is small, or not,
beginning with the former.   

\smallskip\noindent {\em Case 1: $|1-z|$ small.}\/
By \refeq{EFzk} and Lemma~\ref{lem-Flbmain},
\eqarray
	|\hat{F}_{z}(k)| & \geq & |\hat{F}^{(0)}_{z}(k)|
	- |E_F(z,k)|
	\nonumber \\ 
	& \geq &
	B_1\frac{k^{2}}{2d} + \frac{B_{2}}{\sqrt{2}}|1-z|^{1/2} 
	- K_1 k^{2 +4\epsilon} 
	- K_1 |1-z|^{\epsilon+1/2} . 
\enarray
Thus we have
\eq
	|\hat{F}_{z}(k)| \geq 
	B_1\frac{k^{2}}{4d} + \frac{B_{2}}{2}|1-z|^{1/2}
\en 
for $k^2 \leq \left[B_1(4dK_1)^{-1}\right]^{1/2\epsilon}$ and 
$|1-z| \leq \left[ B_2 K_1^{-1} 
\left( \frac{1}{\sqrt{2}} - \frac{1}{2} \right) \right]^{1/\epsilon}
\equiv \delta$.  The constant $\delta$ is bounded away from zero as
$d \to \infty$, because $B_2$ is.  Also, $B_1$ is bounded away from 
zero, by Lemma~\ref{lem-Pidisk}.

\smallskip\noindent {\em Case 2: $|1-z|$ large.}\/
Now consider the case $|1-z| \geq \delta$, so that 
$z \in W \equiv \{z \in \Cbold: |z| \leq 1, |1-z| \geq \delta\}$.  
We begin with the inequality \refeq{FzpDlbd}, which states that 
\eq
	\left| \hat{F}_{z}(k) \right| \geq
	-K d^{-1} 
	+ \frac{1}{2e}{\rm Re}[1-z\hat{D}(k)].
\en  
We further reduce our limit on $k$, if necessary, to ensure $\Dhat(k) \geq 0$.  
Then for $z \in W$,
\eq
	{\rm Re} \left [ 1 - z \Dhat(k) \right ]  
	\geq \alpha(\delta) > 0 , 
\en  
for some geometrical constant $\alpha(\delta)$ 
which is bounded away from zero as $d \to \infty$,
because $\delta$ is.
Therefore,
\eq 
	\left | \Fhat_{z} (k) \right | 
	\geq   -Kd^{-1} + \frac{1}{2e} \alpha(\delta)
	\geq \frac{1}{3e} \alpha(\delta)
\en  
for sufficiently large $d$.   
We now require that $k$ be sufficiently small that
$\frac{k^{2}}{2d}  \leq 2 - \sqrt{2}$, to ensure that 
$\frac{k^{2}}{2d} + | 1 - z |^{1/2} \leq 2$.
The lemma then follows, since  
\eq
	 \frac{1}{3e} \alpha(\delta)
	\geq \left ( \frac{k^{2}}{2d} + | 1 - z |^{1/2} 
	\right ) \frac{1}{6e} \alpha(\delta). 
\en  
\qed

\medskip \noindent
{\bf Proof of \protect\refeq{Fprime}--\protect\refeq{EbdsEFprime}.}
Having proved \refeq{EbdsF}, we have also proved Lemma~\ref{lem-chizub}.
It follows from Lemma~\ref{lem-chizub}, together with the fact
that $\chi_z$ has a simple zero at $z=0$ by its definition in
\refeq{chidef}, that 
\eq
\lbeq{zinvchi}
	|z^{-1} \chi_z| \leq c|1-z|^{-1/2}
\en 
uniformly in
$|z|<1$.  With \refeq{FPsi} and Lemma~\ref{lem-Psidisk}, 
this upper bound gives \refeq{Fprime}.
With \refeq{Gammadef} and Lemma~\ref{lem-Psidisk}, it gives the bound 
\eq
\lbeq{dg}
	\left| \frac{d}{dz} \hat{g}_z(k) \right| 
	\leq  c|1-z|^{-1/2} 
\en 
for $|z| < 1$, which implies \refeq{EbdsEgprime}.

To prove \refeq{EbdsEFprime}, we first write
\eq
\lbeq{dEF1}
	\frac{d}{dz} E_F(z,k)  =  \frac{d}{dz} \left[ \hat{\Pi}_z(0) -
	\hat{\Pi}_z(k) \right] + \frac{d}{dz} \left[ \hat{F}_z(0) -
	B_2 \sqrt{1-z}  \right].
\en 
The first term on the right side is
\eq
	z^{-1}\chi_z \left[ \hat{\Psi}_z(0) - \hat{\Psi}_z(k) \right].
\en 
By Lemma~\ref{lem-Psidisk} and \refeq{zinvchi}, 
this is $O(|1-z|^{-1/2}k^2)$.  
Also,
\eq
	\frac{d}{dz} \left[ \hat{F}_z(0) - B_2 \sqrt{1-z}  \right]
	= -z^{-1}\chi_z \hat{\Psi}_z(0)  + \frac{B_2}{2\sqrt{1-z}}.
\en 
By \refeq{Efbd}--\refeq{EFzk},
\eq
	\chi_z = \frac{\hat{g}_z(0)}{\hat{F}_z(0)} 
	= \frac{\hat{g}_1(0)+E_g(z,0)}{B_2\sqrt{1-z}+E_F(z,0)}
	= \frac{\hat{g}_1(0)}{B_2\sqrt{1-z}} + O(|1-z|^{\epsilon -1/2})
\en 
and by \refeq{Psi1bd} and Lemma~\ref{lem-fd2step},
\eq
	\hat{\Psi}_z(0) = \hat{\Psi}_1(0) +O(|1-z|^\epsilon).
\en 
Thus, by the definition of $B_2$ in \refeq{B2def}, 
the second term in \refeq{dEF1} is $O(|1-z|^{\epsilon -1/2})$.
\qed


\section{Reduction of proof of Theorem~\protect\ref{thm-3pt}}
\label{sec-3pt}
\setcounter{equation}{0}

In this section, we reduce the proof of Theorem~\ref{thm-3pt} to
Lemmas~\ref{lem-Pidisk}--\ref{lem-Psidisk}, supplemented
with the related bounds of Lemma~\ref{lem-Psi3disk} below.
The proof of Lemma~\ref{lem-Psi3disk} will be given in 
Sections~\ref{sec-Psidef2} and \ref{sec-Psider}.  

Throughout this section, we fix $p=p_c$ and drop subscripts $p_c$
from the notation.  The basic object
of study is the Fourier transform of the three-point function, defined 
for complex $z$ by
\eq 
\lbeq{t3hatdef}
        \hat{\tau}^{(3)}_z(k,l)
	= \sum_{n=1}^\infty \hat{\tau}^{(3)}(k,l;n) z^n
        = \sum_{n=1}^\infty z^n
	\sum_{x,y} \tau^{(3)}(0,x,y; n) e^{ik\cdot x}
         e^{il\cdot y} , \quad |z| < 1.
\en 
Given $k,l \in [-\pi,\pi]^d$, we will write
\eq
	k^{(1)} = k+ l, \quad k^{(2)} = k, \quad k^{(3)}=l.
\en
These variables are arranged in \refeq{t3hatdef} schematically as:
\begin{center}
\setlength{\unitlength}{0.0060000in}%
\begingroup\makeatletter\ifx\SetFigFont\undefined
\def\x#1#2#3#4#5#6#7\relax{\def\x{#1#2#3#4#5#6}}%
\expandafter\x\fmtname xxxxxx\relax \def\y{splain}%
\ifx\x\y   
\gdef\SetFigFont#1#2#3{%
  \ifnum #1<17\tiny\else \ifnum #1<20\small\else
  \ifnum #1<24\normalsize\else \ifnum #1<29\large\else
  \ifnum #1<34\Large\else \ifnum #1<41\LARGE\else
     \huge\fi\fi\fi\fi\fi\fi
  \csname #3\endcsname}%
\else
\gdef\SetFigFont#1#2#3{\begingroup
  \count@#1\relax \ifnum 25<\count@\count@25\fi
  \def\x{\endgroup\@setsize\SetFigFont{#2pt}}%
  \expandafter\x
    \csname \romannumeral\the\count@ pt\expandafter\endcsname
    \csname @\romannumeral\the\count@ pt\endcsname
  \csname #3\endcsname}%
\fi
\fi\endgroup
\begin{picture}(225,109)(105,644)
\thinlines
\put(120,660){\line( 1, 0){200}}
\put(210,660){\line( 0, 1){ 80}}
\put(105,660){\makebox(0,0)[lb]{\smash{$\scriptstyle 0$}}}
\put(330,660){\makebox(0,0)[lb]{\smash{$\scriptstyle x$}}}
\put(205,745){\makebox(0,0)[lb]{\smash{$\scriptstyle y$}}}
\put(150,665){\makebox(0,0)[lb]{\smash{$\scriptstyle k^{(1)}$}}}
\put(215,700){\makebox(0,0)[lb]{\smash{$\scriptstyle k^{(3)}$}}}
\put(260,665){\makebox(0,0)[lb]{\smash{$\scriptstyle k^{(2)}$}}}
\end{picture}
\end{center}
We will use the notation
\eq
\lbeq{kaplam}
	\kappa_n = k D^{-1}n^{-1/4}, \quad \lambda_n = l D^{-1}n^{-1/4},
	\quad \kappa_n^{(i)} = k^{(i)} D^{-1}n^{-1/4}.
\en

The proof involves showing that 
\eq
\lbeq{tau3mE1}
	\hat{\tau}^{(3)}_z(k,l)
	= 4C^{-2} \prod_{i=1}^3 \frac{C}{D^2 (k^{(i)})^2 + 2^{3/2}(1-z)^{1/2}} 
	+ E^{(3)}_z(k,l),
\en
where $E^{(3)}_z(k,l) = \sum_{n=0}^\infty e^{(3)}_n(k,l) z^n$ 
is an error term in the sense that 
\eq
\lbeq{tau3mE}
	|E^{(3)}_z(\kappa_n,\lambda_n)| \leq O \left(
	\frac{1}{|1-z|^{3/2-\epsilon}}
	+ \frac{n^{-1/2}}{|1-z|^{3/2}} 
	+ \frac{n^{-1/2-\epsilon}}{|1-z|^2}\right)
\en
for any $\epsilon \in (0,\frac{1}{2})$, uniformly in $|z|<1$.  
It then follows from
Lemma~\ref{lem-RLbound}(i) that
$|e^{(3)}_n(\kappa_n,\lambda_n)| \leq
O(n^{1/2-\epsilon})$.  The first term on the right side of \refeq{tau3mE1}
is the main term.  An elementary extension of \cite[(2.15)]{DS98}, to
include an error estimate, gives as the main term's coefficient of 
$z^n$ the value
\eq
	\frac{4C}{2\pi i} \oint_{|z|=r<1} 
	\prod_{i=1}^3 \frac{1}{D^2 (\kappa_n^{(i)})^2 + 2^{3/2}(1-z)^{1/2}}
	 \frac{dz}{z^{n+1}}
	= \frac{C}{\sqrt{8\pi}} n^{1/2} A^{(3)}(k,l) + O(n^{-1/2}).
\en
This yields Theorem~\ref{thm-3pt}.  Thus it is sufficient to obtain
\refeq{tau3mE}.  

The factorization in the main term of \refeq{tau3mE1}
is a product of the leading behaviour 
$C[D^2 (k^{(i)})^2 + 2^{3/2}(1-z)^{1/2}]^{-1}$
of three two-point functions,
multiplied by a vertex factor $4C^{-2}$.  The figure above \refeq{kaplam}
illustrates this factorization.

To prove \refeq{tau3mE}, we will make use of a generalization of
\refeq{taugPix} to site-dependent $z$-variables.  More precisely,
we associate to each site $u \in \Zd$ a variable $z_u \in [0,1]$.  
We write $\vec{z}$ for the collection of $z_u$, $u \in \Zd$.
The probability that a site $u$ is green becomes
$1-z_u$ in this setting.  Then we define
\eq
	\tau_{\vec{z}}(0,x) = P(C(0) \ni x \AND C(0) \cap G = \emptyset)
	= \langle I[C(0) \ni x] \prod_{u \in C(0)} z_u \rangle .
\en
The derivation of the expansion for this more general two-point
function is not changed in any significant way, and we
will prove in Section~\ref{sec-full.exp.bd} an identity
\eq
\lbeq{vecexp}
	\tau_{\vec{z}}(0,x) = g_{\vec{z}}(0,x) + 
	\sum_{v \in \Zd} \Pi_{\vec{z}} (0,v) \tau_{\vec{z}}(v,x)
\en
generalizing \refeq{taugPix}, when $z_u \in [0,a)$ for all $u \in \Zd$, 
for any $a\in [0,1)$.

Now we apply the operator $z_y \frac{\partial}{\partial z_y}$ to \refeq{vecexp}.
For the left side, we have
\eq
	z_y \frac{\partial}{\partial z_y} \tau_{\vec{z}}(0,x)
	= \tau_{\vec{z}}^{(3)}(0,x,y).
\en
Using this also for one contribution to the right side, we obtain
\eq
\lbeq{tauvec0xy}
	\tau_{\vec{z}}^{(3)}(0,x,y) = 
	z_y \frac{\partial}{\partial z_y} g_{\vec{z}}(0,x)  
	+  \sum_{v \in \Zd} 
	\left( z_y \frac{\partial}{\partial z_y} \Pi_{\vec{z}} (0,v) \right)
	\tau_{\vec{z}}(v,x) 
	+ \sum_{v \in \Zd} \Pi_{\vec{z}} (0,v)\tau_{\vec{z}}^{(3)}(v,x,y).
\en
For the two derivatives appearing explicitly on the right side, we will
use the following lemma.

\begin{lemma} 
\label{lem-Psi3disk}
(i)
Let $a \in [0,1)$.  
For $x,y \in \Zd$ and all $z_u \in [0,a]$, there exist 
$\Gamma^{(3)}_{\vec{z}}(0,x,v')$ and $\Psi^{(3)}_{\vec{z}}(0,v,v')$ such
that
\eq
\lbeq{Gamma3def}
	z_y \frac{\partial}{\partial z_y} g_{\vec{z}}(0,x)
	= \sum_{v'}\Gamma^{(3)}_{\vec{z}}(0,x,v') \tau_{\vec{z}}(v',y)
\en
and
\eq
\lbeq{Psi3def}
	z_y \frac{\partial}{\partial z_y} \Pi_{\vec{z}} (0,v)
	= \sum_{v'}\Psi^{(3)}_{\vec{z}}(0,v,v') \tau_{\vec{z}}(v',y).
\en
For $z_u \equiv z$, the Fourier transforms $\hat{\Gamma}^{(3)}_z(k,l)$
and $\hat{\Psi}^{(3)}_z(k,l)$ extend to complex $z$ with $|z|\leq 1$.
In this disk, $\|\hat{\Psi}^{(3)}_z(k,l) \|$ is bounded, as is the
norm of any second derivative of $\hat{\Psi}^{(3)}_z(k,l)$ with respect
to $k$ and/or $l$.  
The above bounds also apply to $\hat{\Gamma}^{(3)}$.
In addition, $\hat{\Psi}_1^{(3)}(0,0)=\hat{\Psi}_1(0)$.
\\
(ii)
For $|z|<1$ and $k,l \in [-\pi,\pi]^d$,  
$\| \frac{d}{dz} \hat{\Psi}^{(3)}_z(k,l) \| \leq \mbox{const.} \chi_{|z|}$.
The same bound is obeyed by $\| \frac{d}{dz} \hat{\Gamma}^{(3)}_z(k,l) \|$.
\end{lemma}

Substituting \refeq{Gamma3def} and \refeq{Psi3def} into \refeq{tauvec0xy} gives
\eq
	\tau_{\vec{z}}^{(3)}(0,x,y) =
	\sum_{v'}\Gamma^{(3)}_{\vec{z}}(0,x,v') \tau_{\vec{z}}(v',y)
	+ \sum_{v,v'}\Psi^{(3)}_{\vec{z}}(0,v,v') \tau_{\vec{z}}(v',y)
	\tau_{\vec{z}}(v,x)
	+ \sum_{v} \Pi_{\vec{z}} (0,v)\tau_{\vec{z}}^{(3)}(v,x,y).
\en
We now take $z_u \equiv z \in [0,1)$,
multiply by $e^{ik\cdot x + il \cdot y}$,
sum over $x,y \in \Zd$, and solve for $\hat{\tau}_z^{(3)}(k,l)$.
The result is
\eq
\lbeq{tau3k}
	\hat{\tau}^{(3)}_z(k,l) =
	\frac{1}{1-\hat{\Pi}_z(k+l)} 
	\hat{\Gamma}^{(3)}_z(k,l) \hat{\tau}_z(l)
	+ \frac{1}{1-\hat{\Pi}_z(k+l)} 
	\hat{\Psi}^{(3)}_z(k,l) 
	\hat{\tau}_z(k) \hat{\tau}_z(l)
\en 
for $z \in [0,1)$.
Since both sides of \refeq{tau3k}
extend to complex $z$ with $|z|<1$, according to Lemmas~\ref{lem-Pidisk}
and \ref{lem-Psi3disk}, \refeq{tau3k} therefore holds for complex $z$
with $|z|<1$.

The first term on the right side can be placed immediately into
the error term $E^{(3)}_z(k,l)$, since 
by \refeq{Flbgood} and Lemmas~\ref{lem-Pidisk} and \ref{lem-Psi3disk},
it is bounded above by a multiple of
\eq
	\frac{1}{|1-z|^{1/2}} 1 \frac{1}{|1-z|^{1/2}} = \frac{1}{|1-z|}.
\en 

To extract the  main contribution of
the second term on the right side of \refeq{tau3k}, we first write this term as
\eq
	\frac{\hat{\Psi}^{(3)}_z(k,l)}{\hat{g}_z(k+l)} 
	\hat{\tau}_z(k+l) \hat{\tau}_z(k) \hat{\tau}_z(l).
\en 
Now
\eq
	\hat{\tau}_z(k) = \frac{C}{D^2k^2 + 2^{3/2} \sqrt{1-z}} + E_z(k),
\en 
and by Theorem~\ref{thm-E} (which we have shown to be a consequence of
Lemmas~\ref{lem-Pidisk}--\ref{lem-Psidisk}),
\eq
	|E_z(\kappa_n)| \leq O(|1-z|^{\epsilon -1/2} + n^{-1/2}|1-z|^{-1/2}
	+ n^{-1/2-\epsilon} |1-z|^{-1}).
\en 
Also,
\eq
	\frac{\hat{\Psi}^{(3)}_z(\kappa_n,\lambda_n)}
	{\hat{g}_z(\kappa_n + \lambda_n)}
	= \frac{\hat{\Psi}_1(0)}{\hat{g}_1(0)} 
	+ O(|1-z|^\epsilon + n^{-1/2})
	= 4C^{-2} + O(|1-z|^\epsilon + n^{-1/2}),
\en 
by \refeq{EbdsEg}, a combination of Lemma~\ref{lem-Psi3disk} with the
methods of Section~\ref{sec-reduction},
and the fact that $4C^{-2} = \hat{\Psi}_1(0)/\hat{g}_1(0)$ by
\refeq{B2def}--\refeq{Cdef}.
Thus we have
\eq
	\hat{\tau}^{(3)}_z(\kappa_n,\lambda_n) =
	4C \prod_{i=1}^3 \frac{1}{D^2 (\kappa_n^{(i)})^2 + 2^{3/2}(1-z)^{1/2}}
	+ O\left( \frac{1}{|1-z|^{3/2-\epsilon}} 
	+ \frac{n^{-1/2}}{|1-z|^{3/2}} 
	+ \frac{n^{-1/2-\epsilon}}{|1-z|^2} \right).
\en 
as required.

\section{The first expansion}
\label{sec-expansion}

Our method makes use of a double expansion.  In this section,
we derive the first of the two expansions, to finite order.  
In contrast to the expansions from I referred to as the one-$M$ 
and two-$M$ schemes, the expansion derived here will be more fully
expanded, and could be called the ``infinite-$M$'' scheme, although
we will not use this terminology.  Using the two-$M$ scheme, we were
able in I to extract the leading behaviour of the Fourier transform
of the critical two-point function, but we obtained minimal control
on the error terms, and this was only for real $z$.  
In I, this was carried out for sufficiently spread-out models in
any dimension $d>6$, and for the nearest-neighbour model in sufficiently
high dimensions.  Using the more
complete expansion derived in this section, we will be able to obtain
stronger power law bounds on error terms, and these bounds will be
valid for complex $z$.  
However, our bounds on the expansion 
will not apply in dimensions near 6, even for
the spread-out model, and we will obtain results only for the
nearest-neighbour model in sufficiently high dimensions. 
For $p<p_c$ and $z=1$, the expansion derived here 
reduces to the expansion of \cite{HS90a}.  We will derive
the expansion for the nearest-neighbour model, but it holds more generally.

In Section~\ref{sec-3pt}, we generalized the magnetic field variable
$z$ to a site dependent field $z_u \in [0,1]$, $u \in \Zd$.  The expansion
will be derived in this general setting.  
For $p \leq p_c$ and $0 \leq z_u \leq 1$, the two-point function is given
by
\eq
	\tau_{\vec{z},p}(0,x) 
	= P(C(0) \ni x, C(0) \cap G = \emptyset)
	= \langle I[C(0) \ni x] \prod_{u \in \Zd} z_u \rangle.
\en
This is the quantity for which we want an expansion.  
The angular brackets denote the joint expectation
with respect to the bond and site variables.
There is, of course, no contribution from any infinite cluster when $p<p_c$,
and for the high-dimensional models we are considering,
the absence of an infinite cluster at $p_c$
is proven in the combined results of \cite{BA91,HS90a}.

Before beginning the derivation of the expansion, 
we first repeat some definitions and lemmas from I that will play
important roles.

\subsection{Definitions and two basic lemmas}
\label{sec-2ndexpdefs}

The following definitions will be used repeatedly in
what follows.

\begin{defn}
\label{def-percterms}
(a)  A \emph{bond} is an unordered pair of distinct sites $\{x,y\}$ with
$\| y-x \|_1 =1$.  A \emph{directed bond} is an ordered pair $(x,y)$ of 
distinct sites with $\| y-x \|_1 =1$.  A \emph{path}
from $x$ to $y$ is a self-avoiding walk from $x$ to $y$, considered to be
a set of bonds.  Two paths are \emph{disjoint} if they have no bonds in
common (they may have common sites).  
Given a bond configuration, an \emph{occupied path} is a path
consisting of occupied bonds.
\newline
(b)  Given a bond configuration, two sites $x$ and $y$ are \emph{connected},
denoted $x \conn y$,
if there is an occupied path from $x$ to $y$ or if $x=y$. 
We write $x \nc y$ when it is not the case that $x \conn y$. 
We denote by $C(x)$ the random set of sites
which are connected to $x$.  Two sites $x$ and $y$ are 
\emph{doubly-connected}, denoted $x \dbc y$,
if there are at least two disjoint occupied paths from $x$ to $y$ or
if $x=y$.  
Given a bond $b = \{u,v\}$ and a bond configuration, we define 
$\tilde{C}^b(x)$ 
to be the set of sites which remain connected to $x$ in the new configuration
obtained by setting $n_b=0$.  Given a set of sites $A$, we say $x \conn A$
if $x \conn y$ for some $y \in A$, and we define 
$C(A) = \cup_{x \in A} C(x)$ and
$\tilde{C}^b(A) = \cup_{x \in A} \tilde{C}^b(x)$.
\newline
(c)  Given a set of sites $A \subset \Zd$ and a bond configuration, 
we say
$x \conn y$  \emph{in} $A$ if there is an occupied path from 
$x$ to $y$ having all of its sites in $A$ (so in particular it is required
that $x,y \in A$), or if $x=y \in A$.
Two sites $x$ and $y$ are \emph{connected
through} $A$, denoted $x \ct{A} y$, if they are connected  
in such a way that every occupied path from $x$ to $y$ has at least one bond
with an endpoint in $A$, or if $x=y \in A$.
\newline
(d)
Given an event $E$ and a bond/site configuration, 
a bond $\{u,v\}$ (occupied or not) is called
\emph{pivotal}
for $E$ if (i) $E$ occurs in the possibly modified configuration in
which $\{u,v\}$ is occupied, and (ii) 
$E$ does not occur in the possibly modified configuration in
which $\{u,v\}$ is vacant.  We say that 
a directed bond $(u,v)$ is pivotal for the connection from $x$ to
$y$ if $x \in \tilde{C}^{\{u,v\}}(u)$, $y\in \tilde{C}^{\{u,v\}}(v)$ and 
$y \, \nin \, \tilde{C}^{\{u,v\}}(x)$.
If $x \conn A$ then there is a natural order to the set of
occupied pivotal bonds for the connection from $x$ to $A$ (assuming there
is at least one occupied pivotal bond), and each of these pivotal bonds is
directed in a natural way, as follows.  The \emph{first pivotal bond from}
$x$ \emph{to} $A$ is the directed occupied pivotal bond $(u,v)$ such that
$u$ is doubly-connected to $x$.  If $(u,v)$ is the first pivotal bond
for the connection from $x$ to $A$, then the second pivotal bond is the
first pivotal bond for the connection from $v$ to $A$, and so on.
\newline
(e)
Given a bond configuration in which $x \dbc y$, we refer to $C(x)$ as
a {\em sausage}.  If $x \conn y$ but $x$ is not doubly connected to $y$,
denote the pivotal bonds for the connection, in order, by
$(u_0,v_0)$, $(u_1,v_1)$, \ldots, $(u_l,v_l)$.
We define the {\em first sausage}\/ for the connection to be 
$\tilde{C}^{\{u_{0},v_{0}\}}(x)$, 
and the last sausage to be $\tilde{C}^{\{u_{l},v_{l}\}}(y)$.  
For $1<j<l$, the $j^{{\rm th}}$ {\em sausage}\/ is 
$\tilde{C}^{\{u_{j},v_{j}\}}(x) \cap \tilde{C}^{\{u_{j-1},v_{j-1}\}}(y)$.
We also define the {\em left}\/ and {\em right endpoints}\/ of the 
$j^{{\rm th}}$ sausage to be, respectively, $v_{j-1}$ and $u_{j}$, 
with $v_{-1}=x$ and $u_{l+1}=y$.
These definitions give rise to a picture in which the connection from
$x$ to $y$ is represented by a string of sausages.
\newline
(f)  
We say that an event $E$ is {\em increasing}\/ if, given a bond/site
configuration $\omega \in E$, and a configuration $\omega'$ having the
same site configuration as $\omega$ and for which each occupied bond in
$\omega$ is also occupied in $\omega'$, then $\omega' \in E$.\end{defn}

\begin{defn}
\label{def-event-on}
(a)
Given a set of sites $S$, we refer to bonds with both endpoints in $S$
as \emph{bonds in $S$}.  
A bond having at least one endpoint in $S$ 
is referred to as a \emph{bond touching $S$}.  
We say that a site $x \in S$ is \emph{in $S$} or \emph{touching $S$}.
We denote by $S_{I}$ the set of bonds and sites in $S$.
We denote by $S_{T}$ the set of bonds and sites touching $S$.
\newline
(b)
Given a bond/site configuration $\omega$ and a set of sites $S$,
we denote by $\omega|_{S_I}$ the bond/site configuration which agrees
with $\omega$ for all bonds and sites in $S$, and which has all
other bonds vacant and all other sites non-green.  Similarly,
we denote by $\omega|_{S_T}$ the bond/site configuration which agrees
with $\omega$ for all bonds and sites touching $S$, and which has all
other bonds vacant and all other sites non-green.
Given an event $E$ and a deterministic set of sites $S$, the event
$\{E$ \emph{occurs in} $S\}$ is defined to 
consist of those configurations $\omega$ for which
$\omega|_{S_I} \in E$.
Similarly, we define the event $\{E$ \emph{occurs on} $S\}$ 
to consist of those configurations $\omega$ for which
$\omega|_{S_T} \in E$.
Thus we distinguish between ``occurs on'' and ``occurs in.''
\newline 
(c) 
The above definitions will now be extended to certain random sets 
of sites.  Suppose $A \subset \Zd$.
For $S=C(A)$ or $S=\Zd \backslash C(A)$, 
we have $\omega |_{S_T} = \omega |_{S_I}$,
since bonds touching but not in $C(A)$ are automatically vacant.  
For such an $S$, we therefore
define $\{ E$ occurs on $S\} = \{ E$ occurs in $S\} = \{ \omega :
\omega |_{S_T} \in E\}$.
For $S= \tilde{C}^{\{u,v\}}(A)$ (see Definition~\ref{def-percterms}(b))
or $S= \Zd \backslash \tilde{C}^{\{u,v\}}(A)$, 
we define $\tilde{S}_T = S_T \backslash \{u,v\}$
and $\tilde{S}_I = S_I \backslash \{u,v\}$, and denote by
$\omega|_{\tilde{S}_T}$ and $\omega |_{\tilde{S}_I}$ 
the configurations obtained by setting $\{u,v\}$ vacant in 
$\omega|_{S_T}$ and $\omega |_{S_I}$ respectively.
Then $\omega |_{\tilde{S}_T} = \omega |_{\tilde{S}_I}$ for these two
choices of $S$, and we define 
$\{ E$ occurs on $S\} 
= \{ E$ occurs in $S\} = \{ \omega : \omega |_{\tilde{S}_T} \in E\}$.
\end{defn}

The above definition of $\{E$ occurs on $S\}$ is intended to capture
the notion that if we restrict attention to the status of
bonds and sites touching $S$, then $E$ is seen to occur.
A kind of asymmetry has been introduced, intentionally, by our
setting bonds and sites not touching $S$ to be respectively vacant
and non-green, as a kind of ``default'' status.  Some
examples are: 
(1) $\{v \conn x$ occurs in $S\}$,
for which Definitions~\ref{def-percterms}(c) and \ref{def-event-on}(b) agree,
(2) $\{ x \conn G$ occurs on $S\} = \{ x \in S\} \cap
\{C(x) \cap S \cap G \neq \emptyset\}$,
and (3) $\{ x \nc G$ occurs on $S\} = \{ x \nin S\} \cup
\{C(x) \cap S \cap G = \emptyset\}$. 

According to Lemma~I.2.3, 
the notion of ``occurs on''
or ``occurs in'' preserves the basic operations of set theory.
Namely, given events $E,F$ and random or deterministic sets 
$S,T$ of sites, we have 
$\{E$ occurs on $S\}^c = \{E^c$ occurs on $S\}$,
$\{(E \cup F)$ occurs on $S\} = \{E$ occurs on $S\} \cup \{F$ occurs on $S\}$,
and $\{\{E$ occurs on $S\}$ occurs on $T\} = \{E$ occurs on $S\cap T\}$.

We now recall the statement of Lemma~I.2.4, 
which is our basic
factorization lemma.  It will be used several times.  
Lemma~I.2.4 
appears in I for constant $z$ but the generalization to site-dependent
$z_u$ is immediate.

\begin{lemma}
\label{lem-cond.0}
Let $p \leq p_c$.  For $p=p_c$, assume there is no infinite cluster.
Given a bond $\{u, v\}$, a finite set of sites $A$, and events $E$, $F$,
we have
\eqarray 
        && \Bigl \langle \Ind 
        \left[ E \textnormal{ occurs on } \tilde{C}^{\{u, v\}}(A) \AND 
        F \textnormal{ occurs in } \Zd \backslash \tilde{C}^{\{u, v\}}(A)
        \AND \{u,v\} \textnormal{ occupied} \right] 
                \Bigr \rangle 
        \nnb
        && 
\lbeq{cond.0lem}
	\qquad = p
        \Bigl \langle \Ind [ 
        E \textnormal{ occurs on } \tilde{C}^{\{u, v\}}(A) ]  
        \langle 
        \Ind [ F \textnormal{ occurs in } 
        \Zd \backslash \tilde{C}^{\{u, v\}}(A) ] 
        \rangle 
        \Bigr \rangle ,
\enarray  
where, in the second line, $\tilde{C}^{\{u, v\}}(A)$
is a random set associated with the outer expectation. 
In addition, the analogue of \refeq{cond.0lem}, in which ``$\{u,v\}$ occupied''
is removed from the left side and ``$p$'' is removed from the right side,
also holds.
\end{lemma}

As an example of a situation
in which an event of the type appearing on the left side of \refeq{cond.0lem}
arises, we recall Lemma~I.2.5, 
which states the following.

\begin{lemma}
\label{lem-pivotal2}
Given a deterministic set $A \subset \Zd$, a directed bond $(a',a)$, 
and a site $y \nin A$, the event $E$ defined by
\eq
	E= \{(a',a) \textnormal{  is a pivotal bond for } y \to A \} 
\en
is equal to the event $F$ defined by
\eq
	F= \bigl \{ a \conn A \ON \tilde{C}^{\{a, a'\}}(A) 
        \AND 
        y \conn a' \INSIDE \Zd\backslash \tilde{C}^{\{a, a'\}}(A) 
	\bigr \} .
\en
\end{lemma}

\subsection{Derivation of the expansion}
\label{sec-exp.rem}

In this section, we generate the expansion, which is essentially a convolution 
equation for $\tau_{\vec{z},p}(0,x)$.
Throughout the discussion, we fix $p,z$ with either $p < p_c$ and $|z|\leq 1$
or $p=p_c$ and $|z|<1$.
The starting point for the expansion is to regard a cluster contributing
to $\tau_{\vec{z},p}(0,x)$, which is 
$P(0 \conn x, 0 \nc G)$, as a string of sausages
joining $0$ to $x$ and not connected to $G$.   We regard these
sausages as interacting with each other, in the sense that they
cannot intersect each other.  In high dimensions, the interaction
should be weak, and our goal is to make an approximation in which
the sausages are treated as independent.  The approximation will
introduce error terms, but these can be controlled in high dimensions. 

We begin by defining some events.  
Given a bond $\{ u', v' \}$, let 
\eqarray
        E_0(0,x) & = & \left\{ 0 \conn x \AND 0 \nc G \right\} ,
        \\
        E_0'(0, x) & = & 
        \left \{ 0 \dbc x \AND 0 \nc G \right \} ,
        \\
        E_0''(0, u', v') & = & E_0'(0, u') \ON \Ctilde^{\{u', v'\}}(0)  ,
        \\
        E_0(0,u',v',x) & = & E_0'(0, u') \cap \left \{  
        (u',v') \mbox{ is occupied and pivotal for } 0 \conn x \right \} .
\enarray
Given a set of sites $A \subset \Zd$, we also define
\eq
        \tau^A_{\vec{z},p}(0,x) = \langle I[(0 \conn x \AND 0 \nc G) \INSIDE
        \Zd \backslash A] \rangle.
\en 

The first step in the expansion is to write
\eq
        \tau_{\vec{z},p}(0,x) = \langle I[E_0(0, x)] \rangle
        = \langle I[E_0'(0, x)] \rangle
        + \sum_{(u_0,v_0)} \langle I[E_0(0, u_0, v_0, x)] \rangle.
\en 
We now wish to apply Lemma~\ref{lem-cond.0} to factor the expectation
in the last term on the right side.  For this, we note that by
definition, $E_0(0, u_0,v_0,x)$ is the event that
$E_0'(0,u_0)$ occurs, that $(u_0,v_0)$ is occupied and pivotal for $0 \conn x$,
and that $\tilde{C}^{\{u_0,v_0\}}(x) \cap G = \emptyset$.  This can be written as
the intersection of the events that
$E'_{0}(0, u_0) \ON \Ctilde^{\{u_0, v_0\}}(0)$, 
that $\{u_0,v_0\}$ is occupied, and 
that $(v_0 \conn x \AND v_0 \nc G) \INSIDE
\Zd \backslash \Ctilde^{\{u_0, v_0\}}(0)$. Applying
Lemma~\ref{lem-cond.0} then gives
\eq
        \langle I[E_0(0, u_0, v_0, x)] \rangle =
        p \langle I[E_0''(0, u_0, v_0)] 
        \tau^{\tilde{C}^{\{u_0,v_0\}}(0)}_{\vec{z},p}(v_0,x) \rangle .
\en 
Therefore,
\eq
\lbeq{expan0}
        \tau_{\vec{z},p}(0,x) = \langle I[E_0'(0, x)] \rangle
        + p \sum_{(u_0,v_0)} \langle I[E_0''(0, u_0, v_0)] 
        \tau^{\tilde{C}^{\{u_0,v_0\}}(0)}_{\vec{z},p}(v_0,x) \rangle.
\en 

To leading order, we would like to replace 
$\tau^{\tilde{C}^{\{u_0,v_0\}}(0)}_{\vec{z},p}(v_0,x)$ by 
$\tau_{\vec{z},p}(v_0,x)$, which
would produce a simple convolution equation for $\tau_{\vec{z},p}$ 
and would effectively treat the first sausage in the cluster joining
$0$ to $x$ as independent of the other sausages.
Such a replacement
should create a small error provided the backbone 
(see Section~\ref{sec-backbone}) joining
$v_0$ to $x$ typically does not intersect the cluster 
$\tilde{C}^{\{u_0,v_0\}}(0)$.  Above the upper critical dimension, where
at $p_c$ 
we expect the backbone to have the character of Brownian motion and the
cluster $\tilde{C}^{\{u_0,v_0\}}(0)$ to have the character of an ISE
cluster, this lack of intersection demands the mutual avoidance of a
2-dimensional backbone and a 4-dimensional cluster.  This is a weak
demand when $d>6$, and this leads to the interpretation of the
critical dimension 6 as $4+2$.  As was pointed out in \cite{AN84},
and as we will show below, bounding errors in the above replacement leads
to the triangle diagram, whose convergence at the critical point is also
naturally associated with $d>6$.  
For simplicity, suppose for the moment that $z_u=z$ for all $u$.
When $z=1$, all diagrams that emerge
in estimating the expansion can be bounded in terms of the triangle diagram,
as was done in \cite{HS90a}, but for $z \neq 1$ other diagrams,
including the square, also arise.  The critical square diagram will diverge in
dimensions $d\leq 8$ when $z \to 1$, but we believe that square diagrams
arise only in conjunction with factors of the
magnetization $M_{z,p} = P(0 \conn G)$, with the product vanishing in
the limit $z \to 1$ for $d>6$.  
Although we believe it to be the case, as we discussed in
Section~\ref{sec-discussion}, we are not able to implement
this mechanism in dimensions larger than but close to 6.
In I, we avoided the difficulty by restricting ourselves to real $z$,
and by employing the 2-$M$ scheme of the expansion.

Let $A$ be a set of sites.
To effect the replacement mentioned in the previous paragraph, we write
\eq
\lbeq{tautauA}
        \tau^{A}_{\vec{z},p}(v,x)  
        =  \tau_{\vec{z},p}(v,x) - 
        [ \tau_{\vec{z},p}(v,x) - \tau^{A}_{\vec{z},p}(v,x) ]     
\en 
and proceed to derive an expression for the difference in square brackets
on the right side.  
Recall the notation $v \ct{A} x$ from Definition~\ref{def-percterms}.  
Similarly, $v \ct{A} G$ will be used to denote the event
that every occupied path from $v$ to any green site must contain
a site in $A$, or that $v \in G \cap A$.
The above difference in square brackets is then given by
\eqarray
        \tau_{\vec{z},p}(v,x) - \tau^{A}_{\vec{z},p}(v,x) & = & 
        \langle I[ v \conn x \AND v \nc G] \rangle
        - \langle I[(v \conn x \AND v \nc G) \INSIDE
        \Zd \backslash A] \rangle
        \nonumber \\
        & = & \langle I[ v \conn x \AND v \nc G] \rangle
        - \langle I[v \conn x \INSIDE
        \Zd \backslash A \AND v \nc G ] \rangle
        \nonumber \\ 
        && + \langle I[v \conn x \INSIDE
        \Zd \backslash A \AND v \nc G ] \rangle
        \nonumber \\ &&
        - \langle I[(v \conn x \AND v \nc G) \INSIDE
        \Zd \backslash A] \rangle
        \nonumber \\
        & = & \langle I[ v \ct{A} x \AND v \nc G] \rangle
        - \langle I[v \conn x \mbox{ in }
        \Zd \backslash A \AND v \ct{A} G ] \rangle  .
\lbeq{F14exp}
\enarray
This can be rewritten as
\eqarray
	\tau_{\vec{z},p}(v,x) - \tau^{A}_{\vec{z},p}(v,x)
        & = &  \langle I[ v \ct{A} x \AND v \nc G] \rangle
        - \langle I[v \conn x \AND v \ct{A} G ] \rangle
	\nonumber \\ &&
        + \langle I[v \ct{A} x \AND v \ct{A} G ] \rangle.
\lbeq{F12exp}
\enarray
Defining
\eqarray
\lbeq{F134def}
        F_1(v, x;A) & = & \left \{ v \ct{A} x  \AND v \nc G \right \}  ,
        \\ 
        F_3 (v, x; A) & = & \left \{v \conn x   \AND v \ct{A} G \right \} ,
        \\ 
        F_4(v,x;A) & = & \left\{ v \ct{A} x \AND v \ct{A} G \right\} ,
\enarray
(the definitions of $F_3$ and $F_4$ are different from those in I)
this gives
\eq
\lbeq{tA123}
        \tau_{\vec{z},p}(v,x) - \tau^{A}_{\vec{z},p}(v,x) =
        \langle I[F_1(v, x;A)] \rangle 
        - \langle I[F_3 (v, x; A)] \rangle
        + \langle I[F_4 (v, x; A)] \rangle.
\en 

Using the terminology of Definition~\ref{def-percterms}(e),
we will decompose the event $F_4 (v, x; A)$ as
a disjoint union 
\eq
\lbeq{F312}
        F_4 (v, x; A) = F_{4,1} (v, x; A) \cup F_{4,2} (v, x; A),
\en  
and combine
$F_{4,1} (v, x; A)$ with $F_1 (v, x; A)$ and 
$F_{4,2} (v, x; A)$ with $F_3 (v, x; A)$.  
The event $F_{4,1} (v, x; A)$ is defined 
to be the event that: (i) $F_4 (v, x; A)$ occurs, (ii)
the first sausage (say the $l^{\rm th}$ sausage)
for $v \conn x$, whose left endpoint
is connected to its right endpoint through $A$, is $G$-free,
and (iii) all sausages following the $l^{\rm th}$ sausage are $G$-free.
The event $F_{4,2} (v, x; A)$ is defined
to be the event that $F_4 (v, x; A)$ occurs
and in addition, the last sausage connected to $G$ has its right endpoint
connected to $v$ through $A$.
In particular, $F_{4,1}(v,x;A) \cap F_1(v,x;A) = \emptyset$ and
$F_{4,2}(v,x;A) \subset F_3(v,x;A)$.
With these definitions, \refeq{F312} holds.  Now we define
\eqarray
\lbeq{EFF1}
        E_1(v,x;A) & = & F_1(v,x;A) \cup F_{4,1} (v, x; A) ,
        \\
\lbeq{EFF2}
        E_2(v,x;A) & = & F_3(v,x;A) \cap F_{4,2} (v, x; A)^c .
\enarray
Then \refeq{tA123} becomes
\eq
\lbeq{tA12}
\tau_{\vec{z},p}(v,x) - \tau^{A}_{\vec{z},p}(v,x) =
        \langle I[E_1(v, x;A)] \rangle 
        - \langle I[E_2 (v, x; A)] \rangle.
\en 

The events $E_1$, $E_2$ can be described in words as follows:
\begin{description}
\item
$E_1(v,x;A)$ is the event that $v \ct{A} x$ and the following holds.
Let $(b,b')$ be the pivotal bond, if there is one, leading into the 
first sausage for $v \conn x$ whose left and right endpoints are
connected through $A$.  If there is such
a pivotal bond, then we require that 
$\tilde{C}^{\{b,b'\}}(x) \cap G = \emptyset$,
while in $\tilde{C}^{\{b,b'\}}(v)$ either $v \nc G$ or $v \ct{A} G$.
If there is no such pivotal bond, then we require that $v \nc G$.
\item
$E_2(v,x;A)$ is the event that $v \conn x$, $v \ct{A} G$, and the following
holds.  Let $S$ be the last sausage for $v \conn x$ that is connected to $G$.
Then we require that $v$ be connected to the right endpoint of $S$ in 
$\Zd \backslash A$.
\end{description}

\begin{figure}
\begin{center}
\includegraphics[scale = 0.4]{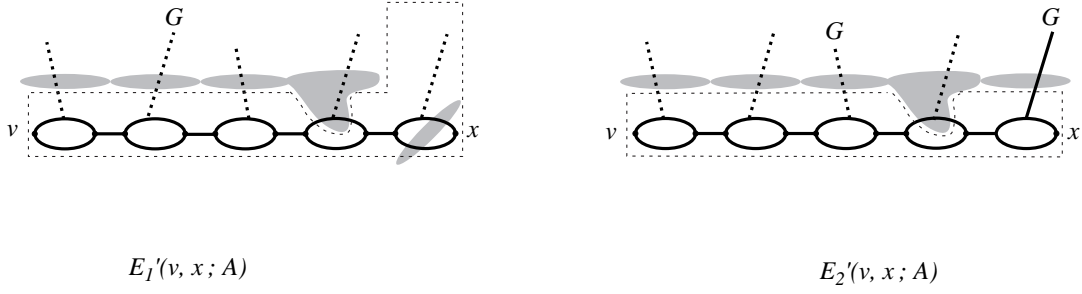} 
\end{center}
\caption{Schematic depiction of the events $E_1'$ and $E_2'$.  Shaded regions
represent the set $A$ and dotted lines represent possible but not mandatory
connections in $C(v)$.  The dashed lines indicate that the inner regions
are $G$-free.}
\label{fig-E12p.def}
\end{figure}

Combining \refeq{tA12}, \refeq{tautauA} and \refeq{expan0} gives 
\eqarray
        \tau_{\vec{z},p}(0,x) & = & \langle I[E_0'(0, x)] \rangle
        + p \sum_{(u_0,v_0)} \langle I[E_0''(0, u_0, v_0)] 
        \rangle \tau_{\vec{z},p}(v_0,x)
        \nonumber \\ &&
        - p \sum_{(u_0,v_0)} \langle I[E_0''(0, u_0, v_0)] 
        \langle I[E_1(v_0, x;\tilde{C}_0^{\{u_0,v_0\}}(0))] \rangle_1 \rangle_0
        \nonumber \\ &&
        +  p \sum_{(u_0,v_0)} \langle I[E_0''(0, u_0, v_0)] 
        \langle 
        I[E_2(v_0, x;\tilde{C}_0^{\{u_0,v_0\}}(0))] 
        \rangle_1 \rangle_0 .
        \lbeq{taueq4}
\enarray
Here, we have tacitly assumed that the series on the right side converge.
We will return to this point at the end of the section.  
Also, we have introduced subscripts on
expectations and random sets to coordinate the two.  In particular,
the random set $\tilde{C}_0^{\{u_0,v_0\}}(0)$ corresponds to the expectation
$\langle \cdot \rangle_0$.
Now we define
\eqarray
        E'_{1}(v, x; A) & = & E_1(v, x;A) \cap
        \left \{ \nexists \mbox{ pivotal
        $(u',v')$ for } v \conn x \mbox{ such that } v \ct{A} u'  \right \} ,
        \hspace{8mm}
        \\ 
\lbeq{E1primeprime}
        E''_{1}(v, u', v'; A) & = & 
        E'_{1}(v, u'; A) \ON \Ctilde^{\{u', v'\}}(v)  ,
        \\
        E_1(v,u',v',x;A) & = & E_1'(v,u';A) \cap 
        \Big \{ (u',v') \mbox{ is occupied and pivotal for } 
        v \conn x 
        \nonumber \\ &&   \AND
        \tilde{C}^{\{u',v'\}}(x) \cap G = \emptyset  \Big \} ,
        \\
        E_2'(v, x; A) & = & E_2 (v, x; A) \cap
        \left \{ x \ct{A} G 
        \mbox{ in (last sausage of } v \to x)\right \} ,
        \\
        E_2''(v, u', v'; A) & = & E_2'(v, u'; A) \ON \Ctilde^{\{u'v'\}}(v) ,
        \\
        E_2(v,u',v',x;A) & = & E_2'(v,u';A) \cap 
        \Big \{ (u',v') \mbox{ is occupied and pivotal for } v \conn x 
        \nonumber \\ &&  \AND
        \tilde{C}^{\{u',v'\}}(x) \cap G = \emptyset  \Big \} .
\enarray
The events $E_1'$ and $E_2'$ are depicted schematically in
Figure~\ref{fig-E12p.def}. 

Next we observe that for $j=1,2$,
\eq
\lbeq{Ej1}
        \langle I[E_j(v,x;A)] \rangle 
        = \langle I[E_j'(v,x;A)] \rangle
        + \sum_{(u',v')} \langle I[E_j(v, u', v',x;A)] 
        \rangle.
\en 
We now claim that Lemma~\ref{lem-cond.0} can be applied to conclude that
for $j=1,2$, 
\eq
\lbeq{Ej2}
        \langle I[E_j(v, u',v',x;A)] \rangle = 
        p \langle I[E_j''(v,u',v';A)] 
        \tau^{\tilde{C}^{\{u',v'\}}(v)}_{\vec{z},p}(v',x)
        \rangle .
\en 
This can be seen as follows.
By definition, $E_j(v, u',v',x;A)$ is the event that
$E_j'(v,u';A)$ occurs, that $(u',v')$ is occupied and pivotal for $v \conn x$,
and that $\tilde{C}^{\{u',v'\}}(x) \cap G = \emptyset$.  It can be written as
the intersection of the events that
$E'_{j}(v, u'; A) \ON \Ctilde^{\{u', v'\}}(v)$, 
that $\{u',v'\}$ is occupied, and 
that $(v' \conn x \AND v' \nc G) \INSIDE
\Zd \backslash \Ctilde^{\{u', v'\}}(v)$. Applying
Lemma~\ref{lem-cond.0}, we get
\eq
        \langle I[E_j(v, u',v',x;A)] \rangle = 
        p \langle I[E_j''(v,u',v';A)] 
        \tau^{\tilde{C}^{\{u',v'\}}(v)}_{\vec{z},p}(v',x)
        \rangle ,
\en 
as required.

We are now in a position to generate an expansion.
First, we introduce some 
abbreviated notation.  Let 
$\tilde{C}_n = \tilde{C}_n^{\{u_n,v_n\}}(v_{n-1})$, with $v_{-1}=0$,
and let
\eqarray
\lbeq{Xndef}
	X_n & = &   I[E_1  (v_{n-1},x; \tilde{C}_{n-1})] 
		 -  I[E_2  (v_{n-1},x; \tilde{C}_{n-1})] , \\
\lbeq{Xnpdef}
	X_n' & = &  I[E_1' (v_{n-1},x; \tilde{C}_{n-1})] 
		 -  I[E_2' (v_{n-1},x; \tilde{C}_{n-1})] , \\
\lbeq{Xnppdef}
	X_n'' & = & I[E_1''(v_{n-1}, u_{n}, v_{n}; \tilde{C}_{n-1})] 
		  - I[E_2''(v_{n-1}, u_{n}, v_{n}; \tilde{C}_{n-1})].
\enarray
In terms of this new notation, \refeq{taueq4} 
can be written as
\eq
\lbeq{exptoX1}
	\tau_{\vec{z},p}(0,x) = \langle I[E_0'(0,x)]\rangle_0 
	+ p \sum_{(u_0,v_0)}\langle I[E_0''(0,u_0,v_0)]\rangle_0
	\tau_{\vec{z},p}(v_0,x)
	- p \sum_{(u_0,v_0)}\langle I[E_0''(0,u_0,v_0)]
	\langle X_1 \rangle_1 \rangle_0 .
\en
In view of \refeq{Ej1} and \refeq{Ej2},  
\eq
\lbeq{Xit}
        \langle X_{n} \rangle_{n}
	= \langle X'_{n} \rangle_{n} 
	+ p \sum_{(u_{n},v_{n})}
	 \langle X''_{n} \tau_{\vec{z},p}^{\tilde{C}_{n}}(v_{n}, x) \rangle_{n} .
\en 
By \refeq{tautauA} and \refeq{tA12},
\eq
\lbeq{tC}
        \tau_{\vec{z},p}^{\tilde{C}_{n}} (v_n, x) 
	= \tau_{\vec{z},p} (v_n, x) 
	- \langle X_{n+1} \rangle_{n+1} .
\en 
With \refeq{Xit}, this gives
\eq
\lbeq{1Cn}
        \langle X_{n} \rangle_{n}
	= \langle X'_{n}\rangle_{n} 
	+ p \sum_{(u_{n},v_{n})} 
	\langle X''_{n} \rangle_{n} \tau_{\vec{z},p}(v_{n},x)
	- p \sum_{(u_{n},v_{n})}
	\langle X''_{n} \langle X_{n+1} \rangle_{n+1} \rangle_{n}.
\en 

An expansion can now be generated by recursively substituting
\refeq{1Cn} into \refeq{exptoX1}.
To further abbreviate the notation, in generating the expansion we omit
all arguments related to sites and omit the products of $p$
with summations over $(u_{n}, v_{n})$ that are associated with
each product.
The first few iterations then yield
\eqarray 
	\tau & = & 
	\langle I[E_0'] \rangle_0 + \langle I[E_0''] \rangle_0 \tau 
	- \langle I[E_0''] \langle X_{1} \rangle_1 \rangle_0
	\nnb 
	& = & 
	\Big( \langle I[E_0'] \rangle_0 
	- \langle I[E_0''] \langle X_{1}' \rangle_1 \rangle_0 \Big)
	+ \Big( \langle I[E_0''] \rangle_0 
	- \langle I[E_0''] \langle X''_{1} \rangle_1 \rangle_0 \Big) \tau 
	+ \langle I[E_0''] \langle X''_{1} \langle X_{2} 
	\rangle_2 \rangle_1 \rangle_0
	\nnb 
	& = & 
	\Big( \langle I[E_0'] \rangle_0
	 - \langle I[E_0''] \langle X_{1}' \rangle_1 \rangle_0 
	+ \langle I[E_0''] \langle X''_{1} \langle X_{2}' 
	\rangle_2 \rangle_1 \rangle_0 \Big)
	\nnb &&
	+ \Big ( \langle I[E_0''] \rangle_0
	 - \langle I[E_0''] \langle X''_{1} \rangle_1 \rangle_0
	+ \langle I[E_0''] \langle X''_{1} \langle X_{2}'' 
	\rangle_2 \rangle_1 \rangle_0  \Big) 
	\tau 
	\nnb &&
\lbeq{expto3}
	- \langle I[E_0''] \langle X''_{1} \langle X_{2}'' 
	\langle X_3 \rangle_3
	\rangle_2 \rangle_1 \rangle_0, 
\enarray 
and so on.   

To simplify expressions involving nested expectations, we will
sometimes use $\Ebold$ rather than $\langle \cdot \rangle$ to
denote expectations.
To organize the terms in the expansion, we introduce the following
notation, for $n \geq 1$:
\eqarray
	g_{\vec{z},p}^{(0)}(0,x) & = & \langle I[E_0'(0,x)]\rangle 
	= \Ebold I[E_0'(0,x)] , \\
	g_{\vec{z},p}^{(n)}(0,x) & = & (-1)^n \Ebold_0 I[E_0''] 
	\Ebold_1 X_1'' 
	\cdots \Ebold_{n-1} X_{n-1}'' \Ebold_n X_{n}' , \\
	\Pi_{\vec{z},p}^{(0)}(0, v_0) & = & 
	p \sum_{u_0} \langle I[E_0''(0, u_0, v_0)]\rangle 
	= p \sum_{u_0} \Ebold I[E_0''(0, u_0, v_0)] , \\
\lbeq{Pindef}
	\Pi_{\vec{z},p}^{(n)}(0, v_{n}) & = & 
	(-1)^n
	p \sum_{u_{n}} \Ebold_0 I[E_0''] \Ebold_1 X_1'' 
	\cdots \Ebold_{n-1} X_{n-1}'' \Ebold_n X_{n}''  , \\
\lbeq{UNdef}
	U_{\vec{z},p}^{(n)}(0,x) & = & (-1)^n \Ebold_0 I[E_0''] \Ebold_1 X_1'' 
	\cdots \Ebold_{n-1} X_{n-1}'' \Ebold_n X_{n} .
\enarray
In the above, the notation continues to omit sums over $(u_j,v_j)$
and factors of $p$ associated with each product.
The terms $g^{(n)}$ account for the terms in \refeq{expto3}
whose innermost expectation involves an $X$ bearing a single prime,
the terms $\Pi^{(n)}$ account for the terms in \refeq{expto3}
whose innermost expectation involves an $X$ bearing a double prime,
and $U^{(n)}$ accounts for the single term whose innermost expectation
involves an unprimed $X$.  For each $N \geq 0$, the expansion can 
then be written as
\eq
\lbeq{taugPiU}
	\tau_{\vec{z},p}(0,x) = \sum_{n=0}^N g_{\vec{z},p}^{(n)}(0,x) + 
	\sum_{n=0}^N \sum_{v_{n}}
	\Pi_{\vec{z},p}^{(n)}(0, v_{n}) \tau_{\vec{z},p}(v_{n}, x) + 
	U_{\vec{z},p}^{(N+1)}(0,x).
\en

For $z_u \equiv z=1$ the set $G$ of green sites is empty, and
the event $E_2$, which requires connection to $G$, cannot occur.
Therefore all terms involving $E_2$ events vanish for $z=1$.
In this special case, \refeq{taugPiU} becomes the expansion of \cite{HS90a}
and agrees also with the expansion of I.  

Taking the Fourier transform of \refeq{taugPiU} and solving for
$\hat{\tau}_{\vec{z},p}(k)$ gives, for each $N \geq 0$,
\eq
\lbeq{taugPiUk}
	\hat{\tau}_{\vec{z},p}(k) = 
	\frac{\sum_{n=0}^N \hat{g}_{\vec{z},p}^{(n)}(k)  
	+ \hat{U}_{\vec{z},p}^{(N+1)}(k)}
	{1- \sum_{n=0}^N \hat{\Pi}_{\vec{z},p}^{(n)}(k)}.
\en
Existence of all Fourier transforms appearing in the right side of
\refeq{taugPiUk} will be shown in
Section~\ref{sec-Pidiski}, for $p=p_c$, $z_u \in [0,a]$ with $a<1$, and for
$d$ sufficiently large.  The bounds of Section~\ref{sec-Pidiski} will also
provide the convergence of summations mentioned below \refeq{taueq4} 
and tacitly assumed in the subsequent analysis.  The bounds apply also to
the simpler subcritical case of $p<p_c$, $z_u \in [0,1]$, but we omit
any explicit discussion of this case.

\section{Proof of Lemmas~\protect\ref{lem-Pidisk} and \protect\ref{lem-Fklb}}
\label{sec-full.exp.bd}
\setcounter{equation}{0}

Throughout this section, we restrict attention to $p=p_c$, and drop subscripts
$p_c$ from the notation.  We will
define $g_{\vec{z}}(0,x)$ and $\Pi_{\vec{z}}(0,x)$
obeying \refeq{vecexp} for $z_u \in [0,a]$, for any
$a \in [0,1)$.  This also gives \refeq{taugPix} for $z \in [0,1)$.
For constant $z_u=z$, $\hat{g}_{z}(k)$ and $\hat{\Pi}_{z}(k)$
will be shown to obey \refeq{taugPi}, and to extend
to complex $z$ with $|z|\leq 1$.  This involves taking the limit
$N \to \infty$ in \refeq{taugPiUk}, with the term 
$\hat{U}_{\vec{z}}^{(N+1)}(k)$ vanishing in the limit.
This will then give \refeq{taugPix} and \refeq{taugPi} for $|z|<1$.
We will prove Lemmas~\ref{lem-Pidisk} and \ref{lem-Fklb} 
in Sections~\ref{sec-Pidiski} and \ref{sub-prf.lem-2.1ii}
respectively.

\subsection{Proof of Lemma~\protect\ref{lem-Pidisk}}
\label{sec-Pidiski}

We begin by stating two lemmas providing the estimates needed
to obtain the identities
\refeq{taugPix}, \refeq{taugPi} and \refeq{vecexp}, and 
to prove Lemma~\ref{lem-Pidisk}.  After stating the lemmas, we
will show that these identities hold and prove Lemma~\ref{lem-Pidisk},
assuming the two lemmas.
Then the two lemmas will be proved.  Throughout this section, we assume
without further mention that the dimension $d$ is sufficiently large.
We remark that the use of $m_2=4$ in \refeq{Pinbd} requires us to
take at least $d \geq 8$, but since we are not attempting to obtain
estimates valid for all $d>6$ (for spread-out models), we have not tried
to do without $m_2=4$.

\begin{lemma}
\label{lem-Pinbd}
(i)
For $z_u \in [0,1]$ and $n \geq 0$, 
$|\hat{g}_{\vec{z}}^{(n)}(k)| \leq \sum_x |g_{\vec{z}}^{(n)}(0,x)|
\leq O(d^{-n})$
and $|\hat{\Pi}_{\vec{z}}^{(n)}(k)| \leq \sum_x |\Pi_{\vec{z}}^{(n)}(0,x)|
\leq O(d^{-n})$.
\newline
(ii)
For $n \geq 0$ and constant $z_u \equiv z$,
$\hat{g}_z^{(n)}(k)$ and $\hat{\Pi}_z^{(n)}(k)$ can be extended
to complex $z$ with $|z| \leq 1$.
For $|z| \leq 1$, $n \geq 0$, and $m_1 = 0, 2$, $m_2 = 0, 2, 4$
their norms obey the bounds 
\eqarray  
	\| \nabla_k^{m_1} \hat{g}_z^{(n)}(k) \| & \leq &
        \sum_{x} |x|^{m_1} \, \|  g^{(n)}_{z}(0,x) \| 
	 \leq  O(d^{-n} )  ,
	\lbeq{gnbd}
	\\ 
	\| \nabla_k^{m_2} \hat{\Pi}_z^{(n)}(k) \| & \leq &
 	\sum_{x} |x|^{m_2} \, \|  \Pi^{(n)}_{z}(0,x) \| 
	 \leq  O(d^{-n} ) . 
	\lbeq{Pinbd}
\enarray
\end{lemma}

\begin{lemma}
\label{lem-UNbd}
Let $a \in [0,1)$.  
For $z_u \in [0,a]$, $k \in [-\pi,\pi]^d$, and $N \geq 0$, the remainder term 
$\hat{U}^{(N+1)}_{\vec{z}}(k)$ obeys the bound  
\eq
\lbeq{UNbd.1}
	| \hat{U}_{\vec{z}}^{(N+1)} (k) | \leq O(d^{-(N+1)})(1+ \chi_{a}) .  
\en 
\end{lemma}

By Lemmas~\ref{lem-Pinbd}(i) and \ref{lem-UNbd}, \refeq{vecexp} follows
by taking the limit $N \to \infty$ in \refeq{taugPiUk} and then taking
the inverse Fourier transform.  The functions $\ghat_{\vec{z}}(k)$ 
and $\Pihat_{\vec{z}}(k)$
are given by
\eq
	\ghat_{\vec{z}}(k) = \sum_{n=0}^\infty \ghat_{\vec{z}}^{(n)}(k),
	\quad
	\Pihat_{\vec{z}}(k) = \sum_{n=0}^\infty \Pihat_{\vec{z}}^{(n)}(k).
\en
Taking $z_u \equiv z \in [0,1)$ then gives
\eq
\lbeq{tauextend}
	\tauhat_z(k) = \frac{\ghat_z(k)}{1 - \Pihat_z(k)} . 
\en 

As power series in $z$, $\ghat_z(k)$ and $\Pihat_z(k)$
converge absolutely for $|z| \leq 1$,
by Lemma~\ref{lem-Pinbd}(ii).  The left side 
of \refeq{tauextend} is a power series that
converges absolutely, and therefore defines an analytic function,
for  $z$ in $|z|<1$.  The two series are therefore equal for $|z|<1$,
and the right side extends the left side to $|z|\leq 1$.
This proves \refeq{taugPi}, and by taking the inverse Fourier transform,
also proves \refeq{taugPix}. 
Also, since \refeq{taugPiU} agrees with the expansion of I
for $z=1$, the claim at the end of Section~\ref{sec-reqbdexp} that 
$\hat{g}_1(k) = \hat{\phi}_{h=0}(k)$ and 
$\hat{\Pi}_1(k) = \hat{\Phi}_{h=0}(k)$, 
where $\hat{\phi}_{h}(k)$ and $\hat{\Phi}_{h}(k)$ are the functions
occuring in the one-$M$ scheme in (I.3.88), 
then follows.

Next, we prove Lemma~\ref{lem-Pidisk}, assuming Lemmas~\ref{lem-Pinbd} 
and \ref{lem-UNbd}.

\smallskip \noindent
{\bf Proof of Lemma~\ref{lem-Pidisk}}. 
The upper bounds of Lemma~\ref{lem-Pidisk} are immediate consequences
of Lemma~\ref{lem-Pinbd}, and we need only prove that
$\ghat_1(0)$ and $- \nabla^2 \Pihat_1(0)$ are both equal to $1+O(d^{-1})$.

We begin with $\ghat_1(0)$.  
By (I.3.6), 
$\ghat^{(0)}_1(0) 
= \hat{\phi}^{(0)}_{h=0}(0) = 1 + \hat{\phi}_{h=0}^{(01)} (0)$, and
we have 
\eq
	\ghat_1(0) = 1 + \hat{\phi}_{h=0}^{(01)} (0) 
	+ \sum_{n=1}^\infty \ghat_1^{(n)}(0). 
\en 
The bound (I.3.20) 
can easily be improved to
$|\hat{\phi}_{h=0}^{(01)} (0)| \leq O(d^{-1})$.
With Lemma~\ref{lem-Pinbd}(ii), this gives
\eq
	\ghat_1 (0) = 1 + O(d^{-1}) + 
	\sum_{n=1}^\infty O(d^{-n}) 
	= 1 + O(d^{-1})  . 
\en 
Similarly, we have 
\eqsplit
	- \nabla^2 \Pihat_1(0) 
	& 
	= \sum_x \sum_{n=0}^\infty 
	|x|^2 \Pi_{h=0}^{(n)}(0,x) 
	\\ 
	& 
	=   2d p_c
	+ p_c \sum_{(u,v): u \neq 0} |v|^2 \expec{I[E_0''(0,u,v)]}_1 
	+ \sum_{n=1}^\infty \sum_x |x|^2 \Pi_{h=0}^{(n)} (0,x) . 
\ensplit  
The first term is $1+O(d^{-1})$ by \refeq{pc}.  
The second term can be bounded using the weighted bubble diagram
${\sf W}^{(2)}_{h=0,p_c}$ of (I.3.16), 
which was shown to be $O(d^{-1})$ in \cite{HS90a}.
By Lemma~\ref{lem-Pinbd}, the third term is 
$O(d^{-1})$.  This gives the desired result  
$- \nabla^2 \Pihat_1(0) = 1+O(d^{-1})$.
\qed

\medskip \noindent
{\bf Proof of Lemma~\ref{lem-Pinbd}.}
The bounds obtained in part (ii) of the lemma are actually stronger
than those needed for part (i), and we discuss only the proof of part (ii)
here.  Also, since $g$ and $\Pi$ are almost identical, we discuss
only $g$.

To bound the norm in \refeq{gnbd}, we will use the fact that
for a power series $f$ with real coefficients $a_n$ all of the same sign,
the norm \refeq{fnorm} obeys
\eq
	\| f(z) \| = \sum_{n=0}^\infty |a_n| |z|^n = | f(|z|) | .
	\lbeq{dbnorm.1}
\en 
We will explain in more detail below that, as power series in $z$, 
$g_z^{(n)}(0,x)$ and $\Pi_z^{(n)}(0,x)$ have real coefficients 
of varying signs.  To handle a power series $f(z)$ with both positive
and negative coefficients, our strategy will be to decompose it
into a sum of functions $f_j(z)$ with coefficients of definite sign.  
Then we use the triangle inequality and \refeq{dbnorm.1} 
to conclude 
\eq
	\| f(z) \| \leq \sum_{j} \| f_j (z) \| 
	= \sum_{j} | f_j (|z|) | , 
\en 
and thus reduce the problem of estimating $\| f(z) \|$ to that of estimating 
$| f_j (|z|) |$. 

Beginning with $g_z^{(0)}$, by definition 
\eq
	g_z^{(0)} (0,x) = \expec{I[E_0'(0,x)]} 
	= \expec{I[E_{0,b}'(0,x)] z^{|C(0)|} }_{b}, 
\en 
where $E_{0,b}'(0,x)$ is the bond event $\{ 0 \dbc x \}$ and
$\expec{ \cdot }_b$ denotes expectation with respect to bond variables only.  
This can be written as a power series in $z$ with 
positive coefficients, and hence it extends to complex $z$ within the 
disk of convergence.  Using \refeq{dbnorm.1}, we 
argue as in (I.3.36) 
to obtain 
\eq
\lbeq{gz0bdm}
	\left \| g_z^{(0)} (0,x) \right \| 
	= \expec{I[E_{0,b}'(0,x)] |z|^{|C(0)|} }_{b} 
	\leq \tau_1 (0,x)^2 .
\en 
Multiplying by $|x|^m$ and summing over $x$ then gives \refeq{gnbd}
for $n=0$, $|z| \leq 1$, using the methods of \cite{HS90a}.
The case $m=4$ (which is needed for $\Pi_z$) did not occur in 
\cite{HS90a}, but the methods there apply if we associate
a factor $|x|^2$ to each factor $\tau_1(0,x)$ on the right side of
\refeq{gz0bdm}.

\begin{figure}
\begin{center}
\includegraphics[scale = 0.3]{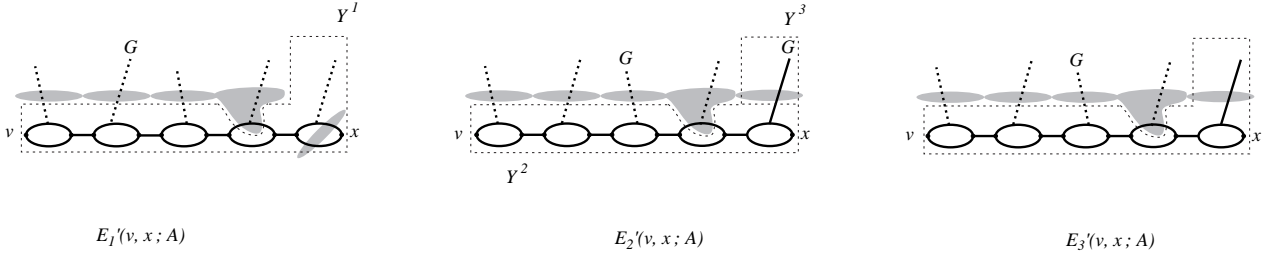} 
\end{center}
\caption{Schematic depiction of  $Y^1$, $Y^2$, $Y^3$, and of the event
$E_3'$ defined under \refeq{E3Q5}.  Shaded regions
represent the set $A$ and dotted lines represent possible but not mandatory
connections in $C(v)$.  For $E_3'$,
the dashed lines indicate that the inner region is $G$-free.}
\label{fig-Q1-5.def}
\end{figure}

For $n \geq 1$, we have 
\eq
\lbeq{gn1etc}
	g_z^{(n)}(0,x)  = (-1)^n 
	\Ebold_0 I[E_0''] \Ebold_1 X_1 '' \cdots 
	\Ebold_{n-1} X_{n-1}'' \Ebold_n X_n'  .
\en 
We insert $X'' = I[E_1''] - I[E_2'']$ in the above, and similarly
for the inner expectation with $X'$, and expand 
the products to obtain a sum of $\pm1$ times expectations involving
$E_j'$ and $E_j''$.  
The bond connections required by the events $E_j'$ and $E_j''$ are given
in terms of the auxiliary {\em bond}\/ events 
\eqarray 	
	E_{1, b}'(v, x; A) 
	& \equiv & 
	\bigl \{ 
	v \ct{A} x \AND \nexists \textrm{ pivotal } (u', v') \textrm{ for }
	v \conn x \textrm{ such that } v \ct{A} x 
	\bigr \} ,
	\lbeq{E1pb-def}
	\\
	E_{1, b}''(v, u', v'; A) 
	& \equiv & 
	E_{1, b}'(v, u';A) \ON \tilde{C}^{\{u', v'\}}(v) ,
	\lbeq{E1ppb-def}
	\\
	E_{2, b}'(v, x; A) 
	& \equiv & 
	\bigl \{ 
	v \conn x \IN \Zd\backslash A 
	\AND \textrm{(last sausage of } v \conn x) \conn A  
	\bigr \} ,
	\lbeq{E2pb-def}
	\\ 
	E_{2, b}''(v, u', v'; A) 
	& \equiv & 
	E_{2, b}'(v, u';A) \ON \tilde{C}^{\{u', v'\}}(v) . 
	\lbeq{E2ppb-def}
\enarray 
The conditions due to the site variables, in $E_j'$ and $E_j''$, can
be expressed in terms of the subclusters $Y^1$, $Y^2$, $Y^3$ depicted
in Figure~\ref{fig-Q1-5.def} and defined as follows:
\eqarray
	Y^1 & = &  (\mbox{last sausage of $v \longrightarrow x$}) 
	\cup \{ y \in C(v)  : y \conn v \mbox{ in } \Zd\backslash A\},
	\\
	Y^{2} & = & \{y \in C(v) : y \conn v \mbox{ in } \Zd\backslash A\},
	\\
	Y^{3} & = & (\mbox{last sausage of $v \longrightarrow x$})
	\backslash Y^{2}.
\enarray
The above definitions are appropriate for $E_1'$ and $E_2'$.
For $E_1''$ and $E_2''$, we define the $Y$'s to be the intersection of the
above with $\tilde{C}^{\{u',v'\}}(v)$.

Using these definitions, and using $\langle \cdot \rangle_s$ and
$\langle \cdot \rangle_b$ for expectations with respect to the site
and bond variables, for $z \in [0,1]$ we have 
\eqarray 
	\expec{I[E_1'(v, x; A)] }_{s} 
	& = & 
	I[ E_{1, b}'(v, x; A)] \, z^{|Y^1| } , 
	\lbeq{E1psite.1}
	\\
	\expec{I[E_2'(v, x; A)] }_{s} 
	& = & 
	I[ E_{2, b}'(v, x; A)] \, z^{|Y^{2}|} (1 - z^{|Y^{3}|}) , 
	\lbeq{E2psite.1}
	\\
	\expec{I[E_1''(v, u', v'; A)] }_{s} 
	& = & 
	I[ E_{1, b}''(v, u', v'; A)] \, z^{|Y^1| } , 
	\lbeq{E1ppsite.1}
	\\
	\expec{I[E_2''(v, u', v'; A)] }_{s} 
	& = & 
	I[ E_{2, b}''(v, u', v'; A)] \, z^{|Y^{2}|} (1 - z^{|Y^{3}|}) . 
	\lbeq{E2ppsite.1}
\enarray 

Having expanded the $X$'s in \refeq{gn1etc}, we insert 
\refeq{E1psite.1}--\refeq{E2ppsite.1}, leaving only bond expectations.
For terms involving $E_2$, we use 
\eq
\lbeq{1-zQ}
	(1-z^{|Y^{3}|})z^{|Y^{2}|} 
	= z^{|Y^{2}|} - z^{|Y^{2}|+|Y^{3}|}, 
\en
and further expand to
obtain a sum of terms in which the $z$-dependence of each term is
simply a power determined by the $Y$'s.  
The result is a sum of terms, with coefficients $\pm1$,
of power series in $z$ having nonnegative coefficients.  
The number of terms resulting
from these two expansions is at most $4^n$.  This provides
a formula which extends $g_z^{(n)}(0,x)$ to complex $z$ within any disk
in which all the series converge.  We will show that, in fact, 
when summed over $x$ the series all converge absolutely for $|z| \leq 1$.

Explicitly, we demonstrate the convergence 
for two examples only.  The other cases are similar, using standard
diagrammatic estimation.   
First, we consider the term involving only $E_1$-events.  To bound
\eq
	\left \| 
	\Ebold_{0,b} I[E_{0,b}''] z^{|\tilde{C}_0(0)|} 
	\Ebold_{1,b} I[E_{1,b}''] z^{|Y_1^1|}  \cdots 
	\Ebold_{n-1,b} I[E_{1,b}''] z^{|Y_{n-1}^1|} 
	\Ebold_{n,b} I[E_{1,b}' ] z^{|Y_n^1|}  
	\right \| ,
\en
we use \refeq{dbnorm.1} to effectively replace $z$ by $|z|$, and then bound 
$|z|$ by 1.  We then bound the result using the BK inequality,
exactly as was done for $\hat\Phi(k)$ in 
Section~I.3.2, 
and obtain Feynman diagrams having propagator 
$\tau_{p_c}(0,x)$.  If we sum over $x$, we get diagrams which can 
be bounded by triangles.  With the presence of $|x|^2$, 
the situation is similar.  For $|x|^4$, we can arrange that two different
two-point functions each receive a factor $|x|^2$, and again standard methods
apply. 

For the diagrams involving at least one $E_2$ event, we proceed
as outlined in the paragraph containing \refeq{1-zQ}, and then bound
all factors of $|z|$ by $1$.  We are then left with bounding nested 
bond expectations.
We illustrate the method for bounding these nested bond expectations 
by considering a specific contribution to the case $n=2$, given by 
\eq
\lbeq{Tx}
	T(x) = p_c \sum_{(u_0,v_0)} p_c \sum_{(u_1,v_1)}
	\langle I[E_{0,b}'' 
	\langle I[E_{2,b}''] 
	\langle I[E_{1,b}' ]  
	\rangle_{2,b} \rangle_{1,b} \rangle_{0,b}.
\en  
The general case is similar, using the methods of \cite{HS90a}.  
The analysis is
simplified by the fact that we are taking the dimension to be large,
so that we can use squares and higher order diagrams in our bounds.

In this approach, the use of \refeq{1-zQ} loses any factors of $M_z$
that could be expected due to connections to $G$, and this prevents
us from obtaining bounds for sufficiently spread-out models in all 
dimensions $d>6$.  We do not know how to avoid the loss of the factors $M_z$
for complex $z$, though we were able to exploit these factors in I for real $z$.
As was mentioned previously, the case $m_2=4$ in \refeq{Pinbd}
also prevents us from handling 
all $d>6$.  In addition, we will be encountering complicated diagrams in
Sections~\ref{sec-Psidef2} and \ref{sec-Psider}, 
which also will prevent us from handling dimensions above but near $6$.
 
Using the BK inequality, we bound the innermost expectation 
$\langle I[E_{1,b}'(v_1, x; \tilde{C}_1^{\{u_1,v_1\}}(v_0)) ]\rangle_2$ 
of \refeq{Tx} by
\begin{equation}
\begin{split}
\lbeq{T2inmost.1}
	&
	\expec{I[\exists w_1 \in \tilde{C}_1^{\{u_1,v_1\}}(v_0), \exists a \in \Zd,
	(v_1 \conn a) \circ (a \conn w_1) \circ (w_1 \conn x)
	\circ (a \conn x)] }_2 
	\\ 
	& \hspace{1cm}
	\leq  \sum_{a,w_1 \in \Zd} I[w_1 \in \tilde{C}_1^{\{u_1,v_1\}}(v_0)] 
	\tau_1(v_1, a) \tau_1(a, w_1) \tau_1(w_1, x) \tau(a,x) .
\end{split}
\end{equation} 
By \refeq{E2ppb-def}, we can bound
$\langle I[E_{2,b}'' (v_0, u_1, v_1; \tilde{C}_0^{\{u_0,v_0\}}(0))] 
I[ w_1 \in \tilde{C}_1(v_0)] \rangle_1$ above by 
\begin{equation}
\begin{split}
\lbeq{T2-1.1}
	& \expec{I[\exists w_0 \in \tilde{C}_0^{\{u_0,v_0\}}(0), 
	(v_0 \conn u_1) \circ (u_1 \conn w_0) ] I[ w_1 \conn v_0] } 
	\\
	& \hspace{1cm} 
	\leq \sum_{w_0,t_1 \in \Zd}
	I[w_0 \in \tilde{C}_0^{\{u_0,v_0\}}(0)] \Big(
	\tau_1(v_0,u_1)\tau_1(u_1,t_1)\tau_1(t_1,w_0)\tau_1(t_1,w_1) 
	\\
	& \hspace{2.5cm} 
	+ \tau_1(v_0,t_1)\tau_1(t_1,w_1)\tau_1(t_1,u_1)\tau_1(u_1,w_0) \Big).
\end{split}
\end{equation} 
Finally, we bound 
$\langle I[E_{0, b}''(0, u_0, v_0) ] 
I[ w_0 \in \tilde{C}_0^{\{u_0,v_0\}}(0)]\rangle_0$ 
above by
\eq
\lbeq{T2-2.1}
	\expec{I[
	(0 \conn u_0) \circ (u_0 \conn 0) ] I[ w_0 \conn 0] } 
	\leq \sum_{t_0 \in \Zd} 
	\tau_1(0, u_0) \tau_1(u_0, t_0) \tau_1(t_0, 0) \tau_1(t_0, w_0)  
\en
Combining \refeq{T2inmost.1}--\refeq{T2-2.1}, we obtain as an upper bound
for $T_2(x)$ the diagrams 
\begin{center}
\setlength{\unitlength}{0.000400in}%
\begingroup\makeatletter\ifx\SetFigFont\undefined%
\gdef\SetFigFont#1#2#3#4#5{%
  \reset@font\fontsize{#1}{#2pt}%
  \fontfamily{#3}\fontseries{#4}\fontshape{#5}%
  \selectfont}%
\fi\endgroup%
\begin{picture}(6225,1374)(1051,-2248)
\thinlines
\put(1201,-1561){\line( 1, 2){300}}
\put(1201,-1561){\line( 1,-2){300}}
\put(1501,-961){\line( 0,-1){1200}}
\thicklines
\put(1501,-886){\line( 0,-1){150}}
\put(1501,-1036){\line( 0, 1){  0}}
\put(1576,-886){\line( 0,-1){150}}
\thinlines
\put(1576,-961){\line( 1, 0){825}}
\put(2401,-961){\line( 0,-1){1200}}
\put(2401,-2161){\line(-1, 0){450}}
\put(1951,-2161){\line(-1, 0){450}}
\thicklines
\put(2401,-886){\line( 0,-1){150}}
\put(2476,-886){\line( 0,-1){150}}
\thinlines
\put(2476,-961){\line( 1, 0){825}}
\put(3301,-961){\line( 1,-2){300}}
\put(3601,-1561){\line(-1,-2){300}}
\put(3301,-2161){\line(-1, 0){900}}
\put(3301,-961){\line( 0,-1){1200}}
\put(4801,-1561){\line( 1, 2){300}}
\put(5101,-961){\line( 0,-1){1200}}
\put(5101,-2161){\line(-1, 2){300}}
\thicklines
\put(5101,-886){\line( 0,-1){150}}
\put(5176,-886){\line( 0,-1){150}}
\thinlines
\put(5176,-961){\line( 1, 0){825}}
\put(6001,-961){\line( 0,-1){1200}}
\put(6001,-2161){\line(-1, 0){900}}
\thicklines
\put(6001,-2236){\line( 0, 1){150}}
\put(6076,-2236){\line( 0, 1){150}}
\thinlines
\put(6076,-2161){\line( 1, 0){825}}
\put(6901,-2161){\line( 1, 2){300}}
\put(7201,-1561){\line(-1, 2){300}}
\put(6901,-961){\line(-1, 0){900}}
\put(6901,-961){\line( 0,-1){1200}}
\put( 900,-1461){\makebox(0,0)[lb]{\smash{\SetFigFont{12}{14.4}{\rmdefault}{\mddefault}{\updefault}$0$}}}
\put(3726,-1461){\makebox(0,0)[lb]{\smash{\SetFigFont{12}{14.4}{\rmdefault}{\mddefault}{\updefault}$x$}}}
\put(4500,-1461){\makebox(0,0)[lb]{\smash{\SetFigFont{12}{14.4}{\rmdefault}{\mddefault}{\updefault}$0$}}}
\put(7326,-1461){\makebox(0,0)[lb]{\smash{\SetFigFont{12}{14.4}{\rmdefault}{\mddefault}{\updefault}$x$}}}
\put(5551,-2161){\circle*{150}}
\put(1951,-2161){\circle*{150}}
\put(4126,-1561){\makebox(0,0)[lb]{\smash{\SetFigFont{12}{14.4}{\rmdefault}{\mddefault}{\updefault}$+$}}}
\end{picture}
\end{center}
Summed over $x$, these diagrams can be bounded using the triangle diagram.
It is then routine to argue that, for $m=0,2,4$,
\eq
        \sum_{x} |x|^m \|  T_2(x) \| \leq O(d^{-2} ) . 
\en  
For the case $m=4$, we require that $d \geq 10$.

General $n$ can be handled similarly.  For example, 
a typical contribution arising in bounding 
$\Ebold_{0,b} I[E_{0,b}''] \Ebold_{1,b} I[E_{2,b}''] \Ebold_{2,b} I[E_{2,b}''] 
\Ebold_{3,b} I[E_{1,b}'' ] \Ebold_{4,b} I[E_{2,b}'' ]$ is the diagram
\begin{center}
\setlength{\unitlength}{0.000400in}%
\begingroup\makeatletter\ifx\SetFigFont\undefined%
\gdef\SetFigFont#1#2#3#4#5{%
  \reset@font\fontsize{#1}{#2pt}%
  \fontfamily{#3}\fontseries{#4}\fontshape{#5}%
  \selectfont}%
\fi\endgroup%
\begin{picture}(4962,1374)(1051,-2848)
\thinlines
\put(1201,-2161){\line( 1, 2){300}}
\put(1501,-1561){\line( 0,-1){1200}}
\put(1501,-2761){\line(-1, 2){300}}
\put(1576,-1561){\line( 1, 0){825}}
\put(2401,-1561){\line( 0,-1){1200}}
\put(2401,-2761){\line(-1, 0){900}}
\put(2476,-1561){\line( 1, 0){825}}
\put(3301,-1561){\line( 0,-1){1200}}
\put(3301,-2761){\line(-1, 0){900}}
\put(3376,-2761){\line( 1, 0){825}}
\put(4201,-2761){\line( 0, 1){1200}}
\put(4201,-1561){\line(-1, 0){900}}
\thicklines
\put(1501,-1486){\line( 0,-1){150}}
\put(1576,-1486){\line( 0,-1){150}}
\put(2401,-1486){\line( 0,-1){150}}
\put(2476,-1486){\line( 0,-1){150}}
\put(3301,-2836){\line( 0, 1){150}}
\put(3376,-2836){\line( 0, 1){150}}
\put(4201,-2836){\line( 0, 1){150}}
\put(4276,-2836){\line( 0, 1){150}}
\thinlines
\put(4201,-1561){\line( 1, 0){1300}}
\put(4701,-1561){\line( 1,-1){400}}
\put(5101,-1961){\line( 1, 1){400}}
\put(5201,-1561){\line(-1, 0){600}}
\put(4276,-2761){\line( 1, 0){825}}
\put(5101,-2761){\line( 0, 1){800}}
\thicklines
\put(5501,-1486){\line( 0,-1){150}}
\put(5576,-1486){\line( 0,-1){150}}
\thinlines
\put(5576,-1561){\line( 1, 0){425}}
\put(6001,-1561){\line( 0,-1){1200}}
\put(6001,-2761){\line(-1, 0){900}}
\put( 900,-2061){\makebox(0,0)[lb]{\smash{\SetFigFont{12}{14.4}{\rmdefault}{\mddefault}{\updefault}$0$}}}
\put(1951,-2761){\circle*{150}}
\put(2851,-2761){\circle*{150}}
\end{picture}
\end{center}
Such diagrams can be bounded using the square diagram.
The result, for $m=2,3,4$, is the bound 
\eq
	\sum_x |x|^m \| g_z^{(n)} (0,x) \| \leq O(d^{-n}) .
\en  
The case $m=4$ is not needed for \refeq{gnbd}, but it is used in \refeq{Pinbd}.
Any value of $m$ can be handled, at the cost of increasing $d$.  For $m=4$,
we require $d \geq 12$.
\qed

\medskip

For the proof of Lemma~\ref{lem-UNbd}, we will need the
cut-the-tail Lemma~I.3.5, 
which is restated as
Lemma~\ref{lem-tail.tau1} below.  
Although stated in I for constant $z$, the proof of the cut-the-tail
lemma extends immediately to site-dependent $z_u$.
In its statement, $M_{\vec{z}} = P_{p_c}(0 \conn G)$.

\begin{lemma}
\label{lem-tail.tau1}
Let $x$ be a site, $\{u,v\}$ a bond, and $E$ an increasing event. 
Then for a set of sites $A$ with $A \ni u$, 
and for $p \leq p_{c}$ (assuming no infinite cluster at $p_c$)
and $z_y \in [0,1]$ for all $y \in \Zd$, 
\eq 
	\expec { \Ind [ E \ON \tilde{C}^{\{u,v\}}(A) ]  \,  
	\tau_{\vec{z}}^{\tilde{C}^{\{u,v\}}(A)} (v, x) }
	\leq \frac{1}{1 - p M_{\vec{z}}} \, \prob{E} \, 
	\tau_{\vec{z}}(v,x) .
\en 
\end{lemma}

\medskip \noindent
{\bf Proof of Lemma~\ref{lem-UNbd}.}
Beginning with the definition of the remainder term in \refeq{UNdef},
and expanding the factors of $X$ using \refeq{Xndef}--\refeq{Xnppdef} 
as in the proof of Lemma~\ref{lem-Pinbd}, $U^{(n)}_{\vec{z}}(0,x)$
can be estimated by 
\eq
	| U_{\vec{z}}^{(n)}(0,x) |   \leq   
	\sum_{\sigma_j = 1, 2} 
	\left | \langle I_0'' \langle I_{1,\sigma_1} '' \langle \cdots 
	\langle I_{n-1, \sigma_{n-1}}'' \langle I_{n, \sigma_n} 
	\rangle_{n} \rangle_{n-1} \cdots
	\rangle_{2} \rangle_{1} \rangle_{0}  \right | 
\en 
where $I_{\ell, \sigma}''$ represents $I[E_{\sigma}'']$ on the level-$\ell$ 
expectation, and $I_{n, \sigma}$ represents $I[E_{\sigma}]$ in the level-$n$ 
expectation.  
Combining \refeq{Ej1} and \refeq{Ej2}, we have
\eq
	\expec{I[E_j(v, x; A)]} 
	= 
	\expec{I[E_j'(v, x; A)]} 
	+ 
	p \sum_{(u', v')} \expec{I[E_j''(v, u', v'; A)] 
	\tau_{\vec{z}}^{\tilde{C}^{\{u', v'\}}(v)} (v', x)
	}  .
	\lbeq{Un.inmost.1}
\en 
This identity is used for the innermost expectation 
$\langle I_{n, \sigma_n}\rangle$. 
 
The first term of \refeq{Un.inmost.1} is bounded as before,
using the analogue of \refeq{E1psite.1} and \refeq{E2psite.1} for 
site-dependent variables.  The result is 
\begin{align} 
	\left | \expec{I[E_1'(v, x; A)]} \right | 
	& = 
	\left | \expec{I[E_{1,b}'(v, x; A)] \, 
	{\textstyle{\prod_{u \in Y^1}}} z_u }_b 
	\right |
	\leq \expec{I[E_{1,b}'(v, x; A)]}_b ,  
	\\
	\left | \expec{I[E_2'(v, x; A)]} \right | 
	& = 
	\left | \expec{I[E_{1,b}'(v, x; A)] \, 
	( {\textstyle{\prod_{u \in Y^{2}}}} z_u  )
	( 1 - {\textstyle{\prod_{y \in Y^{3}}}} z_y) }_b \right |
	\leq 2 \expec{I[E_{1,b}'(v, x; A)]}_b ,  
\end{align} 
where in the second line we used $|(\prod_u z_u )(1-\prod_v z_v)| \leq 2$. 
The remaining diagrammatic estimation is routine, and yields a bound
$O(d^{-n})$. 

For the second term of \refeq{Un.inmost.1}, we use the cut-the-tail
Lemma~\ref{lem-tail.tau1}.  In preparation, we note by analogy with 
\refeq{E1ppsite.1} and \refeq{E2ppsite.1} that 
\eqarray 
	\expec{I[E_1''(v, u', v'; A)] \, 
	\tau_{\vec{z}}^{\tilde{C}^{\{u', v'\}}(v)} (v', x)}  
	& \leq & 
	\expec{I[E_{1,b}''(v, u', v'; A)]  \, 
	\tau_{\vec{z}}^{\tilde{C}^{\{u', v'\}}(v)} (v', x) 
	}_b ,  
	\\
	\expec{I[E_2''(v, u', v'; A)]\, 
	\tau_{\vec{z}}^{\tilde{C}^{\{u', v'\}}(v)} (v', x)}  
	& \leq &
	2 \expec{I[E_{2,b}''(v, u', v'; A)] \, 
	\tau_{\vec{z}}^{\tilde{C}^{\{u', v'\}}(v)} (v', x) }_b .  
\enarray
To obtain increasing events for application of the cut-the-tail lemma, 
we note that
\begin{align}
	& 
	\bigl \{ 
	E_{1,b}'(v, u', v'; A)  \ON \tilde{C}^{\{u', v'\}} 
	\bigr \} 
	\subset 
	\bigcup_{a,b \in \Zd} 
	\bigl \{ 
	\bar{F}_1(v, u', a, b; A) \ON \tilde{C}^{\{u', v'\}} 
	\bigr \} ,
	\\
	& 
	\bigl \{ 
	E_{2,b}' (v, u', v'; A)  \ON \tilde{C}^{\{u', v'\}} 
	\bigr \} 
	\subset \bigcup_{a \in \Zd} 
	\bigl \{ 
	\bar{F}_2(v, u', a; A)  \ON \tilde{C}^{\{u', v'\}} 
	\bigr \} ,
\end{align}	
where  
\eqarray
	\bar{F}_{1}(v, u', a, b; A) 
	& =  &
	\bigl \{
	(v \conn b) \circ (b \conn a) \circ 
	(a \conn u') \circ (b \conn u') 
	\bigr \} \cap 
	\{ a \in A \} ,
	\\
	\bar{F}_2(v, u', a; A) 
	& = & 
	\{ (v \conn u') \circ ( u' \conn a ) \} 
	\cap 
	\{ a \in A \} .  
\enarray
Now in this form, we can use Lemma~\ref{lem-tail.tau1}. 
After applying Lemma~\ref{lem-tail.tau1}, we use $\tau_{\vec{z}}(v',x)
\leq \tau_a(v',x)$.  The remaining diagrammatic estimation is routine. 
Summation of the factor $\tau_a(v',x)$ over $x$ gives the factor $\chi_a$
in the statement of the lemma.
\qed

\subsection{Proof of Lemma~\protect\ref{lem-Fklb}}
\label{sub-prf.lem-2.1ii}

In this section, we prove Lemma~\ref{lem-Fklb}, which asserts that
$\hat{F}_z(k)$ obeys the bound $|\hat{F}_{z}(k)| \geq -K d^{-1} 
+ \frac{1}{2e}{\rm Re}[1-z\hat{D}(k)]$ for high $d$, when $p=p_c$,
uniformly in $|z|<1$, and $k \in [-\pi,\pi]^d$.   

By \refeq{F100}, $\hat{\Pi}_1(0) =1$, and hence
$\hat{F}_{z}(k) = \Pihat_{1}(0) - \Pihat_{z}(k)$.
We write this as $\hat{F}_{z}(k) = A_{1} + A_{2}$, with 
$A_1$ and $A_2$ defined by
\eqarray 
        A_{1} & =  & p_c \sum_{(0,v)} \left [ 
        \expec{I[E''_{0}(0, 0, v)]}_{1} 
        - 
        \expec{I[E''_{0}(0, 0, v) ]}_{z} 
        e^{ik\cdot v}
        \right ],	\\ 
\lbeq{A2def}
        A_{2} & =  &
         p_c \sum_{(u,v): u \neq 0} \left [ 
        \expec{I[E''_{0}(0, u, v)]}_{1} 
        - 
        \expec{I[E''_{0}(0, u, v) ]}_{z} 
        e^{ik\cdot v} 
	\right ] +
        \sum_{n=1}^{\infty}  \left [ 
        \Pihat_{1}^{(n)}(0) - \Pihat_{z}^{(n)}(k) \right ]  .
\enarray 

The first term on the right side of \refeq{A2def} is $O(d^{-1})$ by
Lemma~I.3.4 
and \refeq{pc}, as follows.  
In the notation of Lemma~I.3.4, 
this term is 
$\hat{\Phi}^{(01)}_0(k) - \hat{\Phi}^{(01)}_h(k)$, which is bounded above
in absolute value
by $|\hat{\Phi}^{(01)}_0(k)| + |\hat{\Phi}^{(01)}_h(k)|$.
The $z$-dependence of the second term is dominated
by its value when $z=1$, and the bound
of Lemma~I.3.4
is easily seen to be $O(d^{-1})$, so that
$|\hat{\Phi}^{(01)}_0(k)| + |\hat{\Phi}^{(01)}_h(k)| \leq O(d^{-1})$.  
Therefore, by Lemma~\ref{lem-Pinbd}, 
\eq
\lbeq{Fzlower-A1-1}
        | A_{2} | 
        \leq O(d^{-1}) + \sum_{n=1}^{\infty} O(d^{-n}) 
	= O(d^{-1}).
\en 

The main term $A_{1}$ is, by definition, given by
\eq
\lbeq{Fzlower-A2-1}
        A_{1}  =  p_c \sum_{(0,v)} \sum_{n=1}^{\infty} 
        \expec{I[| \Ctilde^{\{0,v\}}(0)| = n]}_b  
        \left( 1 - z^{n} e^{ik\cdot v}
        \right) .
\en 
First we bound $| A_{1} |$ below by ${\rm Re}A_1$.
Since ${\rm Re} (1 - z^{n} e^{ik\cdot v}) \geq 0$,
we obtain a further lower bound by discarding the terms $n \geq 2$.  
The result is 
\eq 
        | A_{1} | \geq  {\rm Re} A_{1} 
        \geq p_c \sum_{(0,v)} 
        \prob{| \Ctilde^{\{0,v\}}(0)| = 1}  
        {\rm Re} \left [ 1 - z e^{ik\cdot v} \right ] 
\lbeq{Fzlower-A2-2}
        = 2dp_c  (1 - p_c ) ^{2d -1} 
        {\rm Re} \left [ 1 - z \Dhat(k) \right ] .
\en   
By \refeq{pc}, $2dp_c  (1 - p_c ) ^{2d -1} \geq \frac{1}{2e}$.
Combined with \refeq{Fzlower-A2-2} and 
\refeq{Fzlower-A1-1}, this proves Lemma~\ref{lem-Fklb}.
\qed

\section{The $z$-derivative of $\Pi_{z}$:  the second expansion}
\label{sec-Psidef}

The quantities $\Gamma_z$ and $\Psi_z$ were defined
in \refeq{FPsi}--\refeq{Gammadef} to obey, for $z \in [0,1)$,  
\eq  
\lbeq{Psidef5}
	z\frac{d}{dz} g_{z}(0,x) = \chi_z \Gamma_z(0,x),
	\quad \quad
        z\frac{d}{dz} \Pi_{z}(0,x) = \chi_z \Psi_z(0,x).
\en
In addition, \refeq{Gamma3def}--\refeq{Psi3def} require that
we identify functions $\Gamma_z^{(3)}$ and $\Psi_z^{(3)}$ which obey
\eq
\lbeq{Psi3def7}
	z_y \frac{\partial}{\partial z_y} g_{\vec{z}}(0,x)
	= \sum_{v}\Gamma^{(3)}_{\vec{z}}(0,x,v) \tau_{\vec{z}}(v,y),
	\quad\quad
	z_y \frac{\partial}{\partial z_y} \Pi_{\vec{z}} (0,x)
	= \sum_{v}\Psi^{(3)}_{\vec{z}}(0,x,v) \tau_{\vec{z}}(v,y).
\en
Because $g_{\vec{z}}$ and $\Pi_{\vec{z}}$ are almost identical, 
we consider only $\Pi_{\vec{z}}$ and $\Psi_{\vec{z}}$ explicitly.  
A similar analysis will apply
for $g_{\vec{z}}$ and $\Gamma_{\vec{z}}$.

In this section, we will derive an expression for 
$\Psi^{(3)}_{\vec{z}}(0,x,v)$.
This will also provide an expression for $\Psi_z(0,x)$.  In fact,
using \refeq{Psidef5} and
\refeq{Psi3def7}, it follows by summing over $y$ and setting
$z_y \equiv z$ that
\eq
	\sum_y z_y \left. \frac{\partial}{\partial z_y} \Pi_{\vec{z}} (0,x)
	\right|_{z_y=z}
	= z\frac{d}{dz} \Pi_{z}(0,x)
	= \chi_z \Psi_z(0,x)
	= \chi_z \sum_{a}\Psi^{(3)}_{\vec{z}}(0,x,a) .
\en
Therefore,
\eq
\lbeq{PsiPsi3}
	\Psi_z(0,x) = \sum_{a}\Psi^{(3)}_{\vec{z}}(0,x,a),
	\qquad \hat{\Psi}_z(k) = \hat{\Psi}^{(3)}_z(k,0).
\en
In particular, this yields the claim in Lemma~\ref{lem-Psi3disk}(i)
that $\hat{\Psi}^{(3)}_1(0,0)=\hat{\Psi}_1(0)$.  

The quantities $\Gamma$ and $\Psi$ will be defined by means
of a second expansion, as in the differentiation of $\Phi_z$ in
Section~I.5. 
However, in the differentiation of $\Phi_z$ the second expansion was
performed using the one-$M$ scheme, whereas here we will employ
the more extensive expansion of Section~\ref{sec-expansion}.
Throughout this section, we consider $z_u \in [0,a]$ with $a<1$,
and we fix $p=p_c$.

\subsection{The differentiation}
\label{sec-Psidiff}
 
In this section, we will derive an expression for the derivatives 
of $\Pi$ in \refeq{Psidef5} and \refeq{Psi3def7}.  
To differentiate $\Pi_{\vec{z}}(0,x)$, we recall from \refeq{Pindef}
that this is given as the sum over $N$ of 
\eq
	\Pi^{(N)}_{\vec{z}}(0,x) = (-1)^N p_c \sum_{u_N}
	\Ebold_0 I[E_0''] \Ebold_1 X_1'' \cdots
	\Ebold_{N-1} X_{N-1}'' \Ebold_N X_N'' 
	\quad (N=0,1,2,\ldots) .
\en  
Here the nested expectations are performed from right to left,
$X_n'' = I[E_1''(v_{n-1}, u_n, v_n; \tilde{C}_{n-1})]
- I[E_2''(v_{n-1}, u_n, v_n; \tilde{C}_{n-1})]$, summations
$p_c \sum_{(u_0,v_0)} \cdots p_c \sum_{(u_{N-1},v_{N-1})}$ are tacitly
understood on the right side, and we write $\Ebold$ in place of 
$\langle \cdot \rangle$ to denote a joint bond/site expectation.  
Therefore, $\Pi^{(N)}_{\vec{z}}(0,x)$
is given by a sum of $2^N$ terms involving
\eq
\lbeq{PiNnestEpp}
	\Ebold_0 I[E_0''] \Ebold_1 I[E_{\sigma_1}''] \cdots
	\Ebold_{N-1} I[E_{\sigma_{N-1}}'']  \Ebold_N I[E_{\sigma_{N}}''] ,
\en
with each $\sigma_n$ equal to $1$ or $2$.

Before proceeding further, we recall an observation from
Example~I.4.5 
that will
be used repeatedly in what follows.
By the definition of ``occurs on $\tilde{C}$'' in
Definition~\ref{def-event-on}(c),
an expectation 
involving $E_\sigma''$ can be written as a \emph{conditional} expectation of 
$E_\sigma'(v_{n-1}, u_n; A)$, under the 
condition that $\{u_n, v_n\}$ is vacant.  
We will write $\tilde{\Ebold}$ or $\expec{\, \cdot \,}^{\tilde{}}$ 
for this conditional expectation. 
In the conditional expectations, we may replace $E_\sigma''$ events
by $E_\sigma'$, and we may drop the tilde from 
$\tilde{C}_n^{\{u_{n}, v_{n}\}}(v_{n-1})$, 
since $C_n=C_n(v_{n-1})$ is then equal to 
$\tilde{C}_n^{\{u_{n}, v_{n}\}}(v_{n-1})$.  Therefore \refeq{PiNnestEpp}
can be rewritten as
\eq
\lbeq{PiNnestEp}
	\Ebold_0 I[E_0'']  \cdots \Ebold_{n-1} I[E_{\sigma_{n-1}}'']
	\tilde{\Ebold}_{n} I[E_{\sigma_n}'] 
	\tilde{\Ebold}_{n+1} I[E_{\sigma_{n+1}}']
	\Ebold_{n+2} I[E_{\sigma_{n+2}}'']  \cdots
	\Ebold_N I[E_{\sigma_{N}}''] .
\en

Our goal is to differentiate \refeq{PiNnestEp} with respect to $z_y$.
Its $z_y$-dependence resides in the $z_y$-dependence
of each of the nested expectations.  For any individual expectation,
the form of this dependence is different for $\sigma =1$ and
$\sigma =2$, and can easily be given explicitly with the help of
the clusters $Y^{1}, Y^2, Y^3$ depicted in 
Figure~\ref{fig-Q1-5.def}.  By definition, $Y^2$ and $Y^3$ are disjoint.
To simplify the notation, given a finite
set $Y \subset \Zd$, we will write
\eq
	z^Y = \prod_{u \in Y} z_u.
\en 
The $z$-dependence of an expectation involving $E_{1}'$ is
$z^{Y^{1}}$, and that of an expectation involving $E_{2}'$ is
$z^{Y^{2}} ( 1 - z ^{Y^{3}})$. The overall $z$-dependence of
the nested expectation is thus a product of such factors, and the
differentiation can be performed using the product rule, with one
term arising from differentiation of the $z$-dependence associated 
with each of the $N+1$ expectations.  Each of the $2^N$ terms
\refeq{PiNnestEp} thus gives rise, after application of 
$z_y \frac{\partial}{\partial z_y}$, 
to $N+1$ terms of the form
\eq
\lbeq{PiNnestdiff}
	\Ebold_0 I[E_0''] \cdots \Ebold_{n-1} I[E_{\sigma_{n-1}}''] 
	z_y \frac{\partial}{\partial z_y} 
	\tilde{\Ebold}_n I[E_{\sigma_{n}}'] 
	\tilde{\Ebold}_{n+1} I[E_{\sigma_{n+1}}']
	\Ebold_{n+2} I[E_{\sigma_{n+2}}'']  \cdots
	\Ebold_N I[E_{\sigma_{N}}''] ,
\en
with only the $z$-dependence of the $n^{\rm th}$ expectation being
differentiated.  The case $n=0$ involves no new difficulties compared
to the other values of $n$, and will not be discussed explicitly in
what follows.

When $E_{\sigma_n}' = E_1'$, differentiation gives 
\eq
	z_y \frac{\partial}{\partial z_y}  z^{Y^{1}} 
	=  I[y \in Y^{1}] \, z^{Y^{1}}, 
\en   
while for $E_{2}'$, differentiation gives 
\eq 
\lbeq{22.after.diff}
        z_y \frac{\partial}{\partial z_y} 
	z^{Y^{2}} ( 1 - z ^{Y^3})
        = I[y \in Y^2]  z^{Y^2} ( 1 - z ^{Y^3}) 
        - I[y \in Y^3] z^{Y^2}z^{Y^3} .
\en  
The factors $I[y \in Y]$ appearing in the above
entail a connection to $y$ within the $n^{\rm th}$
expectation.   We wish to ``cut off'' this connection at a suitable
pivotal bond $(a',a)$, to extract a 
factor $\tau_{\vec{z}}(a,x)$ from \refeq{PiNnestdiff}.
However, to do so requires dealing with the fact that this connection 
to $y$ is
not an independent event, and we handle this by means of
a second application of the expansion.  This second application
of the expansion is in principle similar to the first application of
the expansion in Section~\ref{sec-expansion}, but in practice 
it is more technical and complicated.
The derivation of the second expansion will be completed in
Section~\ref{sec-Psi}. 
The second expansion will then be bounded in Section~\ref{sec-Psidef2}.
The analysis is similar to that of Section~I.5. 

After differentiation of the
$n^{\rm th}$ expectation, three possibilities occur for this expectation.
When $\sigma_n =1$, the $n^{\rm th}$ expectation contains 
\eq
\lbeq{E1Q12}
	I[E_{1}']I[y \in Y^{1} ].
\en 
When $\sigma_n =2$, two terms result.
The first term on the right side of \refeq{22.after.diff} produces
\eq
\lbeq{E2Q34}
	I[E_{2}']I[y \in Y^2].  
\en
The second term on the right side of \refeq{22.after.diff} produces
(with a minus sign) 
\eq
\lbeq{E3Q5}
	I[E_3']I[y \in Y^3],
\en
for a new event $E_{3}'(v_{n-1}, u_{n}; C_{n-1})$
which is defined as follows.  Let $S$ denote the last sausage for
the connection from $v_{n-1}$ to $u_n$.  Then
\begin{description}
\item $E_{3}'(v_{n-1}, u_{n}; C_{n-1})$ is the intersection of the
events:  (i) $v_{n-1} \conn u_{n} \in \Zd \backslash C_{n-1}$,
(ii) $S \cap G = \emptyset$, (iii) $S \cap C_{n-1} \neq \emptyset$ 
({\em i.e.}\/ $Y^3$ is not empty), and (iv) if 
$\exists g \in (C(v_{n-1}) \backslash S) \cap G$, then 
$v_{n-1} \ct{C_{n-1}} g$. 
\end{description}
The event $E_3'$ was depicted in Figure~\ref{fig-Q1-5.def}.
This unifies \refeq{E1Q12}--\refeq{E3Q5} and allows us to write 
\refeq{PiNnestdiff} as a sum of terms of the form
\eq
\lbeq{EnuYnu}
	\Ebold_0 I[E_0''] \cdots \Ebold_{n-1} I[E_{\sigma_{n-1}}'']
	\tilde{\Ebold}_n I[E_\nu '] I[y \in Y^\nu ] 
	\tilde{\Ebold}_{n+1} I[E_{\sigma_{n-1}}']
	\Ebold_{n+2} I[E_{\sigma_{n+2}}'']  \cdots
	\Ebold_N I[E_{\sigma_{N}}''] 
	\quad (\nu =1,2,3).
\en

\subsection{The cutting lemma}
\label{sec-secondcut}

Our task is to perform an expansion to cut off a factor of 
$\tau_{\vec{z}}(a,y)$
corresponding to the connection to $y \in Y^\nu$ in \refeq{EnuYnu}.  
This requires the identification of a ``cutting bond,'' where the connection
to $y$ will be severed.  We wish to cut off a $G$-free connection to $y$,
which will be possible since by definition $Y^\nu \cap G = \emptyset$
for $\nu =1,2,3$.

We begin by narrowing the focus in \refeq{EnuYnu} to the
two relevant expectations
\eq
\lbeq{cutfocus1}
	\cdots \tilde{\Ebold}_n I[E_\nu '] I[y \in Y^\nu ] 
	\tilde{\Ebold}_{n+1} I[E_{\sigma_{n+1}}'] \cdots 
	\quad (\nu =1,2,3; \sigma_{n+1}=1,2) .
\en
Note that the event $\event{y \in Y^\nu}$ only makes sense when
it occurs in conjunction with the event $E_\nu'$.  
The $(n+1)^{\rm st}$ expectation is relevant, because $E_{\sigma_{n+1}}'$
depends on the cluster $C_n$.  We apply Fubini's theorem 
to interchange the $n^{{\rm th}}$ and 
$(n+1)^{{\rm st}}$ bond/site expectations, to write \refeq{cutfocus1} as
\eq
\lbeq{cutfocus}
	\cdots \tilde{\Ebold}_{n+1} 
	\tilde{\Ebold}_n I[E_\nu '] I[y \in Y^\nu ] 
	I[E_{\sigma_{n+1}}'] \cdots .
\en
We will work within the expectation $\tilde{\Ebold}_n$, regarding
the clusters of levels $0, \ldots, n-1$ and $n+1$ as being fixed.

The event $E_{\sigma_{n+1}}'$ can be decomposed as an intersection
of events on levels $n$ and $n+1$, as
\eq
\lbeq{Esigdecomp}
	E_{\sigma_{n+1}}'(v_{n}, u_{n+1} ; C_{n})_{n+1} 
	= E_{\sigma_{n+1},0}(v_{n}, u_{n+1} )_{n+1} 
	\cap J_{\rm cut}^{(\sigma_{n+1})}(C_{n+1})_{n}, 
\en 
where we have introduced an event $E_{\sigma_{n+1},0}$ which ensures
that $C_{n+1}$ contains certain bond connections and an event
$J_{\rm cut}^{(\sigma_{n+1})}(C_{n+1})_{n}$ which forces $C_n$ to
be compatible with $C_{n+1}$.  More precisely, the events $E_{\sigma_{n+1},0}$
are defined by
\begin{align}
   	E_{1,0}(v_{n}, u_{n+1})_{n+1} 
	= &
	\event{v_{n} \conn u_{n+1} \AND 
	\tilde{C}^{\{u_{n+1}',v_{n+1}'\}}(u_{n+1}) \cap G_{n+1} = \emptyset} ,
	\\  \nonumber
	E_{2,0}(v_{n}, u_{n+1})_{n+1} 
	= &
	\event{v_{n} \conn u_{n+1} \AND 
	\tilde{C}^{\{u_{n+1}',v_{n+1}'\}}(u_{n+1}) \cap G_{n+1} \neq \emptyset} ,
\end{align}
with $(u_{n+1}',v_{n+1}')$ the last pivotal bond for the level-$(n+1)$
connnection from $v_n$ to $u_{n+1}$.  If there is no such pivotal bond,
then $\tilde{C}^{\{u_{n+1}',v_{n+1}'\}}(u_{n+1})$ is to be interpreted
as $C(u_{n+1})$.
The events 
$J_{\rm cut}^{(\sigma_{n+1})}(C_{n+1})_{n}$
are defined by
\begin{align}
	J_{\rm cut}^{(1)}(C_{n+1})_{n} 
	= &
	\Bigl \{ 
	C_{n} \textrm{ intersects } C_{n+1} 
	\textrm{ such that the level-$(n+1)$ 
	connections satisfy:} 
	\nonumber \\ \nonumber
	& 
	\left( v_{n+1}' \dbc u_{n+1} \textrm{ through } C_n \right)
	\AND 
	\left(v_n \conn u_{n+1}' \IN \Zd \backslash C_n \right)
	\\
	& \AND 
	\left( \textrm{if } \exists g \in 
	\tilde{C}^{\{u_{n+1}',v_{n+1}'\}}(v_{n}) \cap G_{n+1}
	\textrm{ then } v_{n+1}' \ct{C_n} g  \right)
	\Bigr \}
	, 
	\\ \nonumber
	J_{\rm cut}^{(2)}(C_{n+1})_{n} 
	= &
	\Bigl \{ 
	C_{n} \textrm{ intersects } C_{n+1} 
	\textrm{ such that the level-$(n+1)$ 
	connections satisfy:} 
	\\ \nonumber
	& 
	\left( 
	(u_{n+1}  \ct{C_n} G_{n+1}) \IN  
	(\textrm{ last sausage of } v_{n} \conn u_{n+1} )
	\right)
	\AND 
	\left( v_n \conn u_{n+1} \IN \Zd \backslash C_n \right)
	\\
\lbeq{Jcut2def}
	& \AND 
	\left( \textrm{if } \exists g \in 
	\tilde{C}^{\{u_{n+1}',v_{n+1}'\}}_{n+1}(v_{n}) \cap G_{n+1}
	\textrm{ then } v_{n+1}' \ct{C_n} g \right)
	\Bigr \}
	. 
\end{align}
Thus we can rewrite \refeq{cutfocus} as
\eq
\lbeq{cutfocusJ-1}
	\cdots \tilde{\Ebold}_{n+1} 
	I[E_{\sigma_{n+1},0}(v_{n}, u_{n+1} )_{n+1}]
	\tilde{\Ebold}_n I[E_\nu ']_n I[y \in Y^\nu ]_n
	I[J_{\rm cut}^{(\sigma_{n+1})}(C_{n+1})_{n}]
	\cdots .
\en
Defining
\eq
\lbeq{Jnusigdef}
	J_{\nu, \sigma_{n+1}}(y) =
	E_{\nu}'(v_{n-1}, u_{n} ; C_{n-1}) \cap \{y \in Y^\nu \} \cap
	J_{\rm cut}^{(\sigma_{n+1})}(C_{n+1}),
\en
\refeq{cutfocusJ-1} can be rewritten as
\eq
\lbeq{cutfocusJ}
	\cdots \tilde{\Ebold}_{n+1} 
	I[E_{\sigma_{n+1},0}(v_{n}, u_{n+1} )_{n+1}]
	\tilde{\Ebold}_n I[J_{\nu, \sigma_{n+1}}(y)_{n}]
	\cdots .
\en

We now choose the ``cutting bond'' $(a',a)$.
We recall the existence of an ordering of the pivotal bonds for 
the connection from a site to a set of sites, as defined in 
Definition~\ref{def-percterms}(d).
The \emph{cutting bond} is defined to be 
the last pivotal bond $(a',a)$ for the connection 
$y \to \{v_{n-1}, u_{n}\} \cup C_{n+1} \cup C_{n-1}$. 
It is possible that no such pivotal bond exists, and in 
that case, no expansion will be required.

\begin{figure}
\begin{center}
\includegraphics[scale = 0.4]{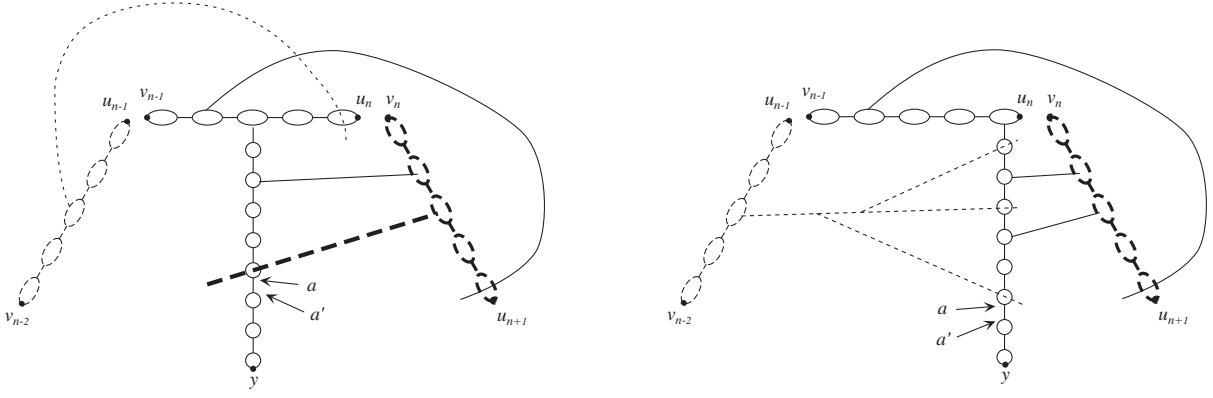} 
\end{center}
\caption{Two examples of the choice of cutting bond $(a',a)$.
Solid lines represent $C_n$, dashed lines represent $C_{n-1}$, and bold
dashed lines represent $C_{n+1}$.}
\label{fig-gen-cut}
\end{figure}

In choosing the cutting bond, 
we require it to be pivotal for $\{v_{n-1}, u_{n}\}$ 
to preserve (on the $a$ side of the cluster $C_{n}$) 
the backbone structure of the cluster $C_{n}$ which is 
required by $E_\nu '(v_{n-1}, u_{n} ; C_{n-1})$.   
We require the cutting bond to be pivotal for
$C_{n+1}$ to ensure that we do not 
cut off as a tail something which may be needed 
to ensure that the last sausage
of $C_{n+1}$ is connected through $C_{n}$, or that a previous
sausage of $C_{n+1}$ is connected to $G_{n+1}$ through $C_{n}$.  
Finally, we require the cutting bond to be pivotal
for the connection to $C_{n-1}$ because connections to $G$ must
pass through $C_{n-1}$ and hence this realizes our goal to cut off
a $G$-free connection to $y$.  The above choice of the cutting bond is
simpler than that of Section~I.5.3. 
We do not expect our choice here to allow for dimensions near $d=6$
to be handled, but it will simplify our analysis.

Recall the definition of $\tilde{C}^{(a,a')}(A)$ given in 
Definition~\ref{def-percterms}.
We now define several events, 
depending on $y,v_{n-1}, u_{n}, C_{n-1}, C_{n+1}$,
but to simplify the notation we make only the $y$-dependence explicit
in the notation.  
We also make the abbreviation
\eq
\lbeq{Andef}
	A_n = \{v_{n-1}, u_{n}\} \cup C_{n+1} \cup C_{n-1}.
\en
Let
\begin{align}
\lbeq{J1def}
	J_{\nu, \sigma_{n+1}}'(y)_{n}
	= \,  & J_{\nu, \sigma_{n+1}}(y)_n \cap 
	\event{y \dbc A_n }  ,
	\\ 
\lbeq{J1ppdef}
	J_{\nu, \sigma_{n+1}}''(a, a')_{n} 
	= \, & 
	\{ J_{\nu, \sigma_{n+1}}'(a)_n \ON  
	\tilde{C}_{n}^{\{a,a'\}}(A_n) 
	\} ,
	\\
	J_{\nu, \sigma_{n+1}}(a, a', y)_{n}
	= \, & 
	J_{\nu, \sigma_{n+1}}(y)_{n} \cap 
	\event {(a', a) \textrm{ is the last occupied pivotal bond for }
	y \to A_n} .
\end{align}
The overall level-$n$ event $J_{\nu, \sigma_{n+1}}(y)_{n}$ can be 
written as the disjoint union
\eq
\lbeq{J1decomp}
	J_{\nu, \sigma_{n+1}}(y)_{n} = 
	J_{\nu, \sigma_{n+1}}'(y)_{n} \bigcup^{\cdot}
	\biggl ( \bigcup_{(a,a')}^{\cdot} 
	J_{\nu, \sigma_{n+1}}(a, a', y)_n \biggr ) . 
\en  
In \refeq{J1decomp}, configurations in $J_{\nu, \sigma_{n+1}}(y)_n$ 
have been classified
according to the last pivotal bond $(a', a)$. The appearance of
$J_{\nu, \sigma_{n+1}}'$ corresponds to the possibility that 
there is no such pivotal bond.   
For the configurations in which there is a pivotal bond, we will use
the following important lemma. 

\begin{lemma}
\label{lem-11cut}
For $\nu = 1,2,3$ and $\sigma_{n+1} =1,2$,
\begin{align}
	J_{\nu, \sigma_{n+1}}(a, a', y)_n
	= \, & J_{\nu, \sigma_{n+1}}''(a, a')_{n}
	\cap 
	\event{ ( y \conn a' \AND y \nc G) \INSIDE \Zd \backslash 
         \tilde{C}_n^{\{a,a'\}}(A_n) } 
	 \nnb 
	 & 
	 \cap \event{\{a, a'\} \textnormal{ is occupied}}.
\lbeq{11cuteq}
\end{align}
\end{lemma}

Before proving the lemma, we note that together with \refeq{J1decomp} and
Lemma~\ref{lem-cond.0} it implies the identity
\eq
	\expec{I[J_{\nu, \sigma_{n+1}}(y)_{n}]}_{n}^{\tilde{}} = 
	\expec{I[J_{\nu, \sigma_{n+1}}'(y)_{n}]}_{n}^{\tilde{}} +
	p \sum_{(a,a')}  
	\langle I[ J_{\nu, \sigma_{n+1}}''(a, a')_{n}] \, 
	\tau_{z}^{\tilde{C}_n^{\{a,a'\}}
	(A_n)}(a', y) \rangle_{n}^{\tilde{}} . 
	\lbeq{J1-cut.2}
\en 
This will be the point of departure for the second expansion.
We have actually employed a minor modification of Lemma~\ref{lem-cond.0}
to the conditional expectation $\langle \cdot \rangle^{\tilde{}}$, and
initially the restricted two-point function appearing in the above equation
should be with respect to the conditional, rather than the usual expectation.
However, there is no difference between the two.  To see this,
note that the event that $a' \conn y$ in  
$\Zd \backslash \tilde{C}_n^{\{a,a'\}}(A_n)$
is independent of the bond $\{u_{n}, v_{n}\}$, since this bond
touches the set $\tilde{C}_n^{\{a,a'\}}(A_n)$.  Therefore either expectation
can be used for the restricted two-point function, 
and for simplicity, we will use the ordinary unconditional expectation.

\bigskip
\noindent 
\textbf{Proof of Lemma~\ref{lem-11cut}.} 
To abbreviate the notation, we define
\eq
	E_{\rm piv} = \{(a', a) \textrm{ is pivotal for }y \to A_n \}.
\en
We will show below that
\eqsplit 
\lbeq{Jclaim}
	J_{\nu, \sigma_{n+1}}(a, a', y)_n 
	= &
	\{ J_{\nu, \sigma_{n+1}}(a)_n \ON \tilde{C}^{\{a,a'\}}(A_n)\} 
	\\ &
	\cap \{(y \conn a' \AND y \nc G) 
	\INSIDE \Zd\backslash \tilde{C}^{\{a,a'\}}(A_n)\} 
	\\
	& \cap \event{ \{a,a'\} \textrm{ is occupied }} 
	\cap E_{\rm piv} . 
\ensplit
By Lemma~\ref{lem-pivotal2}, 
$E_{\rm piv} = \{ a \conn A_n \ON \tilde{C}^{\{a, a'\}}(A_n) \} \cap
\{ y \conn a' \mbox{ in } \Zd\backslash \tilde{C}^{\{a, a'\}}(A_n) \}$.
Thus \refeq{Jclaim} is equivalent to
\begin{align}
	J_{\nu, \sigma_{n+1}}(a, a', y)_n 
	= &
	\bigl \{ J_{\nu, \sigma_{n+1}}(a)_n \ON \tilde{C}^{\{a, a'\}}(A_n) 
	\bigr \} \cap \event{ \{a,a'\} \textrm{ is occupied }}
	\nnb 
	& \cap
        \bigl \{ 
        (y \conn a' \AND y \nc G) \INSIDE 
        \Zd\backslash \tilde{C}^{\{a, a'\}}(A_n) 
	\bigr \}    , 
\end{align}
which is the desired identity \refeq{11cuteq}. 

It remains to prove \refeq{Jclaim}.
By definition of $J_{\nu, \sigma_{n+1}}(a, a', y)_n$, 
\eq
\lbeq{J1is}
	J_{\nu, \sigma_{n+1}}(a, a', y)_n 
	= \event{ \{a,a'\} \textrm{ is occupied }}
	\cap J_{\nu, \sigma_{n+1}}(y)_n \cap 
	\event{a \dbc A_n} 
	\cap E_{\rm piv} . 
\en 
Combining \refeq{J1is} and \refeq{Jnusigdef}, we have
\eqalign
	J_{\nu, \sigma_{n+1}}(a, a', y)_n 
	=  & \event{ \{a,a'\} \textrm{ is occupied }}
	\cap
	\event{y \in Y^\nu} \cap 
	E_{\nu}'(v_{n-1}, u_{n}; C_{n-1})_{n} 
	\nnb 
	&
	\cap 
	J_{\rm cut}^{(\sigma_{n+1})}(C_{n+1})_{n}
	\cap 
	\event{a \dbc A_n} 
	\cap E_{\rm piv} .
\enalign
To see that this can be written in the form \refeq{Jclaim}, we will
analyze the various events in the above expression.

First, we claim that 
\eq
\lbeq{Jcutpf}
	J_{\rm cut}^{(\sigma_{n+1})}(C_{n+1})_{n}  \cap E_{\rm piv}
	=  \{ J_{\rm cut}^{(\sigma_{n+1})}(C_{n+1})_{n} 
	\ON \tilde{C}^{\{a, a'\}}(A_n) \}
         \cap E_{\rm piv} . 
\en
In fact, if the left side occurs, then the right side occurs because
$(a',a)$ is pivotal for the connection
$y \to C_{n+1}$ (since $E_{\rm piv}$ occurs)
and hence all $C_n$'s intersections with $C_{n+1}$
are independent of bonds not touching $\tilde{C}^{\{a, a'\}}(A)$.
Conversely, the right side is contained in the left side for the same reason.

Next, we claim that
\eq
\lbeq{adbcApf}
	\event{a \dbc A_n}  \cap E_{\rm piv}
	=  \{ a \dbc A_n \ON \tilde{C}^{\{a, a'\}}(A_n) \} \cap E_{\rm piv}. 
\en
In fact, 
because $\event{a \dbc A_n}$ is increasing, the right side is contained in
the left side.  Conversely, for a configuration on the left side,
it must be the case that $\{ a \dbc A_n \ON \tilde{C}^{\{a, a'\}}(a) \}$,
and since $\tilde{C}^{\{a, a'\}}(a) \subset \tilde{C}^{\{a, a'\}}(A_n)$,
the right side occurs.

Finally, we will prove that
\eqalign
\lbeq{F1pf.1}
	&
	E_{\nu}'(v_{n-1}, u_{n}; C_{n-1})_{n}
	\cap \event{y \in Y^\nu}
	\cap E_{\rm piv}
	\cap \event{\{a, a'\} \textnormal{ is occupied }}
	\nnb 
	& \hspace{1cm}
	=  \left\{ E_{\nu}'(v_{n-1}, u_{n} ; C_{n-1}) 
        \ON  \tilde{C}^{\{a,a'\}}(A_n) \right\} 
        \cap 
	\left\{ y \nc G \IN 
	\Zd\backslash \tilde{C}^{\{a,a'\}}(A_n) \right\} 
	\nnb 
	& \hspace{1.5cm}
        \cap
	\{ y \conn a' \textnormal{ in } 
	\Zd\backslash \tilde{C}^{\{a, a'\}}(A_n) 
	\}
	\cap
	\{ a \in Y^\nu \ON  \tilde{C}^{\{a, a'\}}(A_n) \} 
	\nnb 
	& \hspace{1.5cm} 
	\cap 
	\event{\{a, a'\} \textnormal{ is occupied }}.
\enalign  
The event $J_{\nu, \sigma_{n+1}}(a,a',y)_n$ 
is the intersection of the events occurring
on the left sides of \refeq{Jcutpf}, \refeq{adbcApf}
and \refeq{F1pf.1}.  Therefore it
is the intersection of the events occurring 
on the right sides of these equations.  
By \refeq{Jnusigdef} and \refeq{J1def}--\refeq{J1ppdef},
a rearrangement of these right side
events then gives \refeq{Jclaim} and completes the proof.   It remains
to prove \refeq{F1pf.1}.

In preparation for this, we first observe that
\eqarray
\lbeq{yvpf.0}
	\event{y \conn v_{n-1}} 
	\cap E_{\rm piv}
	&= & 
	\event{y \conn a'}  \cap
	\event{ \{a,a'\} \textrm{ is occupied }} 
	\nonumber \\ &&
	\cap
	\{ a \conn v_{n-1} \ON \tilde{C}^{\{a, a'\}}(A_n) \}
	\cap E_{\rm piv} .
\enarray
In fact, the right side is clearly contained in the left side.  Conversely,
for a configuration on the left side, since $v_{n-1} \in A_n$,
the bond $(a',a)$ must also
be pivotal for $y \conn v_{n-1}$, and this implies that
$\event{y \conn a'}$, that $\{a',a\}$ is occupied, and that
$a \conn v \ON \tilde{C}^{\{a, a'\}}(v_{n-1})$.  Since
$\tilde{C}^{\{a, a'\}}(v_{n-1}) \subset \tilde{C}^{\{a, a'\}}(A_n)$,
this implies 
$\{ a \conn v_{n-1} \ON \tilde{C}^{\{a, a'\}}(A_n) \}$.
This proves \refeq{yvpf.0}.
Now, by Lemma~\ref{lem-pivotal2}, $E_{\rm piv}$
is the intersection of the events
$\{ y \conn a' \INSIDE \Zd\backslash \tilde{C}^{\{a, a'\}}(A_n) \}$ and
$\{ a \conn A_n \ON \tilde{C}^{\{a, a'\}}(A_n) \}$, and hence
it follows from \refeq{yvpf.0} that
\eqarray
\lbeq{yvpf}
	\event{y \conn v_{n-1}} 
	\cap E_{\rm piv}
	& = & 
	\event{ \{a,a'\} \textrm{ is occupied }}
	\cap
	\{ a \conn v_{n-1} \ON \tilde{C}^{\{a, a'\}}(A_n) \}
	\nnb && 
        \cap
	\{ y \conn a' \INSIDE \Zd\backslash \tilde{C}^{\{a, a'\}}(A_n) 
	\} .
\enarray

Returning to \refeq{F1pf.1}, we first note that 
on the right side of \refeq{F1pf.1} we may add an intersection with 
$\{a \conn v_{n-1} \ON \tilde{C}^{\{a, a'\}}(A_n) \}$ since this is 
a subset of $\{ a \in Y^\nu \ON  \tilde{C}^{\{a, a'\}}(A_n) \}$.
Hence by \refeq{yvpf}, \refeq{F1pf.1} is
equivalent to
\eqalign
\lbeq{F1pf.2}
	&
	E_{\nu}'(v_{n-1}, u_{n}; C_{n-1})_{n}
	\cap \event{y \in Y^\nu }
	\cap E_{\rm piv}
	\cap \event{\{a, a'\} \textnormal{ is occupied }}
	\nnb 
	& \hspace{1cm}
	=  \left\{ (E_{\nu}'(v_{n-1}, u_{n} ; C_{n-1}) 
	\cap \{ a \in Y^\nu \} )
        \ON  \tilde{C}^{\{a,a'\}}(A_n) \right\} 
        \cap 
	\left\{ y \nc G \IN 
	\Zd\backslash \tilde{C}^{\{a,a'\}}(A_n) \right\} 
	\nnb 
	& \hspace{1.5cm}
	\cap E_{\rm piv}
	\cap \event{y \conn v_{n-1}}
	\cap 
	\event{\{a, a'\} \textnormal{ is occupied }}.
\enalign
To obtain \refeq{F1pf.1}, it therefore suffices to prove \refeq{F1pf.2}.

To prove \refeq{F1pf.2}, 
we begin by supposing we have a configuration in the left side.  To show
that it is in the right side, it suffices to 
show that the first two events on the right side must then occur.
For a configuration on the left side, we must have $a \in Y^\nu
\ON \tilde{C}_n^{\{a,a'\}}(a)$, and therefore $a \in Y^\nu
\ON \tilde{C}_n^{\{a,a'\}}(A_n)$.  Also,
because $(a',a)$ is pivotal for $y \to C_{n-1}$, it follows that
$\Ctilde^{\{a',a\}}(y) \cap G = \emptyset$.  This in turn
implies that $y \nc G$ in $\Zd\backslash \tilde{C}_{n}^{\{a,a'\}}(A_n)$.
It remains to show that 
$\{ E_{\nu}'(v_{n-1}, u_{n} ; C_{n-1}) \ON \tilde{C}^{\{a,a'\}}(A_n) \}$.

For this, consider first $\nu=1$.
The connections $v_{n}' \dbc u_n$ through $C_{n-1}$
and $v_{n-1} \conn u'_{n}$ in $\Zd \backslash C_{n-1}$
required by $E_1'$ are conditions on the backbone of $C_n$, which
is not affected by turning off all bonds not touching 
$\tilde{C}^{\{a,a'\}}(A_n)$.  Hence these connections occur on
$\tilde{C}^{\{a,a'\}}(A_n)$.  Also, because $(a',a)$ is pivotal for
$y \to G$ (since connections to $G$ are through $C_{n-1} \subset A_n$),
the restrictions on
connections to $G$ imposed by $E_1'$ are unaffected by the status
of bonds or sites not touching $\Ctilde^{\{a',a\}}(A_n)$.  This implies
$\{ E_{1}'(v_{n-1}, u_{n} ; C_{n-1})$ occurs on 
$\tilde{C}^{\{a,a'\}}(A_n) \}$.
The cases $\nu =2,3$ are similar.  Therefore the
left side of \refeq{F1pf.2} is contained in the right side.

Conversely, given a configuration on the right side of \refeq{F1pf.2}, 
we need to show
that $E_\nu'$ occurs and that $y \in Y^\nu$.  
The fact that $y \in Y^\nu$ is a consequence of 
$\{a \in Y^\nu \} \cap E_{\rm piv}
\cap \event{\{a, a'\} \textnormal{ is occupied }}$.
To see that $E_\nu '$ occurs, consider first $\nu=1$.
Given that $\{ E_{1}'$ occurs on 
$\tilde{C}^{\{a,a'\}}(A_n) \}$,
the only way that $E_1'$ could fail to occur would be if
$v_{n-1} \conn u'_{n}$ through $C_{n-1}$, or if
$v_{n-1} \conn G$ in $\Zd \backslash C_{n-1}$, 
or if it were not the case that $v_n' \dbc u_n$ through $C_{n-1}$, or
if the last sausage were connected to $G$.  However, none of these
connections can be made to occur by the connections present due
to the event $y \nc G$ in $\Zd\backslash \tilde{C}^{\{a,a'\}}(A_n)$,
on account of $E_{\rm piv}$.
Similarly, for $\nu=2$ or $\nu =3$, 
given that $\{ E_{\nu}'$ occurs on 
$\tilde{C}^{\{a,a'\}}(A_n) \}$, the pivotal nature of $(a',a)$ implies
that $E_{\nu}'$ occurs.
The connections present due
to the event $y \nc G$ in $\Zd\backslash \tilde{C}^{\{a,a'\}}(A_n)$
cannot prevent the occurrence of $E_\nu'$,
on account of $E_{\rm piv}$.

This completes the proof of
\refeq{F1pf.2}, and hence of \refeq{F1pf.1}.
\qed

\subsection{Definition of $\Psi_{\vec{z}}^{(3)}$: the second
expansion}
\label{sec-Psi}

In this section, we derive an expression for $\Psi^{(3)}_{\vec{z}}(0,x,a)$.
As pointed out in \refeq{PsiPsi3}, we then have
$\Psi_{z}(0,x) = \sum_{a \in \Zd} \Psi^{(3)}_{\vec{z}}(0,x,a)$. 
We begin with \refeq{cutfocusJ}, which gives 
\eq
	\cdots \tilde{\Ebold}_{n+1} 
	I[E_{\sigma_{n+1},0}(v_{n}, u_{n+1} )_{n+1}]
	\tilde{\Ebold}_n I[J_{\nu, \sigma_{n+1}}(y)_{n}]
	\cdots 
\en
as a typical term arising in the derivative of $\Pi_z$.  
Using \refeq{J1-cut.2}, we can rewrite the above expression as
\eqarray
\lbeq{2nest1}
	&& \cdots \tilde{\Ebold}_{n+1} 
	I[E_{\sigma_{n+1},0}(v_{n}, u_{n+1} )_{n+1}]
	\tilde{\Ebold}_n I[J_{\nu, \sigma_{n+1}}' (y)_n]
	\cdots 
	\\ \nonumber && \hspace{1cm}
	+p_c \sum_{(u_{n,0},v_{n,0})}
	\cdots \tilde{\Ebold}_{n+1} 
	I[E_{\sigma_{n+1},0}(v_{n}, u_{n+1} )_{n+1}]
	\tilde{\Ebold}_n I[J_{\nu, \sigma_{n+1}}'' (u_{n,0},v_{n,0})_n]
	\tau_{\vec z}^{\tilde{C}^{\{u_{n,0},v_{n,0}\}}(A_n)}(v_{n,0},y)
	\cdots .
\enarray

The main term here is the second term, and we proceed just as
in the derivation of the first expansion in Section~\ref{sec-expansion}.
For this, we define
\eqarray
	\tilde{C}_{n,0} & = & \tilde{C}_n^{\{u_{n,0},v_{n,0}\}}(A_n), \\
	\tilde{C}_{n,m} & = & \tilde{C}_{n,m}^{\{u_{n,m},v_{n,m}\}}(u_{n,m-1}),
	\quad (m \geq 1),
\enarray
and, for $m \geq 1$,
\eqarray
\lbeq{Xnmdef}
	X_{n,m} & = &   I[E_1  (v_{n,m-1},y; \tilde{C}_{n,m-1})] 
		 -  I[E_2  (v_{n,m-1},y; \tilde{C}_{n,m-1})] , \\
\lbeq{Xnmpdef}
	X_{n,m}' & = &  I[E_1' (v_{n,m-1},y; \tilde{C}_{n,m-1})] 
		 -  I[E_2' (v_{n,m-1},y; \tilde{C}_{n,m-1})] , \\
\lbeq{Xnmppdef}
	X_{n,m}'' & = & I[E_1''(v_{n,m-1}, u_{n,m}, v_{n,m}; \tilde{C}_{n-1,m})] 
		  - I[E_2''(v_{n,m-1}, u_{n,m}, v_{n,m}; \tilde{C}_{n-1,m})].
\enarray
Then we use \refeq{tC}, which gives
\eq 
\lbeq{tCagain}
	\tau_{\vec{z}}^{\tilde{C}_{n,0}}(v_{n,0},y) 
	= \tau_{\vec{z}}(v_{n,0},y) - \Ebold_{n,1} X_{n,1}.
\en  
The second term on the right side is  
treated iteratively by employing \refeq{1Cn}, to obtain 
\eqarray
	\tau^{C} & = & \tau - \Ebold_{n,1} X_{n,1}' 
	 - \Ebold_{n,1} X_{n,1}''  \tau 
	+ \Ebold_{n,1} X_{n,1}'' \Ebold_{n,2} X_{n,2}  
	\\ & = & \nonumber 
	\tau - \Ebold_{n,1} X_{n,1}' - \Ebold_{n,1} X_{n,1}''  \tau 
	+ \Ebold_{n,1} X_{n,1}'' \Ebold_{n,2} X_{n,2}' 
	+ \Ebold_{n,1} X_{n,1}'' \Ebold_{n,2} X_{n,2}'' \tau 
	- \Ebold_{n,1} X_{n,1}'' \Ebold_{n,2} X_{n,2},
\enarray 
and so on.  It will be a consequence of bounds we obtain subsequently
that this iteration can be carried on indefinitely to obtain an
infinite series on the right side.  We then insert the result into
\refeq{2nest1}.   This leads to two kinds of terms.  The main terms
are those ending with a factor of the two-point function $\tau$,
and a secondary group of terms contain no factor $\tau$ and have a singly
primed event in their last expectation $\Ebold_{n,M}$.

Thus we obtain an identity 
\eq
\lbeq{Psi3tail}
	z_y \frac{\partial}{\partial z_y} \Pi_{\vec{z}} (0,x)
	= \sum_{a \in \Zd}\Psi^{(3)}_{\vec{z}}(0,x,a) \tau_{\vec{z}}(a,y)
\en
in which $\Psi^{(3)}_{\vec{z}}(0,x,a)$ consists of two kinds of terms.
The main terms are of the form
\eqarray
	\Psi_{\vec{z}}^{\vec{\alpha}}(0,x,a)
	&=&
	\Ebold_{0} I[E_{0}''] \Ebold_{1} I[E_{\sigma_1}'']
	\cdots \Ebold_{n-1} I[E_{\sigma_{n-1}}'']
	\tilde\Ebold_{n+1} I[E_{\sigma_{n+1},0}']
	\nonumber \\
\lbeq{PsiNM}
	&& \times \left[ \tilde\Ebold_{n} I[J_{\nu,\sigma_{n+1}}'']
	\Ebold_{n,1} I[E_{\sigma_{n,1}}''] \cdots
	\Ebold_{n,M} I[E_{\sigma_{n,M}}''] \right]
	\Ebold_{n+2} I[E_{\sigma_{n+2}}''] \cdots
	\Ebold_{N} I[E_{\sigma_{N}}''] , \hspace{1cm}
\enarray
for $N \geq 1$, $M \geq 0$.  Here $\vec{\alpha}$ represents the
dependence on $N,M,n,$ as well as $\nu$ and the $\sigma_j$.
The number of possible values for $n$, $\nu$ and the $\sigma$'s is bounded
by a constant to the power $N+M$ and is no source of concern.
The variable $x$ is equal to the variable
$v_{N}$ associated with $E_{\sigma_{N}}'$, while $a$ is the variable
$u_{n,M}$ associated with $E''_{\sigma_n,M}$.
The secondary terms are essentially
of the same form, but contain also a factor $\delta_{a,y}$,
where $a = u_{n,M}$.  This Kronecker delta neutralizes the effect
of the two-point function $\tau_z(a,y)$, so that no factor $\chi_z$
arises when \refeq{Psi3tail} is summed over $y$.  We will concentrate
in the remainder of the paper on \refeq{PsiNM}, since the other terms
involve no new ideas.

\section{Bounds on $\Psi_z^{(3)}$:
Proof of Lemmas~\protect\ref{lem-Psidisk}(i)
and \protect\ref{lem-Psi3disk}(i)}
\label{sec-Psidef2}

In this section, we prove Lemmas~\ref{lem-Psidisk}(i)
and \ref{lem-Psi3disk}(i).  These two lemmas involve bounds on
both $\Gamma$ and $\Psi$, but because these two quantities are
almost identical and can be treated using the same methods, 
we discuss only the bounds on $\Psi$.
Throughout the section, we fix $p=p_c$ and take the dimension $d$ to be large.
 
To prove the upper bounds of 
Lemma~\ref{lem-Psi3disk}(i), it is sufficient to obtain
bounds on $\sum_{x,y}  \| \Psi_z^{(3)}(0,x,y)\|$ and on similar sums
involving an additional factor $|x|^2$ or $|y|^2$ under the summation,
with these bounds uniform in complex $z$ with $|z|<1$.   We will obtain
such bounds uniform also in high $d$.  The uniformity in $z$ permits the
extension of $\hat\Psi_z^{(3)}(k,l)$ to $|z|=1$, by continuity.
The cases with $|x|^2$ or $|y|^2$
can be handled in the same way as when these factors are not present
(at the cost of requiring higher dimension), and for simplicity we will
only discuss the bound on $\sum_{x,y}  \| \Psi_z^{(3)}(0,x,y)\|$ in what
follows.  In view of \refeq{PsiPsi3}, having such bounds uniform in $d$
then implies the upper bounds of Lemma~\ref{lem-Psidisk}(i).
Also by \refeq{PsiPsi3}, the identity 
$\hat{\Psi}_z^{(3)}(0,0) = \hat{\Psi}_z(0)$
follows.  This will complete the proof of Lemma~\ref{lem-Psi3disk}(i).
We will then complete the proof of Lemma~\ref{lem-Psidisk}(i) by showing
that $\hat{\Psi}_1(0)= K_1 + K_2$,  where $K_1$ and $K_2$ are the
constants of Propositions~I.5.1 
and I.5.2.  
These two propositions then imply that 
$\lim_{d \to \infty} \hat{\Psi}_1(0) =1$,
which implies the lower bound on $\hat{\Psi}_1(0)$ of 
Lemma~\ref{lem-Psidisk}(i).

To obtain the required bound on $\sum_{x,y}  \| \Psi_z^{(3)}(0,x,y)\|$,
it suffices to obtain a bound on the summed norm of \refeq{PsiNM}, namely 
$\sum_{x,y} \|\Psi_{\vec{z}}^{\vec{\alpha}}(0,x,y) \|$,
of the form $c^{N+M} d^{-(N+M)}$  
with $c$ independent of $d$.  The norm is handled exactly as in 
the proof of Lemma~\ref{lem-Pinbd}.  In that proof, the $z$-dependence
of an expectation involving an $E_1$ event was noted to be of the form
$z^{|Y^1|}$, while that of an $E_2$ event was written in the form
$(1-z^{|Y^3|})z^{|Y^2|} = z^{|Y^2|}-z^{|Y^2|+|Y^3|}$.  
However, before using this inequality, we insert a factor 
$I[Y^3 \neq \emptyset]$
to ensure that important bond connections are not lost --- this led to
the condition (last sausage of $v \conn x) \conn A$ in \refeq{E2pb-def}.
Explicitly, for $\sigma =1,2$ the bond connections within the
event $E_\sigma(v_{k-1}, u_k,v_k;A)$ are included in the event
\eq
\lbeq{EbdwA}
	\{ \exists w \in A : (v_{k-1} \conn u_k) \circ (u_k \conn w)\}.
\en
Analogous expressions
can be written for the $z$-dependence of the expectation involving
the event $J_{\nu,\sigma_{n+1}}''$.  
We then multiply out all these $z$-dependent factors and
bound the resulting sum term by term, setting $z=1$ for an upper bound.
This introduces a harmless factor $2^{N+M}$ in the bounds. 
We are left with bounding a nested bond expectation.  

To bound such
a nested expectation, we proceed as in Section~I.5, 
although
here there is the simplification that we are content to obtain bounds
valid in sufficiently high dimensions, rather than for all $d>6$,
and we can therefore employ diagrams having higher critical dimension
than $6$, such as the square diagram.  For a discussion of the notion
of the critical dimension of a diagram, as well as general power counting
techniques for bounding diagrams, see Appendix~I.A. 
We will make use of the constructions and terminology of 
Appendix~I.A 
in what follows.

As a very rough bound on a contribution to 
$\|\Psi_{\vec{z}}^{\vec{\alpha}}(0,x,y) \|$, after having dealt with the
$z$-dependence as outlined above,
we can bound all the nested expectations at levels-$(n,1)$ to $(n,M)$ of 
\refeq{PsiNM} by $1$.  Glancing at
\refeq{Esigdecomp}--\refeq{Jcut2def} and \refeq{Jnusigdef}, it is
apparent that the important connections that were originally present in the
events $E_{\nu}'$ and $E_{\sigma_{n+1}}'$
remain present.  Thus a crude upper bound would
be to use the diagrams employed to bound $\hat{\Pi}_z(0,x)$ also for
$\Psi_z^{(3)}(0,x,y)$.  However, this crude bound is completely inadequate,
as a sum over $y \in \Zd$ will lead to a divergent volume factor.
We require better bounds, summable in $y$.  

For this, we examine the additional bond connections that are required
to be present both by $J_{\nu,\sigma_{n+1}}''$ and by the events occuring
at levels-$(n,1)$ to $(n,M)$.
First we note that $J_{\nu,\sigma_{n+1}}''$ additionally requires the
existence of disjoint connections from $u_{n,0}$ to the set $A_n$.
For the case $M=0$, we would have $u_{n,0}=y$.  Diagrammatically, we
would therefore have an additional pair of lines, with one
leading from $y$ to level-$n$ and another leading to  
somewhere on levels-$(n-1)$, $n$, or $(n+1)$.  In the worst case, this will
correspond to two applications of construction~1 and one application of
construction~2 of Appendix~I.A, 
and hence leads to a convergent diagram in sufficiently high dimensions.
Explicitly, the line from $y$ to level-$n$ terminates at 
an additional vertex on a
level-$n$ line.  Also, a connection from $y$ to level-$(n-1)$ 
(say) entails a level-$n$ path from $y$ to some (new)
vertex on level-$(n-1)$, which
itself would be connected generically within level-$(n-1)$
to the previously existing 
level-$(n-1)$ lines by a new line joined to a new vertex.  
Overall, these three new lines plus three new vertices correspond to
construction~1 followed by construction~2 followed by construction~1.

The case $M \geq 1$ can be handled in a similar fashion.  The event at
level-$(n,1)$ is special, due to the definition of $\tilde{C}_{n,0}$.
We use \refeq{EbdwA} as our upper bound, but here 
$A=\tilde{C}_{n,0}= \tilde{C}_{n,0}^{\{u_{n,0},v_{n,0}\}}(A_n)$.
This implies disjoint connections $\{u_{n,1} \conn v_{n,0}\} \circ \{v_{n,0}
\conn w\}$, with $w \in \tilde{C}_{n,0}^{\{u_{n,0},v_{n,0}\}}(A_n)$.  
This latter connection can be bounded by
a level-$n$ connection from $w$ to some $w'$ lying on one of the diagrammatic
lines corresponding to level-$(n-1)$, $n$, or $(n+1)$.  This corresponds
to performing construction~2 followed by two applications of construction~1.  

For $M >1$, the connections at levels-$(n,m)$ for $2 \leq m \leq M$ lead
to connections like those encountered in the diagrammatic estimates on
$\Pi$, and there is nothing new to comment on.  A typical diagram,
which would arise in bounding 
\eq
\lbeq{E42}
	\Ebold_{0} I[E_0''] \Ebold_{1} I[E_{\sigma_1}''] 
	\tilde\Ebold_{3} I[E_{\sigma_3,0}' ] 
	\left[ \tilde\Ebold_{2} I[J_{\nu,\sigma_{3}}''] 
	\Ebold_{2,1}I[E_{\sigma_{2,1}}'']
	\Ebold_{2,2}I[E_{\sigma_{2,2}}''] \right]
	\Ebold_{4} I[E_{\sigma_4}'' ],
\en
is
\begin{center}
\setlength{\unitlength}{0.000400in}%
\begingroup\makeatletter\ifx\SetFigFont\undefined
\def\x#1#2#3#4#5#6#7\relax{\def\x{#1#2#3#4#5#6}}%
\expandafter\x\fmtname xxxxxx\relax \def\y{splain}%
\ifx\x\y   
\gdef\SetFigFont#1#2#3{%
  \ifnum #1<17\tiny\else \ifnum #1<20\small\else
  \ifnum #1<24\normalsize\else \ifnum #1<29\large\else
  \ifnum #1<34\Large\else \ifnum #1<41\LARGE\else
     \huge\fi\fi\fi\fi\fi\fi
  \csname #3\endcsname}%
\else
\gdef\SetFigFont#1#2#3{\begingroup
  \count@#1\relax \ifnum 25<\count@\count@25\fi
  \def\x{\endgroup\@setsize\SetFigFont{#2pt}}%
  \expandafter\x
    \csname \romannumeral\the\count@ pt\expandafter\endcsname
    \csname @\romannumeral\the\count@ pt\endcsname
  \csname #3\endcsname}%
\fi
\fi\endgroup
\begin{picture}(4737,3999)(976,-4048)
\thicklines
\put(3001,-886){\line( 0, 1){825}}
\put(3001,-61){\line( 1, 0){1200}}
\thinlines
\put(4201,-961){\line( 0,-1){900}}
\put(4201,-61){\line( 0,-1){900}}
\put(3001,-961){\line( 1, 0){1200}}
\thicklines
\put(4201,-1861){\line( 0,-1){900}}
\thinlines
\put(3001,-1786){\line( 0, 1){825}}
\thicklines
\put(3001,-1861){\line( 0,-1){900}}
\put(3601,-2761){\line( 0,-1){1200}}
\put(2476,-2761){\line( 1, 0){1125}}
\put(3601,-3961){\line(-1, 0){600}}
\thinlines
\put(3001,-3961){\line(-1, 0){600}}
\put(2401,-3961){\line( 0, 1){1200}}
\put(2401,-2761){\line(-1, 0){825}}
\put(2401,-3961){\line(-1, 0){450}}
\thicklines
\put(1951,-3961){\line(-1, 0){450}}
\put(1501,-3961){\line( 0, 1){1200}}
\put(1501,-2761){\line(-1,-2){300}}
\put(1201,-3361){\line( 1,-2){300}}
\thinlines
\put(3676,-3961){\line( 1, 0){1125}}
\put(4801,-3961){\line( 0, 1){1200}}
\put(4801,-2761){\line(-1, 0){300}}
\thicklines
\put(4501,-2761){\line(-1, 0){900}}
\put(3751,-2761){\line(-5, 6){750}}
\put(4876,-2761){\line( 1, 0){825}}
\put(5701,-2761){\line( 0,-1){1200}}
\thinlines
\put(5701,-3961){\line(-1, 0){900}}
\thicklines
\put(1501,-2686){\line( 0,-1){150}}
\put(1576,-2686){\line( 0,-1){150}}
\put(2401,-2686){\line( 0,-1){150}}
\put(2476,-2686){\line( 0,-1){150}}
\put(3601,-4036){\line( 0, 1){150}}
\put(3676,-4036){\line( 0, 1){150}}
\put(4801,-2686){\line( 0,-1){150}}
\put(4876,-2686){\line( 0,-1){150}}
\put(2926,-1861){\line( 1, 0){150}}
\put(2926,-1786){\line( 1, 0){150}}
\put(2926,-961){\line( 1, 0){150}}
\put(2926,-886){\line( 1, 0){150}}
\put(876,-3361){\makebox(0,0)[lb]{\smash{\SetFigFont{12}{14.4}{\rmdefault}{\mddefault}{\updefault}$0$}}}
\put(5871,-2901){\makebox(0,0)[lb]{\smash{\SetFigFont{12}{14.4}{\rmdefault}{\mddefault}{\updefault}$x$}}}
\put(2701,-61){\makebox(0,0)[lb]{\smash{\SetFigFont{12}{14.4}{\rmdefault}{\mddefault}{\updefault}$y$}}}
\put(1951,-3961){\circle*{150}}
\put(3001,-3961){\circle*{150}}
\put(4501,-2761){\circle*{150}}
\put(4201,-1861){\circle*{150}}
\end{picture}
\end{center}
In the diagram, thick lines are used for connections arising from
expectations at levels-0, 2, 4, (2,2), and thin lines are used for
levels-1, 3, (2,1).
Standard diagrammatic estimates then lead to a bound of the required 
form $c^{N+M}d^{-(N+M)}$ in the general case, with $c$ independent of $d$.
For example, the sum over $x,y$ of the 
above diagram can be bounded by a product of two triangles,
two squares and two pentagons, all of which are $O(d^{-1})$ due to the
presence of pivotal bonds, times a product of two $O(1)$ triangles from
the two loops without pivotal bonds.  There are combinatorial factors 
associated with the number of ways that diagrams can be drawn, but these
can be absorbed into the constant $c^{N+M}$.

Finally, we come to the proof that $\hat{\Psi}_1(0)= K_1 + K_2$,  
where $K_1$ and $K_2$ are the
constants of Propositions~I.5.1 
and I.5.2. 
For this, we first note that $\hat{\Psi}_z(0) = \hat{\Psi}_1(0) +o_z(1)$.
Multiplying this equation by $\chi_z$ and then integrating with respect
to $z$ over the interval $[z,1]$ gives 
\eq
	1-\hat{\Pi}_z(0) = \hat{\Psi}_1(0) M_z [1 + o_z(1)],
\en
which in turn gives
\eq
	\hat{\tau}_z(0) = \frac{\hat{g}_z(0)}{1-\hat{\Pi}_z(0)}
	= \frac{\hat{g}_1(0)}{\hat{\Psi}_1(0) M_z} [1+o_z(1)].
\en
Comparing with (I.5.7), 
and using $\hat{g}_1(0) = \hat{\phi}_{h=0}(0)$,
we conclude that $\hat{\Psi}_1(0) = K_1+K_2$.

\section{Bounds on $\frac{d}{dz}\Psi_z^{(3)}$:  Proof of
Lemmas~\protect\ref{lem-Psidisk}(ii)
and \protect\ref{lem-Psi3disk}(ii)}
\label{sec-Psider}

Throughout this section, we fix $p=p_c$ and take the dimension $d$
to be large.  We will focus on bounding the $z$-derivative of $\Psi$
and omit any discussion of $\Gamma$, which can be handled with the same
methods.  
We will prove the bound
\eq 
\lbeq{Psi1bd5}
        \left\| \frac{d}{dz} \hat{\Psi}_{z,p}^{(3)}(k,l) \right\|
        \leq K \chi_{|z|}(p), \quad |z| < 1
\en 
of Lemma~\ref{lem-Psi3disk}(ii), with $K$ independent of $d$.  
In view of \refeq{PsiPsi3}, the bound
of Lemma~\ref{lem-Psidisk}(ii) then follows immediately.  

We focus our discussion on 
the $z$-derivative of $\Psi_{z}^{\vec{\alpha}}(0,x,y)$ of \refeq{PsiNM}.
The $z$-dependence of $\Psi_{z}^{\vec{\alpha}}(0,x,y)$ 
resides in the $z$-dependence
of each of the (doubly) nested expectations involved in its definition.
To differentiate, we apply Leibniz's rule as we did
in the differentiation of $\Pi$.  This leads to a sum over $\ell$,
with the derivative applied to the $\ell^{\rm th}$ expectation, 
where $\ell = 1, \ldots,n-1, n+1, n, (n,1), \ldots, (n,M),
n+2, \ldots N$.   In each case, the derivative is applied to an expression
involving $z^{|Q|}$ for some cluster of sites $Q$, and again this will lead
to a factor $|Q|z^{|Q|}=\sum_{w \in \Zd} I[w \in Q]z^{|Q|}$.  
Diagrammatically, this corresponds
to a $G$-free ``tail'' leading from a new vertex on
the diagram corresponding to 
$\Psi_{z}^{\vec{\alpha}}(0,x,y)$ to the site $w$.  This time, there is no
need to perform an expansion to cut off this tail, and a bound will
suffice.  
The procedure is similar in spirit to that used in the proof of
Lemma~\ref{lem-UNbd}.
The tail will give rise to the factor $\chi_{|z|}$ of
\refeq{Psi1bd5}.  The remaining diagram gives rise to the constant,
after summation over the variables in $\vec{\alpha}$, including
summation over $N,M$.    

Although conceptually straightforward, the complicated nature of the
definition of $\Psi_{z}^{\vec{\alpha}}(0,x,y)$ makes a detailed proof
quite lengthy.  We therefore content ourselves here with describing
the salient features only, omitting most details.
We will consider the following cases in turn:
(i)  $\ell \in \{0,\ldots,n-2, (n,2), \ldots, (n,M),
n+2, \ldots N \}$,
(ii) $\ell = (n,1)$, (iii) $\ell = n-1$, (iv) $\ell \in \{n,n+1\}$.

\subsubsection*{Case (i):  $\ell \in \{0,\ldots,n-2, (n,2), \ldots, (n,M),
n+2, \ldots N \}$ } 

When the $z$-derivative is applied to the expectation at one of
these levels $\ell$, prior to differentiation 
the $z$-dependence is identical to that
encountered in Section~\ref{sec-Psidiff} for the $z$-dependence of a typical
level of the nested expectations in $\Pi_z$.  We perform the 
differentiation exactly as in Section~\ref{sec-Psidiff}, and choose the
cutting bond in the same way we did in Section~\ref{sec-secondcut}, namely
the last pivotal bond for the connection from $w$ to $\{v_{\ell-1}, u_\ell \}
\cup C_{\ell-1} \cup C_{\ell+1}$.
This choice requires an application of Fubini's Theorem to interchange
the expectations at levels-$\ell$ and $\ell +1$.  For the $\ell$'s
being considered, there is no interference from the expectations that
were modified at levels-$n$ and $n+1$ by the previous differentiation,
and the two differentations do not interfere with each other.
There is no need now to perform an expansion, as our goal is a bound
rather than an identity.  To obtain the desired bound, we will
use the cut-the-tail Lemma~\ref{lem-tail.tau1}.  This requires 
treatment of two issues:  the cut-the-tail lemma applies
only to increasing events, and we are dealing here with complex $z$
and must take the norm.  

We first discuss the issue of the norm.  We proceed exactly as described in 
Section~\ref{sec-Psidef2}, with one exception.  The exception is that
we retain the factor $|z|^{|Q|}$ in bounding the factor 
$|Q|\, |z|^{|Q|} = \sum_{w \in \Zd} I[w \in Q]|z|^{|Q|}$ described above.
We choose a cutting bond $(b,b')$ 
exactly as was done in differentiating $\Pi$
in Section~\ref{sec-secondcut}.  Let $T$ denote the ``tail,'' {\it i.e.}, the
subset of $Q$ which is on the $w$ side of the cutting bond.  We then
use the estimate $|z|^{|Q|} \leq |z|^{|T|}$,
which is then the overall $z$-dependence of the upper bound.  This 
$z$-dependence will allow us to cut off a factor $\tau_{|z|}(b',w)$.

To apply the cut-the-tail Lemma~\ref{lem-tail.tau1}, we proceed
as in Section~\ref{sec-Psidef2} and obtain an increasing event by
demanding only that certain bond connections occur.  To illustrate,
a diagram arising in differentiating the level-$4$ expectation of
\refeq{E42} is
\begin{center}
\setlength{\unitlength}{0.000400in}%
\begingroup\makeatletter\ifx\SetFigFont\undefined
\def\x#1#2#3#4#5#6#7\relax{\def\x{#1#2#3#4#5#6}}%
\expandafter\x\fmtname xxxxxx\relax \def\y{splain}%
\ifx\x\y   
\gdef\SetFigFont#1#2#3{%
  \ifnum #1<17\tiny\else \ifnum #1<20\small\else
  \ifnum #1<24\normalsize\else \ifnum #1<29\large\else
  \ifnum #1<34\Large\else \ifnum #1<41\LARGE\else
     \huge\fi\fi\fi\fi\fi\fi
  \csname #3\endcsname}%
\else
\gdef\SetFigFont#1#2#3{\begingroup
  \count@#1\relax \ifnum 25<\count@\count@25\fi
  \def\x{\endgroup\@setsize\SetFigFont{#2pt}}%
  \expandafter\x
    \csname \romannumeral\the\count@ pt\expandafter\endcsname
    \csname @\romannumeral\the\count@ pt\endcsname
  \csname #3\endcsname}%
\fi
\fi\endgroup
\begin{picture}(4737,3999)(976,-4048)
\thicklines
\put(3001,-886){\line( 0, 1){825}}
\put(3001,-61){\line( 1, 0){1200}}
\thinlines
\put(4201,-61){\line( 0,-1){1800}}
\put(3001,-961){\line( 1, 0){1200}}
\thicklines
\put(4201,-1861){\line( 0,-1){900}}
\thinlines
\put(3001,-1786){\line( 0, 1){825}}
\thicklines
\put(3001,-1861){\line( 0,-1){900}}
\put(2476,-2761){\line( 1, 0){1125}}
\put(3601,-2761){\line( 0,-1){1200}}
\put(3601,-3961){\line(-1, 0){600}}
\thinlines
\put(3001,-3961){\line(-1, 0){600}}
\put(2401,-3961){\line( 0, 1){1200}}
\put(2401,-2761){\line(-1, 0){825}}
\put(2401,-3961){\line(-1, 0){450}}
\thicklines
\put(1951,-3961){\line(-1, 0){450}}
\put(1501,-3961){\line( 0, 1){1200}}
\put(1501,-2761){\line(-1,-2){300}}
\put(1201,-3361){\line( 1,-2){300}}
\thinlines
\put(3676,-3961){\line( 1, 0){1125}}
\put(4801,-3961){\line( 0, 1){1200}}
\put(4801,-2761){\line(-1, 0){300}}
\thicklines
\put(4501,-2761){\line(-1, 0){900}}
\put(3751,-2761){\line(-5, 6){750}}
\put(4876,-2761){\line( 1, 0){825}}
\put(5701,-2761){\line( 0,-1){1200}}
\put(5701,-2761){\line( 1,-2){300}}
\put(5701,-3961){\line( 1, 2){300}}
\put(5701,-3961){\line(-1, 0){400}}
\put(6076,-3361){\line( 1, 0){900}}
\thinlines
\put(5301,-3961){\line(-1, 0){500}}
\thicklines
\put(6001,-3436){\line( 0, 1){150}}
\put(6076,-3436){\line( 0, 1){150}}
\put(1501,-2686){\line( 0,-1){150}}
\put(1576,-2686){\line( 0,-1){150}}
\put(2401,-2686){\line( 0,-1){150}}
\put(2476,-2686){\line( 0,-1){150}}
\put(3601,-4036){\line( 0, 1){150}}
\put(3676,-4036){\line( 0, 1){150}}
\put(4801,-2686){\line( 0,-1){150}}
\put(4876,-2686){\line( 0,-1){150}}
\put(2926,-1861){\line( 1, 0){150}}
\put(2926,-1786){\line( 1, 0){150}}
\put(2926,-961){\line( 1, 0){150}}
\put(2926,-886){\line( 1, 0){150}}
\put(876,-3361){\makebox(0,0)[lb]{\smash{\SetFigFont{12}{14.4}{\rmdefault}{\mddefault}{\updefault}$0$}}}
\put(5301,-2601){\makebox(0,0)[lb]{\smash{\SetFigFont{12}{14.4}{\rmdefault}{\mddefault}{\updefault}$x$}}}
\put(2701,-61){\makebox(0,0)[lb]{\smash{\SetFigFont{12}{14.4}{\rmdefault}{\mddefault}{\updefault}$y$}}}
\put(7101,-3361){\makebox(0,0)[lb]{\smash{\SetFigFont{12}{14.4}{\rmdefault}{\mddefault}{\updefault}$w$}}}
\put(1951,-3961){\circle*{150}}
\put(3001,-3961){\circle*{150}}
\put(4501,-2761){\circle*{150}}
\put(4201,-1861){\circle*{150}}
\put(5301,-2761){\circle*{150}}
\put(5301,-3961){\circle*{150}}
\end{picture}
\end{center}
(There are other possible topologies, but they can be handled similarly.)
The diagram line terminating at $w$ corresponds to a factor $\tau_{|z|}(b',w)$.
Summation over $w$ leads to a factor $\chi_{|z|}$.  The remaining diagram
can be bounded by an $O(1)$ triangle times the diagram encountered previously
in Section~\ref{sec-Psidef2} with an additional added vertex.
The added vertex makes no difference, since $d$ is large.
In this way, all diagrams arising in case (i) can be handled.

\subsubsection*{Case (ii):  $\ell =(n,1)$}

This case is identical to case (i), apart from the fact that the
set $\tilde{C}_{n,0}$ is of a different form.  However, this is irrelevant
for the differentiation and the procedure is as in case (i).  An example
of a diagram contributing to the derivative of the expectation at 
level-$(2,1)$ of \refeq{E42} is 
\begin{center}
\setlength{\unitlength}{0.000400in}%
\begingroup\makeatletter\ifx\SetFigFont\undefined
\def\x#1#2#3#4#5#6#7\relax{\def\x{#1#2#3#4#5#6}}%
\expandafter\x\fmtname xxxxxx\relax \def\y{splain}%
\ifx\x\y   
\gdef\SetFigFont#1#2#3{%
  \ifnum #1<17\tiny\else \ifnum #1<20\small\else
  \ifnum #1<24\normalsize\else \ifnum #1<29\large\else
  \ifnum #1<34\Large\else \ifnum #1<41\LARGE\else
     \huge\fi\fi\fi\fi\fi\fi
  \csname #3\endcsname}%
\else
\gdef\SetFigFont#1#2#3{\begingroup
  \count@#1\relax \ifnum 25<\count@\count@25\fi
  \def\x{\endgroup\@setsize\SetFigFont{#2pt}}%
  \expandafter\x
    \csname \romannumeral\the\count@ pt\expandafter\endcsname
    \csname @\romannumeral\the\count@ pt\endcsname
  \csname #3\endcsname}%
\fi
\fi\endgroup
\begin{picture}(4737,3999)(976,-4048)
\thicklines
\put(3001,-886){\line( 0, 1){825}}
\put(3001,-61){\line( 1, 0){1200}}
\thinlines
\put(4201,-961){\line( 0,-1){900}}
\put(4201,-61){\line( 0,-1){900}}
\put(3001,-961){\line( 1, 0){1200}}
\thicklines
\put(4201,-1861){\line( 0,-1){900}}
\thinlines
\put(3001,-1786){\line( 0, 1){825}}
\thicklines
\put(3001,-1861){\line( 0,-1){900}}
\put(3601,-2761){\line( 0,-1){1200}}
\put(2476,-2761){\line( 1, 0){1125}}
\put(3601,-3961){\line(-1, 0){600}}
\thinlines
\put(3001,-3961){\line(-1, 0){600}}
\put(2401,-3961){\line( 0, 1){1200}}
\put(2401,-2761){\line(-1, 0){825}}
\put(2401,-3961){\line(-1, 0){450}}
\thicklines
\put(1951,-3961){\line(-1, 0){450}}
\put(1501,-3961){\line( 0, 1){1200}}
\put(1501,-2761){\line(-1,-2){300}}
\put(1201,-3361){\line( 1,-2){300}}
\thinlines
\put(3676,-3961){\line( 1, 0){1125}}
\put(4801,-3961){\line( 0, 1){1200}}
\put(4801,-2761){\line(-1, 0){300}}
\thicklines
\put(4501,-2761){\line(-1, 0){900}}
\put(3751,-2761){\line(-5, 6){750}}
\put(4876,-2761){\line( 1, 0){825}}
\put(5701,-2761){\line( 0,-1){1200}}
\thinlines
\put(5701,-3961){\line(-1, 0){900}}
\thicklines
\put(1501,-2686){\line( 0,-1){150}}
\put(1576,-2686){\line( 0,-1){150}}
\put(2401,-2686){\line( 0,-1){150}}
\put(2476,-2686){\line( 0,-1){150}}
\put(3601,-4036){\line( 0, 1){150}}
\put(3676,-4036){\line( 0, 1){150}}
\put(4801,-2686){\line( 0,-1){150}}
\put(4876,-2686){\line( 0,-1){150}}
\put(2926,-1861){\line( 1, 0){150}}
\put(2926,-1786){\line( 1, 0){150}}
\put(2926,-961){\line( 1, 0){150}}
\put(2926,-886){\line( 1, 0){150}}
\put(876,-3361){\makebox(0,0)[lb]{\smash{\SetFigFont{12}{14.4}{\rmdefault}{\mddefault}{\updefault}$0$}}}
\put(5871,-2901){\makebox(0,0)[lb]{\smash{\SetFigFont{12}{14.4}{\rmdefault}{\mddefault}{\updefault}$x$}}}
\put(2701,-61){\makebox(0,0)[lb]{\smash{\SetFigFont{12}{14.4}{\rmdefault}{\mddefault}{\updefault}$y$}}}
\put(1951,-3961){\circle*{150}}
\put(3001,-3961){\circle*{150}}
\put(4501,-2761){\circle*{150}}
\put(4201,-1861){\circle*{150}}
\thinlines
\put(4201,-1411){\line( 1, 0){900}}
\put(5101,-1411){\line( 0,-1){900}}
\put(5176,-1411){\line( 1, 0){825}}
\thicklines
\put(4201,-2311){\line( 1, 0){900}}
\put(5101,-1336){\line( 0,-1){150}}
\put(5176,-1336){\line( 0,-1){150}}
\put(6201,-1411){\makebox(0,0)[lb]{\smash{\SetFigFont{12}{14.4}{\rmdefault}{\mddefault}{\updefault}$w$}}}
\end{picture}
\end{center}
and the methods of case (i) apply.

\subsubsection*{Case (iii):  $\ell =n-1$} 

In differentiating the expectation at level-$(n-1)$, the $z$-dependence
is the same as in $\Pi_z$, and the differentiation proceeds as in
Section~\ref{sec-Psidiff}.  To choose the cutting bond, we interchange
expectations using Fubini's Theorem, but because of the interchange
of the expectations at levels-$n$ and $n+1$ already performed in 
\refeq{E42}, we interchange expectations twice to put the expectations
in the order 
$\tilde{\Ebold}_{n+1} \tilde{\Ebold}_n \tilde{\Ebold}_{n-1}$.  
Then the choice
of cutting bond is as in Section~\ref{sec-secondcut}, namely we choose
the last pivotal bond for the connection from $w$ to 
$\{v_{n-2},u_{n-1}\} \cup C_{n-2} \cup C_n$.  Then the methods of case (i)
apply.

\subsubsection*{Case (iv):  $\ell \in \{ n,n+1 \}$ } 

To prepare for the case where the differentation falls on level-$n$ or
level-$(n+1)$, we write \refeq{PsiNM} in the form
\eqarray
	\Psi_{\vec{z}}^{\vec{\alpha}}(0,x,a)
	&=&
	\Ebold_{0} I[E_{0}''] \Ebold_{1} I[E_{\sigma_1}'']
	\cdots \Ebold_{n-1} I[E_{\sigma_{n-1}}'']
	\tilde\Ebold_{n+1} I[E_{\sigma_{n+1},0}']
	\nonumber \\ 
\lbeq{PsiNM8}
	&& \times \left[ \tilde{\tilde\Ebold}_{n} I[J_{\nu,\sigma_{n+1}}']
	\tilde\Ebold_{n,1} I[E_{\sigma_{n,1}}'] \cdots
	\Ebold_{n,M} I[E_{\sigma_{n,M}}''] \right]
	\\ \nonumber
	&& \times \;
	\tilde\Ebold_{n+2} I[E_{\sigma_{n+2}}'] 
	\Ebold_{n+3} I[E_{\sigma_{n+3}}'']\cdots
	\Ebold_{N} I[E_{\sigma_{N}}''] , 
\enarray 
where the additional tilde at level-$n$ denotes the expectation conditional
also on $\{u_0,v_0\}$ being vacant.  By \refeq{Esigdecomp} and
\refeq{Jnusigdef},
\eq
	E_{\sigma_{n+1},0}' \cap J_{\nu,\sigma_{n+1}}'
	= E_{\sigma_{n+1}}' \cap E_\nu' \cap \{ y \in Y^\nu\}
	\cap \{y \dbc A_n\}.
\en
This entails the usual $z$-dependence at level-$n+1$, while at
level-$n$ the $z$-dependence is as usual for $\nu=1,2$ and is
$z^{|Y^1|}$ for $\nu=3$.  We can then differentiate without difficulty.

We must then choose the cutting bond.  When the differentiation has fallen
on level-$n$, we apply Fubini as before to interchange the expectations
at levels-$n$ and $(n,1)$.  We then choose the cutting bond as the last
pivotal bond for the connection from $w$ to $\{v_{n-1},u_n\} \cup C_{n-1}
\cup C_{n+1} \cup C_{(n,1)}$, and the bound proceeds as in the previous
cases.

When the differentiation has fallen on level-$(n+1)$, we apply Fubini as
usual to move $\tilde\Ebold_{n+2}$ to the left of $\tilde\Ebold_{n+1}$.
We would like also to move the level-$n$ expectation to the left of
$\Ebold_{n+1}$, so as to have the bond configuration at level-$n$ fixed
for the definition of the cutting bond.  To accomplish this, some care
is needed with the event $\{y \dbc A_n\}$, which cannot be moved to the
left of $\Ebold_{n+1}$ because it involves $C_{n+1}$.  However, we
can write this event as the intersection of the level-$n$ event
$\{y \dbc (\{v_{n-1},u_n\} \cup C_{n-1})\}_n$ and the level-$(n+1)$ event
that the bond connections of $C_{n+1}$ are such that $y \dbc A_n$.
Having done so, we can define the cutting bond as the last pivotal bond
for the connection from $w$ to $\{v_n ,u_{n+1}\} \cup C_{n} \cup C_{n+2}$.
The bound then proceeds as in the previous cases.


\section*{Acknowledgements}
This work was supported in part by NSERC. 
It was carried out in part during extensive visits by both authors
to the University of British Columbia in 1997, and to
Microsoft Research and the Fields Institute in 1998.
The work of G.S.\ was also supported in part by an
Invitation Fellowship of the Japan Society for the Promotion of Science,
during a visit to Tokyo Institute of Technology in 1996.
We thank Michael Aizenman 
and Ed Perkins for stimulating conversations and correspondence.

\bibliography{bib}           
\bibliographystyle{plain}

\end{document}